\documentclass[aps,pre,twocolumn,showpacs,floatfix,superscriptaddress]{revtex4-1}
\usepackage{graphicx,amsfonts,amssymb,amsmath,hyperref}
\usepackage{textcomp}
\usepackage{esint}

\newif\ifhyper
\hypertrue
\ifhyper       
\hypersetup{        
  citecolor = {green},
  colorlinks = {true}, 
  urlcolor = {blue} 
}

\begin{document}

\def\Li{\mbox{Li}} 
\def\Il{I_{k,\rm l}}
\def\It{I_{k,\rm t}}
\def\Jtt{J_{k,\rm tt}}
\def\Jll{J_{k,\rm ll}}
\def\Jtl{J_{k,\rm tl}}
\def\Jlt{J_{k,\rm lt}}
\def\tIl{\tilde I_{k,\rm l}}
\def\tIt{\tilde I_{k,\rm t}}
\def\tJtt{\tilde J_{k,\rm tt}}
\def\tJll{\tilde J_{k,\rm ll}}
\def\tJtl{\tilde J_{k,\rm tl}}
\def\tJlt{\tilde J_{k,\rm lt}}
\def\Tkt{T_{\rm BKT}}


\graphicspath{{./figures/}}

\def\rhoeq{\hat\rho_{\rm eq}}

\newcommand{\marge}[1]{\marginpar{\scriptsize #1}}
\newcommand{\remarque}[1]{\marginpar{\scriptsize Remarque}{\it [#1]}}

\def\beq{\begin{equation}}
\def\eeq{\end{equation}}
\def\bleq{\begin{eqnarray}}
\def\eleq{\end{eqnarray}} 
\def\bfig{\begin{figure}}
\def\efig{\end{figure}}
\def\bline{\begin{multline}}
\def\eline{\end{multline}}
\def\bremark{\begin{quotation} \noindent \small }
\def\eremark{\end{quotation}}
\def\llbrace{\left\lbrace}
\def\rrbrace{\right\rbrace}
\def\lbraket{\left[}
\def\rbraket{\right]}

\newcommand{\Tr}{{\rm Tr}} 
\newcommand{\tr}{{\rm tr}} 
\newcommand{\sgn}{{\rm sgn}} 
\newcommand{\mean}[1]{\langle #1 \rangle}
\newcommand{\commu}[2]{[#1,#2]} 
\newcommand{\bra}[1]{\langle#1|}
\newcommand{\ket}[1]{|#1\rangle}
\newcommand{\braket}[2]{\langle #1|#2\rangle}
\newcommand{\dbraket}[3]{\langle #1|#2|#3\rangle}
\newcommand{\tens}[1]{\overleftrightarrow{#1}}  
\newcommand{\vac}{|{\rm vac}\rangle} 
\def\bravac{\langle{\rm vac}|}
\newcommand{\const}{{\rm const}} 
\newcommand{\atanh}{\,{\rm atanh}}

\newcommand{\ie}{i.e. }
\newcommand{\iet}{i.e.}
\newcommand{\eg}{e.g. }
\newcommand{\cc}{{\rm c.c.}} 
\newcommand{\hc}{{\rm h.c.}} 
\def\etal{{\it et al. }}

\newcommand{\jhatbf}{\hat {\textbf \j}} 
\newcommand{\Jhatbf}{\hat {\textbf \J}} 
\newcommand{\jhat}{\hat {\jmath}} 
\newcommand{\Jhat}{\hat {J}} 
\newcommand{\jbf}{\textbf j}
\newcommand{\Jbf}{\textbf J}

\def\chibf{\boldsymbol{\chi}}
\def\down{\downarrow}
\def\eps{\epsilon}
\def\gam{\gamma} 
\def\phibf{\boldsymbol{\phi}}
\def\varphibf{\boldsymbol{\varphi}}
\def\varphibfs{\boldsymbol{\varphi}_<}
\def\varphibfl{\boldsymbol{\varphi}_>}
\def\varphis{\varphi_{<}}
\def\varphil{\varphi_{>}}
\def\psibf{\boldsymbol{\psi}}
\def\Ome{\Omega}
\def\omeD{{\omega_D}} 
\def\bfOme{\boldsymbol{\Omega}} 
\def\Omebf{\boldsymbol{\Omega}} 
\def\lamb{\lambda}
\def\Lamb{\Lambda}
\def\sig{\sigma}
\def\sigp{{\sigma'}} 
\def\bfsig{\boldsymbol{\sigma}} 
\def\sigbf{\boldsymbol{\sigma}} 
\def\The{\Theta} 
\def\up{\uparrow}

\def\epsk{\epsilon_{\bf k}} 
\def\xik{\xi_{\bf k}} 
\def\xikq{\xi_{{\bf k}+{\bf q}}} 
\def\Ek{E_{\bf k}}
\def\Heff{\hat H_{\rm eff}}
\def\Hem{\hat H_{\rm em}}
\def\Hint{\hat H_{\rm int}}
\def\Hloc{\hat H_{\rm loc}}
\def\HMF{\hat H_{\rm MF}}
\def\Sem{S_{\rm em}}
\def\SMF{S_{\rm MF}} 
\def\SRPA{S_{\rm RPA}} 
\def\Sint{S_{\rm int}} 
\def\Sloc{S_{\rm loc}} 
\def\Zloc{Z_{\rm loc}} 
\def\ZMF{Z_{\rm MF}} 
\def\ZRPA{Z_{\rm RPA}} 
\def\RPA{{\rm RPA}}
\def\loc{{\rm loc}} 
\def\pp{{\rm pp}}
\def\ph{{\rm ph}} 
\def\ch{{\rm ch}}
\def\sp{{\rm sp}} 
\def\qtf{q_{\rm TF}}
\def\epstf{\eps^{}_{\rm TF}} 
\def\epsrpa{\eps^{}_{\rm RPA}} 
\def\chinnzpp{\chi_{nn}^{0}{}\!\!\!''}

\def\half{\frac{1}{2}}
\def\dhalf{\dfrac{1}{2}}
\def\third{\frac{1}{3}} 
\def\quarter{\frac{1}{4}}

\def\qr{{\bf q}\cdot{\bf r}}
\def\wt{\omega t} 

\def\a{{\bf a}}
\def\b{{\bf b}}
\def\e{{\bf e}}
\def\f{{\bf f}}
\def\g{{\bf g}}
\def\h{{\bf h}}
\def\k{{\bf k}}
\def\l{{\bf l}}
\def\m{{\bf m}}
\def\n{{\bf n}} 
\def\p{{\bf p}} 
\def\q{{\bf q}}
\def\r{{\bf r}}
\def\t{{\bf t}}
\def\u{{\bf u}}
\def\v{{\bf v}}
\def\x{{\bf x}}
\def\y{{\bf y}} 
\def\z{{\bf z}} 
\def\A{{\bf A}}
\def\B{{\bf B}}
\def\D{{\bf D}} 
\def\E{{\bf E}} 
\def\F{{\bf F}} 
\def\H{{\bf H}}  
\def\J{{\bf J}}
\def\K{{\bf K}} 

\def\G{{\bf G}}
\def\L{{\bf L}}
\def\M{{\bf M}}  
\def\O{{\bf O}} 
\def\P{{\bf P}} 
\def\Q{{\bf Q}} 
\def\R{{\bf R}}
\def\S{{\bf S}}
\def\epsbf{\boldsymbol{\epsilon}}
\def\mubf{\boldsymbol{\mu}}
\def\nablabf{\boldsymbol{\nabla}}
\def\rhobf{\boldsymbol{\rho}}
\def\sigmabf{\boldsymbol{\sigma}} 
\def\Pibf{\boldsymbol{\Pi}}
\def\pibf{\boldsymbol{\pi}}

\def\para{\parallel}
\def\kpara{{k_\parallel}}
\def\kperp{{k_\perp}} 
\def\kperpp{{k_\perp'}} 
\def\qperp{{q_\perp}} 
\def\tperp{{t_\perp}} 

\def\w{\omega}
\def\wn{\omega_n}
\def\wnu{\omega_\nu}
\def\wp{\omega_p} 
\def\dmu{{\partial_\mu}}
\def\dl{{\partial_l}}  
\def\dt{\partial_t} 
\def\tdt{\tilde\partial_t}
\def\dk{\partial_k}
\def\tdk{\tilde\partial_k}
\def\dx{\partial_x}
\def\dy{\partial_y} 
\def\dtau{{\partial_\tau}}  
\def\det{{\rm det}} 
\def\Pf{{\rm Pf}}

\def\dsum{\displaystyle \sum}
\def\dint{\displaystyle \int} 
\def\intt{\int_{-\infty}^\infty dt} 
\def\inttp{\int_{-\infty}^\infty dt'} 
\def\intk{\int_{\bf k}} 
\def\intkd{\int \frac{d^dk}{(2\pi)^d}}
\def\intq{\int_{\bf q}} 
\def\intr{\int d^dr}  
\def\dintr{\displaystyle \int d^dr} 
\def\intrp{\int d^dr'}
\def\dinttau{\displaystyle \int_0^\beta d\tau}
\def\dinttaup{\displaystyle \int_0^\beta d\tau'}
\def\inttau{\int_0^\beta d\tau}
\def\inttaup{\int_0^\beta d\tau'}
\def\intx{\int d^{d+1}x} 
\def\inttaur{\int_0^\beta d\tau \int d^dr}
\def\intinf{\int_{-\infty}^\infty}
\def\dinttaur{\displaystyle \int_0^\beta d\tau \int d^dr}
\def\dintinf{\displaystyle \int_{-\infty}^\infty}
\def\intw{\int_{-\infty}^\infty \frac{d\w}{2\pi}}
\def\sumr{\sum_{\bf r}} 

\def\calA{{\cal A}} 
\def\calC{{\cal C}} 
\def\dt{\partial_t}
\def\calD{{\cal D}}
\def\calF{{\cal F}} 
\def\calG{{\cal G}}
\def\calH{{\cal H}}
\def\calI{{\cal I}}
\def\calJ{{\cal J}}
\def\calK{{\cal K}}
\def\calL{{\cal L}} 
\def\calN{{\cal N}}
\def\calO{{\cal O}}
\def\calP{{\cal P}}  
\def\calR{{\cal R}} 
\def\calS{{\cal S}}
\def\calT{{\cal T}}
\def\calU{{\cal U}}
\def\calY{{\cal Y}} 
\def\calZ{{\cal Z}} 

\def\calFbf{{\bf F}}

\def\tT{{\tilde T}}
\def\talpha{{\tilde\alpha}}
\def\tdelta{{\tilde\delta}}
\def\teta{{\tilde\eta}} 
\def\tlamb{{\tilde\lambda}}
\def\tmu{{\tilde\mu}}
\def\tphibf{{\tilde\phibf}}
\def\trho{{\tilde\rho}}
\def\tvarphibf{{\tilde\varphibf}} 
\def\tw{{\tilde\omega}}
\def\twn{{\tilde\omega_n}}

\def\asinh{{\rm asinh}} 


\title{Thermodynamics in the vicinity of a relativistic quantum critical point in $2+1$ dimensions} 

\author{A. Ran\c{c}on}
\affiliation{James Franck Institute and Department of Physics,
University of Chicago, Chicago, Illinois 60637, USA}
\affiliation{Laboratoire de Physique Th\'eorique de la Mati\`ere Condens\'ee, 
CNRS UMR 7600, Universit\'e Pierre et Marie Curie, 4 Place Jussieu, 
75252 Paris Cedex 05, France}

\author{O. Kodio}
\author{N. Dupuis}
\affiliation{Laboratoire de Physique Th\'eorique de la Mati\`ere Condens\'ee, 
CNRS UMR 7600, Universit\'e Pierre et Marie Curie, 4 Place Jussieu, 
75252 Paris Cedex 05, France}

\author{P. Lecheminant}
\affiliation{Laboratoire de Physique Th\'eorique et Mod\'elisation, CNRS UMR 8089, Universit\'e de Cergy-Pontoise, Site de Saint-Martin, 2 avenue Adolphe Chauvin, 95302 Cergy-Pontoise Cedex, France} 

\date{June 27, 2013} 

\begin{abstract}
We study the thermodynamics of the relativistic quantum O($N$) model in two space dimensions. In the vicinity of the zero-temperature quantum critical point (QCP), the pressure can be written in the scaling form $P(T)=P(0)+N(T^3/c^2)\calF_N(\Delta/T)$ where $c$ is the velocity of the excitations at the QCP and $|\Delta|$ a characteristic zero-temperature energy scale. Using both a large-$N$ approach to leading order and the nonperturbative renormalization group, we compute the universal scaling function $\calF_N$. For small values of $N$ ($N\lesssim 10$) we find that $\calF_N(x)$ is nonmonotonic in the quantum critical regime ($|x|\lesssim 1$) with a maximum near $x=0$. The large-$N$ approach -- if properly interpreted -- is a good approximation both in the renormalized classical ($x\lesssim -1$) and quantum disordered ($x\gtrsim 1$) regimes, but fails to describe the nonmonotonic behavior of $\calF_N$ in the quantum critical regime. We discuss the renormalization-group flows in the various regimes near 
the QCP and make the connection 
with the quantum nonlinear sigma model in the renormalized classical regime. We compute the Berezinskii-Kosterlitz-Thouless transition temperature in the quantum O(2) model and find that in the vicinity of the QCP the universal ratio $\Tkt/\rho_s(0)$ is
very close to $\pi/2$, implying that the stiffness $\rho_s(\Tkt^-)$ at the transition is only slightly reduced with respect to the zero-temperature stiffness $\rho_s(0)$. Finally, we briefly discuss the experimental determination of the universal function $\calF_2$ from the pressure of a Bose gas in an optical lattice near the superfluid--Mott-insulator transition. 
\end{abstract}
\maketitle

\section{Introduction}

Many zero-temperature critical points observed in quantum many-body systems are described by a relativistic effective field theory~\cite{Sachdev_book,Podolsky12}. Bosonic cold atomic gases constitute a very clean experimental realization of such quantum critical points (QCP): a Bose gas in an optical lattice undergoes a quantum phase transition between a Mott insulator and a superfluid state~\cite{Jaksch98,Greiner02,Stoferle04,Spielman07}. When the transition occurs at fixed density, it is described by a relativistic quantum O(2) model~\cite{Fisher89,Rancon11b}. 

Recent works have focused on the excitation spectrum of the relativistic quantum O($N$) model in the vicinity of the QCP and in particular on the spectral function of the amplitude (``Higgs'') mode in the broken-symmetry phase~\cite{Podolsky12,Podolsky11,Pollet12,Gazit12,Chen13}. Signatures of the amplitude mode have recently been observed in a two-dimensional superfluid near the superfluid--Mott-insulator transition~\cite{Endres12}. 

In this paper, we study the thermodynamics of the relativistic quantum O($N$) model in two space dimensions. We extend previous results~\cite{Sachdev_book,Chubukov94} obtained to leading order in the large-$N$ limit by computing the full scaling function $\calF_\infty(x)$ determining the temperature dependence of the pressure near the QCP. Using a nonperturbative renormalization-group (NPRG) approach~\cite{Berges02,Delamotte07,Kopietz_book}, we then calculate $\calF_N(x)$ for finite values of $N$, including $N=2$ and $N=3$. 

We start from the action 
\begin{align}
S[\varphibf] ={}& \int dx \biggl\lbrace \half (\nablabf\varphibf)^2 + \frac{1}{2c_0^2} (\dtau \varphibf)^2 \nonumber \\ & 
+ \frac{r_0}{2} \varphibf^2 + \frac{u_0}{4!N} (\varphibf^2)^2 \biggr\rbrace, 
\label{action1} 
\end{align}
where we use the shorthand notation 
\begin{equation}
x=(\r,\tau), \quad \int dx = \inttau \int d^2r . 
\end{equation}
$\varphibf(x)$ is an $N$-component real field and $\tau\in [0,\beta]$ an imaginary time ($\beta=1/T$ and we set $\hbar=k_B=1$). $r_0$ and $u_0$ are temperature-independent coupling constants and $c_0$ is the (bare) velocity of the $\varphibf$ field. The factor $1/N$ in Eq.~(\ref{action1}) is introduced to obtain a 
meaningful limit $N\to\infty$ (with $u_0$ fixed). The model is regularized by an ultraviolet cutoff $\Lambda$. In order to maintain the Lorentz invariance of the action~(\ref{action1}) at zero temperature, it is natural to implement a cutoff on both momenta and frequencies but we will also sometimes use a cutoff acting only on momenta.

\begin{figure}
\centerline{\includegraphics[width=5.25cm]{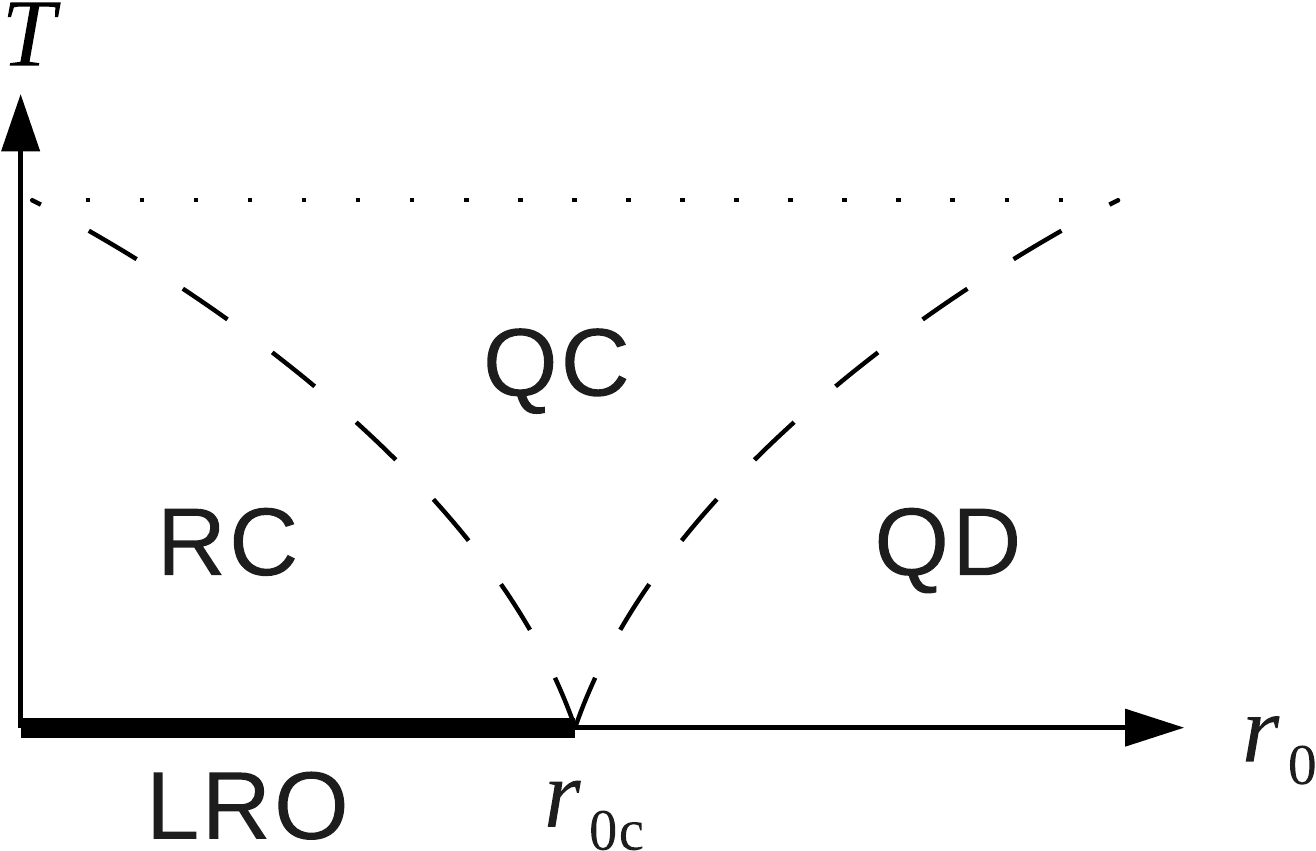}}
\caption{Phase diagram of the relativistic O($N$) model in two space dimensions for $N\geq 3$ [Eq.~(\ref{action1})]. The thick line shows the zero-temperature ordered phase with long-range order (LRO), while the dashed lines are crossover lines between the renormalized classical (RC), quantum critical (QC) and quantum disordered (QD) regimes. The dotted line shows the limit of the high-$T$ region where the physics is not controlled by the QCP anymore. (For $N=2$, there is a finite-temperature BKT transition line for $r_0\leq r_{0c}$, which terminates at $T=0$ for $r_0=r_{0c}$.)}
\label{fig_phase_dia}
\end{figure}

In two space dimensions, the phase diagram of the relativistic quantum O($N$) model is well known (Fig.~\ref{fig_phase_dia})~\cite{Sachdev_book}. At zero temperature, there is a quantum phase transition between a disordered phase ($r_0>r_{0c}$) and an ordered phase ($r_0<r_{0c}$) where the O($N$) symmetry of the action~(\ref{action1}) is spontaneously broken ($u_0$ and $c_0$ are considered as fixed parameters). The QCP at $r_0=r_{0c}$ is in the universality class of the three-dimensional classical O($N$) model with a dynamical critical exponent $z=1$ (this value follows from Lorentz invariance at zero temperature); the phase transition is governed by the three-dimensional Wilson-Fisher fixed point. At finite temperatures, the system is always disordered for $N\geq 2$, in agreement with the Mermin-Wagner theorem, but it is possible to 
distinguish three regimes in the vicinity of the QCP: a renormalized classical regime, a quantum critical regime, and a quantum disordered regime~\cite{Chakravarty89,Sachdev_book}. For $N=2$ and $r_0 < r_{0c}$, there is a finite-temperature Berezinskii-Kosterlitz-Thouless (BKT) phase transition~\cite{Berezinskii70,*Berezinskii71,Kosterlitz73,Kosterlitz74} and the system exhibits algebraic order at low temperatures. The BKT transition temperature line $\Tkt$ terminates at the QCP $r_0=r_{0c}$. 

Below the upper critical dimension $d_c^+=3$ ($d_c^++z=4$) of the quantum phase transition, we expect the hyperscaling hypothesis to hold. In two dimensions, this allows us to write the pressure in the critical regime as~\cite{note4}
\begin{equation}
P(T) = P(0) + N \frac{T^3}{c^2} \calF_N\left(\frac{\Delta}{T}\right) , 
\label{pressure0}
\end{equation}
where $\calF_N$ is a universal scaling function, $c$ the velocity of the critical fluctuations at the QCP~\cite{note2} and $|\Delta|\equiv|\Delta(r_0)|$ a characteristic energy scale at zero temperature. When $r_0>r_{0c}$, the system is disordered and we choose $\Delta$ to be equal to the excitation gap $m_0 \propto (r_0-r_{0c})^{z\nu}$ of the $\varphibf$ field ($\nu$ denotes the correlation-length exponent at the QCP) -- not to be confused with the amplitude (``Higgs'') mode gap. When $r_0<r_{0c}$ it is convenient to take $\Delta$ negative such that $-\Delta$ is the excitation gap in the disordered phase at the point located symmetrically with respect to the QCP, i.e. $|\Delta(r_0)|=m_0(2r_{0c}-r_0)$~\cite{Podolsky12}. $-\Delta$ is then proportional to the stiffness $\rho_s$, the ratio $|\Delta|/\rho_s$ being universal. With these definitions, $\Delta$ varies from negative to positive values as we go across the QCP coming from the ordered phase. The two crossover lines shown in Fig.~\ref{fig_phase_dia} are 
roughly defined by $|\Delta|\sim T$. We stress that the scaling function $\calF_N$ is independent of all microscopic parameters of the model such as $r_0$, $u_
0$ or $c_0$. The latter enter 
the temperature variation of the pressure [Eq.~\eqref{pressure0}] only indirectly {\it via} the values of the renormalized velocity $c$ and the energy scale $\Delta$. 

In the critical regime near the QCP, all thermodynamic quantities can be written in a scaling form. In addition to $\calF_N$, we will compute the universal scaling function $F_N$ which determines the excitation gap  
\begin{equation}
m(T) = T F_N\left( \frac{\Delta}{T} \right) 
\label{gap6}
\end{equation}
at finite temperatures. As we shall see, the knowledge of $F_N$ is necessary to obtain $\calF_N$ in the large-$N$ limit. 

The outline of the paper is as follows. In Sec.~\ref{sec_largeN}, we compute the universal scaling functions $F_N$ and $\calF_N$ to leading order in a $1/N$ expansion. We then use a NPRG approach to calculate $F_N$ and $\calF_N$ for any value $N\geq 2$ (Sec.~\ref{sec_nprg}). The main results are presented in Sec.~\ref{subsec_univ_scal}. Section~\ref{subsec_rgflow} is devoted to a detailed analysis of the RG flows in the renormalized classical, quantum disordered and quantum critical regimes for $N\geq 3$. In the renormalized classical regime, where the physics is dominated by the $N-1$ Goldstone modes of the zero-temperature broken-symmetry phase, we show that the NPRG flow equations yield the one-loop RG equations of the quantum O($N$) nonlinear $\sigma$ model (NL$\sigma$M)~\cite{Chakravarty89}. The BKT transition temperature in the quantum O(2) model is discussed in Sec.~\ref{subsec_bkt}. The implication of our results for cold atomic gases are briefly discussed in the Conclusion. 

\section{Large-$N$ limit}
\label{sec_largeN}

In this section, we use a cutoff $\Lambda$ acting only on momenta, i.e. $|\q|\leq\Lambda$. We do not distinguish between the bare velocity $c_0$ and the renormalized one $c$ since they coincide in the large-$N$ limit.

Following the standard method in the large-$N$ limit (see, e.g., Refs.~\cite{Zinn_book_2,Dupuis11}), we express the partition function as 
\begin{align}
Z ={}& \int \calD[\varphibf,\rho,\lambda] \exp\biggl\lbrace - \int dx \biggl[ \half(\nablabf\varphibf)^2  + \frac{1}{2c^2}(\dtau\varphibf)^2 \nonumber \\ & + \frac{r_0}{2}\rho + \frac{u_0}{4!N} \rho^2 + i \frac{\lambda}{2} (\varphibf^2-\rho) \biggr] \biggr\rbrace .
\end{align}
It can be easily verified that by integrating out $\lambda$ and then $\rho$, one recovers the original action $S[\varphibf]$. If, instead, we first integrate out $\rho$, we obtain
\begin{align}
Z ={}& \int \calD[\varphibf,\lambda] \exp\biggl\lbrace -\int dx \biggl[ \half(\nablabf\varphibf)^2 + \frac{1}{2c^2}(\dtau\varphibf)^2  \nonumber \\ & + i \frac{\lambda}{2} \varphibf^2 \biggr] + \frac{3N}{2u_0} \int dx \,(i\lambda-r_0)^2 \biggr\rbrace .
\end{align}
We then split the $\varphibf$ field into a field $\sigma$ and a $(N-1)$-component field $\pibf$. The integration over the $\pibf$ field gives 
\begin{equation}
\int\calD[\pibf] \,e^{ -\int dx \left[ \half (\nablabf\pibf)^2 + \frac{1}{2c^2} (\dtau\pibf)^2 + i \frac{\lambda}{2} \pibf^2 \right]} = (\det\, g)^{(N-1)/2} ,
\end{equation}
where 
\begin{equation}
g^{-1}(x,x') = [- \nablabf^2 -c^{-2}\dtau^2 + i\lambda(x) ] \delta(x-x')  
\label{ginverse} 
\end{equation}
is the inverse propagator of the $\pi_i$ field in the fluctuating $\lamb$ field. We thus obtain the action 
\begin{multline}
S[\sig,\lambda] = \half \int dx \left[ (\nablabf\sig)^2 + c^{-2} (\dtau\sigma)^2 + i \lambda \sig^2 \right] \\ - \frac{3N}{2u_0} \int dx \,(i\lambda-r_0)^2 + \frac{N-1}{2} \Tr\ln g^{-1} . 
\label{action2} 
\end{multline} 
In the limit $N\to\infty$, the action becomes proportional to $N$ (this is easily seen by rescaling the $\sig$ field, $\sig\to\sqrt{N}\sig$) and the saddle-point approximation becomes exact. For uniform and time-independent fields $\sig(x)=\sig$ and $\lambda(x)=\lambda$, the saddle-point action is given by
\begin{equation}
\frac{1}{\beta V} S[\sig,\lambda] = \frac{i}{2} \lambda \sig^2 - \frac{3N}{2u_0} (i\lambda-r_0)^2 + \frac{N}{2\beta V} \Tr\ln g^{-1}  
\label{action3} 
\end{equation}
(we use $N-1\simeq N$ for large $N$), with $g^{-1}(q)=\q^2+\wn^2/c^2+i\lambda$ in Fourier space. $q=(\q,i\wn)$, $\wn=2\pi Tn$ ($n$ integer) is a bosonic Matsubara frequency, and $V$ denotes the volume of the system. From (\ref{action3}), we deduce the saddle-point equations 
\begin{equation}
\begin{gathered}
\sig m^2 = 0 , \\
\sig^2 = \frac{6N}{u_0}\left( \frac{m^2}{c^2} -r_0 \right) - N \int_q g(q),
\end{gathered}
\label{gap1}
\end{equation}
where we use the notation 
\begin{equation}
\int_q = \frac{1}{\beta} \sum_{\wn} \int_\q = \frac{1}{\beta} \sum_{\wn} \int \frac{d^2q}{(2\pi)^2} 
\end{equation}
and $m^2=i\lambda c^2$ ($i\lambda$ is real at the saddle point). These equations show that the component $\sig$ of the $\varphibf$ field which was singled out plays the role of an order parameter. In the ordered phase, $\sigma$ is nonzero and $m=0$. The propagator $g(q)=1/(\q^2+\wn^2/c^2)$ is gapless, thus identifying the $\pi_i$ fields as the $N-1$ Goldstone modes associated with the spontaneously broken O($N$) symmetry. In the disordered phase, $\sigma$ vanishes and $m$ determines the gap (or ``mass'') of the $\varphibf$ field as well as the correlation length $\xi=c/m$.

\subsection{Zero temperature}

The critical value $r_{0c}$ corresponding to the QCP separating the ordered and disordered phases is obtained by setting $\sigma=m=0$ in Eqs.~\eqref{gap1},
\begin{equation}
r_{0c} = - \frac{u_0}{6} \int_\q \int_\w \frac{c^2}{\w^2+c^2\q^2} = - \frac{u_0 c\Lambda}{24\pi} ,
\label{gap2}
\end{equation}
where $\int_\w = \intw$. 

In the disordered phase $r_0\geq r_{0c}$, $\sigma=0$ and the mass $m_0=m(T=0)$ is determined by Eqs.~(\ref{gap1},\ref{gap2}), 
\begin{align}
& \frac{6}{u_0}\left( \frac{m^2}{c^2} -r_0 + r_{0c} \right) \nonumber \\ - & c^2 \int_\q \int_\w  \left( \frac{1}{\w^2+c^2\q^2+m_0^2} - \frac{1}{\w^2+c^2\q^2} \right) = 0 ,
\label{gap3}
\end{align} 
which gives
\begin{equation}
\frac{6m_0^2}{u_0c^2} + \frac{m_0}{4\pi} = \frac{6}{u_0}(r_0-r_{0c}) . 
\end{equation}
By comparing the two terms on the lhs of this equation, we obtain a characteristic momentum scale, the Ginzburg scale $k_G\sim cu_0/24\pi$, which signals the onset of critical fluctuations~\cite{note3}. In the critical regime, $m_0\ll ck_G$, we obtain 
\begin{equation}
m_0 = \frac{24\pi}{u_0} (r_0-r_{0c}) , 
\end{equation}
which gives $z\nu=1$, i.e. a correlation-length exponent $\nu=1$ since the dynamical critical exponent $z=1$. In the noncritical regime $m_0\gg ck_G$, $m_0\sim (r_0-r_{0c})^{1/2}$ and we recover the classical value $\nu=1/2$. The anomalous dimension $\eta$ vanishes to leading order in the large-$N$ limit. 

In the ordered phase $r_0\leq r_{0c}$, $m_0$ vanishes and $\sigma$ is finite,
\begin{align}
\sigma^2 &= - 6N \frac{r_0}{u_0} - N \int_\q \int_\w \frac{c^2}{\w^2+c^2\q^2} \nonumber \\ 
&=   - \frac{6N}{u_0}(r_0-r_{0c}) . 
\end{align} 
The stiffness is equal to $\rho_s=\sigma^2$~\cite{note6}. 

\subsection{Finite temperatures}

At finite temperatures, the system is always disordered ($\sigma=0$), in agreement with the Mermin-Wagner theorem, and the mass $m$ is obtained from the saddle-point equation 
\begin{align}
0 &= \frac{6}{u_0}\left( \frac{m^2}{c^2} -r_0 \right) - \int_q \frac{c^2}{\wn^2+c^2\q^2+m^2} \nonumber \\ 
&= \frac{6}{u_0}\left( \frac{m^2}{c^2} -r_0 \right) - \frac{T}{2\pi} \ln\left( \frac{\sinh \frac{c\Lambda}{2T}}{\sinh\frac{m}{2T}} \right) \nonumber \\ 
&= \frac{6}{u_0} ( r_{0c} -r_0 ) + \frac{T}{2\pi} \ln\left( 2 \sinh\frac{m}{2T} \right) + \frac{6m^2}{u_0c^2} . 
\end{align}
In the critical regime the last term can be neglected and we obtain 
\begin{equation}
m = 2 T \,\mbox{asinh} \left[\half \exp\left( \frac{\Delta}{2 T} \right) \right] ,
\label{gap4} 
\end{equation}
where we have introduced the characteristic energy scale $|\Delta|$ defined by 
\begin{equation}
\Delta = \frac{24\pi}{u_0} (r_0-r_{0c}) .
\end{equation}
$\Delta$ corresponds to the $T=0$ gap $m_0$ on the disordered side $r_0>r_{0c}$ of the QCP, and to $-4\pi\rho_s/N$ on the ordered side $r_0<r_{0c}$ with $\rho_s$ the zero-temperature stiffness (see the discussion in the Introduction). The critical regime is defined by $T,|\Delta|\ll ck_G$. 

We can rewrite Eq.~\eqref{gap4} as 
\begin{equation}
\frac{m}{T} = F_\infty\left(\frac{\Delta}{T}\right) , 
\label{gap5} 
\end{equation}
with the universal scaling function 
\begin{equation}
F_\infty(x) = 2  \,\mbox{asinh} \left( \half e^{x/2} \right) .
\label{Finf}
\end{equation}
$F_\infty(x)$ satisfies 
\begin{equation}
F_\infty(x) = \llbrace
\begin{array}{lcc}
e^{x/2} & \mbox{if} & x \to -\infty , \\ 
2\,\mbox{asinh}(1/2) & \mbox{if} & x = 0 , \\ 
x & \mbox{if} & x \to \infty ,
\end{array}
\right.
\label{gap7} 
\end{equation}
with $2\,\mbox{asinh}(1/2)\simeq 0.962424$. The three cases in Eq.~(\ref{gap7}) correspond to the renormalized classical ($m\simeq T e^{-|\Delta|/2T}$), quantum critical ($m\simeq 2\,\mbox{asinh}(1/2) T$), and quantum disordered ($m\simeq \Delta$) regimes, respectively (see Fig.~\ref{fig_phase_dia}).

\subsection{Pressure} 

In the large-$N$ limit, the pressure $P=-S[\sigma,\lambda]/\beta V$ is obtained from the saddle-point value of the action, 
\begin{equation}
\frac{P}{N} = - \frac{m^2\sigma^2}{2Nc^2} + \frac{3}{2u_0} \left(r_0 - \frac{m^2}{c^2} \right)^2 - \frac{1}{2\beta V} \Tr\ln g^{-1} .
\end{equation}
Using the results of Appendix~\ref{app_det} for $\Tr\ln g^{-1}$, in the critical regime we can write the pressure in the scaling form~(\ref{pressure0}) with the universal scaling function 
\begin{multline}
\calF_\infty(x) = \frac{1}{2\pi} \biggl[ \frac{x^3}{12} \Theta(x) - \frac{x}{4} F_\infty(x)^2 + \frac{1}{6} F_\infty(x)^3 \\  + F_\infty(x) \Li_2\bigl(e^{-F_\infty(x)}\bigr) + \Li_3\bigl(e^{-F_\infty(x)}\bigr) \biggr] .
\label{calFinf}
\end{multline} 
$\Li_s(z)$ is a polylogarithm,
\begin{equation}
\Li_s(z) = \sum_{k=1}^\infty \frac{z^k}{k^s} , \quad (z \in \mathbb{C}, \;|z|<1) ,
\label{polylog} 
\end{equation}
and $\Theta(x)$ denotes the step function. 

From the definition of $\calF_\infty(x)$, we obtain the limiting cases  
\begin{equation}
\calF_\infty(x) = \llbrace
\begin{array}{ccc}
\dfrac{\zeta(3)}{2\pi} & \mbox{if} & x \to -\infty , \\ 
\dfrac{2 \zeta(3)}{5\pi} & \mbox{if} & x = 0 , \\ 
0 & \mbox{if} & x \to \infty ,
\end{array}
\right.
\label{calF1}
\end{equation}
where $\zeta(z)$ is the Riemann zeta function: $\zeta(3)/2\pi\simeq 0.191313$ and $2\zeta(3)/5\pi\simeq 0.153051$. To obtain $\lim_{x\to-\infty} \calF_\infty(x)$ we use $F_\infty(x)\to e^{x/2}$ for $x\to-\infty$ and $\Li_3(1)=\zeta(3)$. The universal number $\calF_\infty(0)$ is obtained noting that $F_\infty(0)=2 \ln \tau$, with $\tau=(1+\sqrt{5})/2=2-\tau^{-2}$ the Golden mean, and using~\cite{Sachdev93} 
\begin{equation}
\begin{split}
\Li_2(2-\tau) &= \frac{\pi^2}{15} - \quarter \ln^2(2-\tau) , \\
\Li_3(2-\tau) &= \frac{4}{5}\zeta(3) + \frac{\pi^2}{15}\ln(2-\tau) - \frac{1}{12} \ln^3(2-\tau) . 
\end{split}
\end{equation}

It should be noted that the scaling function $F_\infty(x)$ as well as $\calF_\infty(0)$ agree with results obtained from the NL$\sigma$M in the large-$N$ limit~\cite{Chubukov94,Sachdev_book,Sachdev93}. This follows from the fact that the linear and nonlinear O($N$) models are in the same universality class and therefore exhibit the same critical physics. 

\section{NPRG approach}
\label{sec_nprg}

The strategy of the NPRG approach is to build a family of theories indexed by a momentum scale $k$ such that fluctuations are smoothly taken into account as $k$ is lowered from the microscopic scale $\Lambda$ down to 0~\cite{Berges02,Delamotte07,Kopietz_book}. This is achieved by adding to the action (\ref{action1}) the infrared regulator
\begin{equation}
\Delta S_k[\varphibf] = \half \sum_{q,i} \varphi_i(-q) R_k(q) \varphi_i(q) ,
\end{equation}
so that the partition function
\begin{equation}
Z_k[\J] = \int\calD[\varphibf]\, e^{-S[\varphibf]-\Delta S_k[\varphibf] + \int dx \sum_i J_i \varphi_i} 
\end{equation}
becomes $k$ dependent. The $k$-dependent effective action 
\begin{equation}
\Gamma_k[\phibf] = - \ln Z_k[\J] + \int dx  \sum_i J_i \phi_i - \Delta S_k[\phibf] 
\end{equation}
is defined as a modified Legendre transform of $-\ln Z_k[\J]$ which includes the subtraction of $\Delta S_k[\phibf]$. Here $\phibf(x)=\mean{\varphibf(x)}$ is the order parameter (in the presence of the external source). The initial condition of the flow is specified by the microscopic scale $k=\Lambda$ where we assume that the fluctuations are completely frozen by the $\Delta S_k$ term, so that $\Gamma_\Lambda[\phibf]=S[\phibf]$. The effective action of the original model (\ref{action1}) is given by $\Gamma_{k=0}$ provided that $R_{k=0}$ vanishes. For a generic value of $k$, 
the cutoff function $R_k(q)$ suppresses fluctuations with momentum $|\q|\lesssim k$ or frequency $|\wn|\lesssim c_kk$ but leaves unaffected those with $|\q|,|\wn|/c_k \gtrsim k$ (here $c_k$ denotes the (renormalized) velocity of the $\varphibf$ field). The variation of the effective action with $k$ is given by Wetterich's equation~\cite{Wetterich93} 
\begin{equation}
\dt \Gamma_k[\phibf] = \half \Tr\llbrace \dt R_k\left(\Gamma^{(2)}_k[\phibf] + R_k\right)^{-1} \rrbrace ,
\label{rgeq}
\end{equation}
where $t=\ln(k/\Lambda)$. $\Gamma^{(2)}_k[\phibf]$ denotes the second-order functional derivative of $\Gamma_k[\phibf]$. In Fourier space, the trace involves a sum over momenta and Matsubara frequencies as well as the internal index of the $\phibf$ field. We use a regulator function $R_k(q)$ which acts both on momenta and frequencies,
\begin{equation}
R_k(q) = Z_{A,k} \left(\q^2 + \frac{\wn^2}{c_k^2} \right) r \left( \frac{\q^2+\wn^2/c_k^2}{k^2} \right),
\label{R1}
\end{equation}
where $r(Y)=1/(e^Y-1)$. The $k$-dependent constant $Z_{A,k}$ is defined below [Eq.~\eqref{ansatz1}]. 

When $\phibf$ is constant, i.e. uniform and time independent, the effective action coincides with the effective potential, 
\begin{equation}
U_k(\rho) = \frac{1}{\beta V} \Gamma_k[\phibf] \Bigl|_{\phibf\;\const} . 
\end{equation}
Because of the O($N$) symmetry of the effective action $\Gamma_k$, the effective potential $U_k(\rho)$ must be a function of the O($N$) invariant $\rho=\phibf^2/2$. The pressure is then simply defined by 
\begin{equation}
P(T) = - U_{k=0}(\rho_{0}) , 
\label{pressure2} 
\end{equation}
where $\rho_{0,k}$ denotes the position of the minimum of $U_k(\rho)$ and $\rho_0=\lim_{k\to 0} \rho_{0,k}$. 

\subsection{Approximate solution of the flow equation} 

Because of the regulator term $\Delta S_k$, the vertices $\Gamma^{(n)}_{k,i_1\cdots i_n}(q_1,\cdots,q_n)$ are smooth functions of momenta and frequencies and can be expanded in powers of $\q_i^2/k^2$ and $\w_{n_i}^2/c_k^2k^2$. Thus if we are interested only in the long-distance (critical) physics, we can use a derivative expansion of the effective action~\cite{Berges02,Delamotte07}. In the following, we consider the ansatz
\begin{equation}
\Gamma_k[\phibf] = \int dx \biggl\lbrace \frac{Z_{A,k}}{2} (\nablabf\phibf)^2 + \frac{V_{A,k}}{2} (\dtau\phibf)^2 + U_k(\rho) \biggr\rbrace ,
\label{ansatz1}
\end{equation}
which is often referred to as the LPA'. It differs from the local potential approximation (LPA) by the introduction of two field renormalization constants $Z_{A,k}$ and $V_{A,k}$ ($Z_{A,\Lambda}=1$ and $V_{A,\Lambda}=c_0^{-2}$). It is the minimal ansatz beyond the LPA which includes a finite anomalous dimension $\eta$ at the QCP (see below). Moreover the LPA equation for the potential, and therefore the analog equation in the LPA', are exact in the large-$N$ limit~\cite{DAttanasio97}. To further simplify the analysis, we expand $U_k(\rho)$ about the position $\rho_{0,k}$ of its minimum,  
\begin{equation}
U_k(\rho) = \llbrace 
\begin{array}{lcc} 
U_k(\rho_{0,k}) + \frac{\lambda_k}{2}(\rho-\rho_{0,k})^2 & \mbox{if} & \rho_{0,k} > 0 , \\ 
U_k(\rho_{0,k}) + \delta_k \rho + \frac{\lambda_k}{2}\rho^2 & \mbox{if} & \rho_{0,k} = 0 .
\end{array}
\right. 
\label{Ueff}
\end{equation}
Although the RG equations can also be solved for the full effective potential, the determination of the singular part of the pressure turns out to be extremely difficult in that case~\cite{note15}. 

The LPA is known to be very accurate to obtain thermodynamic quantities. It has been used to compute the pressure in the three-dimensional quantum $\varphi^4$ theory with Ising symmetry (i.e. $N=1$)~\cite{Blaizot07a,Blaizot11}. The results compare very well with those of the Blaizot-M\'endez-Wschebor approach (BMW) -- an elaborated NPRG scheme which preserves the full momentum and frequency dependence of the propagator~\cite{Blaizot06,Benitez09,Benitez12}. There are also strong indications that the LPA (or the LPA') is a good approximation even when it is supplemented by a truncation of the effective potential [Eq.~(\ref{Ueff})]~\cite{Canet03a}. As will be shown below, the truncated LPA' remains accurate -- and nearly exact in the renormalized classical regime -- in the limit $N\to\infty$~\cite{note10,DAttanasio97}. Furthermore, it has also been used to determine the phase diagram of the Bose-Hubbard model in two and three dimensions~\cite{Rancon11a,Rancon11b,Rancon12d}: although a truncation of the 
effective potential leads to a loss of 
accuracy, the results remain within 10 percent of the exact ones obtained 
by 
quantum Monte Carlo simulation~\cite{Capogrosso07,Capogrosso10}.  

The derivation of the flow equation for $U_k(\rho)$, $Z_{A,k}$ and $V_{A,k}$ is standard~\cite{Berges02,Delamotte07} (the only difference with the classical O($N$) model comes from the finite size $\beta$ in the imaginary-time direction~\cite{Tetradis93,Reuter93}). The effective potential satisfies the flow equation 
\begin{equation}
\dt U_k(\rho) = \half \int_q \dt R_k(q) [ G_{k,\rm l}(q;\rho) + (N-1) G_{k,\rm t}(q;\rho) ] , 
\label{Uflow}
\end{equation}
where 
\begin{equation}
\begin{split}
G_{k,\rm l}^{-1}(q;\rho) &= Z_{A,k}\q^2 + V_{A,k} \wn^2 + U_k'(\rho) + 2\rho U_k''(\rho) + R_k(q) , \\ 
G_{k,\rm t}^{-1}(q;\rho) &= Z_{A,k}\q^2 + V_{A,k} \wn^2 + U_k'(\rho) + R_k(q)
\end{split}
\label{Glt} 
\end{equation}
determine the longitudinal and transverse parts of the propagator $G_k=(\Gamma_k^{(2)}+R_k)^{-1}$ in a constant field $\phibf$, 
\begin{equation}
G_{k,ij}(q;\phibf) = \frac{\phi_i \phi_j}{2\rho} G_{k,\rm l}(q;\rho) + \left( \delta_{i,j} - \frac{\phi_i \phi_j}{2\rho} \right) G_{k,\rm t}(q;\rho) .
\end{equation}
The contribution of $G_{k,\rm t}$ to $\dt U_k$ comes with a factor $N-1$ corresponding to the number of transverse modes. When $\rho_{0,k}>0$, $U_k'(\rho_{0,k})$ vanishes and these modes become gapless for $R_k(q)\to 0$ (Goldstone modes). The stiffness is given by $\rho_{s,k}=2Z_{A,k}\rho_{0,k}$~\cite{note6}. In the disordered phase, the minimum of $U_k(\rho)$ is located at $\rho_{0,k}=0$ so that all modes exhibit a gap $m_k=\sqrt{U'_k(0)/V_{A,k}}$ (for $R_k(q)\to 0$) corresponding to a finite correlation length $\xi_k=c_k/m_k$ where $c_k=\sqrt{Z_{A,k}/V_{A,k}}$ ($c_\Lambda=c_0$) is the renormalized velocity (see Sec.~\ref{subsubsec_QC} for a further discussion of the velocity). The actual gap $m$ and correlation length $\xi$ in the disordered phase are obtained for $k=0$.  

The flow equations for $Z_{A,k}$ and $V_{A,k}$ are obtained from the flow equation~(\ref{rgeq}) by noting that
\begin{equation}
\begin{split}
Z_{A,k} &= \lim_{q\to 0} \frac{\partial}{\partial\q^2} \Gamma^{(2)}_{k,\rm t}(q;\rho_{0,k}) , \\
V_{A,k} &= \lim_{q\to 0} \frac{\partial}{\partial\wn^2} \Gamma^{(2)}_{k,\rm t}(q;\rho_{0,k}) .
\end{split} 
\label{ZVAdef} 
\end{equation}
At the zero-temperature QCP, $Z_{A,k}\sim k^{-\eta}$ and $V_{A,k}\sim k^{-\eta-2(z-1)}$~\cite{note8}, which allows us to deduce both the anomalous dimension $\eta$ and the dynamical critical exponent $z$. The latter is equal to one due to the Lorentz invariance of the action~\eqref{action1} at $T=0$. The exponent $\nu$ can be obtained from the divergence of the correlation length $\xi\sim (r_0-r_{0c})^{-\nu}$ in the disordered phase as the QCP is approached, or more directly from the escape rate from the fixed point when the system is nearly critical. 

The RG equations are given by~\cite{note9}
\begin{equation}
\begin{split}
\dt \rho_{0,k} &= - \frac{3}{2} \Il  - \frac{N-1}{2} \It \quad \mbox{if} \quad \rho_{0,k}>0, \\ 
\dt \delta_k &= \frac{\lamb_k}{2} (N+2) \Il \quad \mbox{if} \quad \rho_{0,k}=0, \\ 
\dt \lamb_k &= - \lamb_k^2 [9 \Jll(0) + (N-1) \Jtt(0) ] , \\ 
\dt Z_{A,k} &= - 2 \lamb_k^2 \rho_{0,k} \frac{\partial}{\partial \p^2} [\Jtl(p) + \Jlt(p) ] \bigl|_{p=0} , \\ 
\dt V_{A,k} &= - 2 \lamb_k^2 \rho_{0,k} \frac{\partial}{\partial \wn^2} [\Jtl(p) + \Jlt(p) ] \bigl|_{p=0} ,
\end{split}
\end{equation}
while the equation for the thermodynamic potential (per unit volume) $U_k(\rho_{0,k})$ is directly obtained from~(\ref{Uflow}). We have introduced the threshold functions 
\begin{equation}
\begin{split} 
I_{k,\alpha} &= \int_q \tdt G_{k,\alpha} (q;\rho_{0,k}) , \\ 
J_{k,\alpha\beta}(p) &= \int_q [\tdt G_{k,\alpha}(q;\rho_{0,k})] G_{k,\beta}(p+q;\rho_{0,k}) ,  
\end{split}
\end{equation}
with $\alpha,\beta ={\rm l,t}$. The operator $\tdt=(\dt R_k)\partial_{R_k}$ acts only on the $t$ dependence of the cutoff function $R_k$. The propagators $G_{k,\rm l}(p;\rho_{0,k})$ and $G_{k,\rm t}(p;\rho_{0,k})$ are given by~(\ref{Glt}) with $U_k'(\rho_{0,k})=\delta_k$ and $U_k''(\rho_{0,k})=\lamb_k$. 

The flow equations are solved numerically~\cite{note14}. Results related to the thermodynamics are discussed in the following section.

\subsection{Universal scaling functions}
\label{subsec_univ_scal}

\begin{figure}
\centerline{\includegraphics[width=6.5cm]{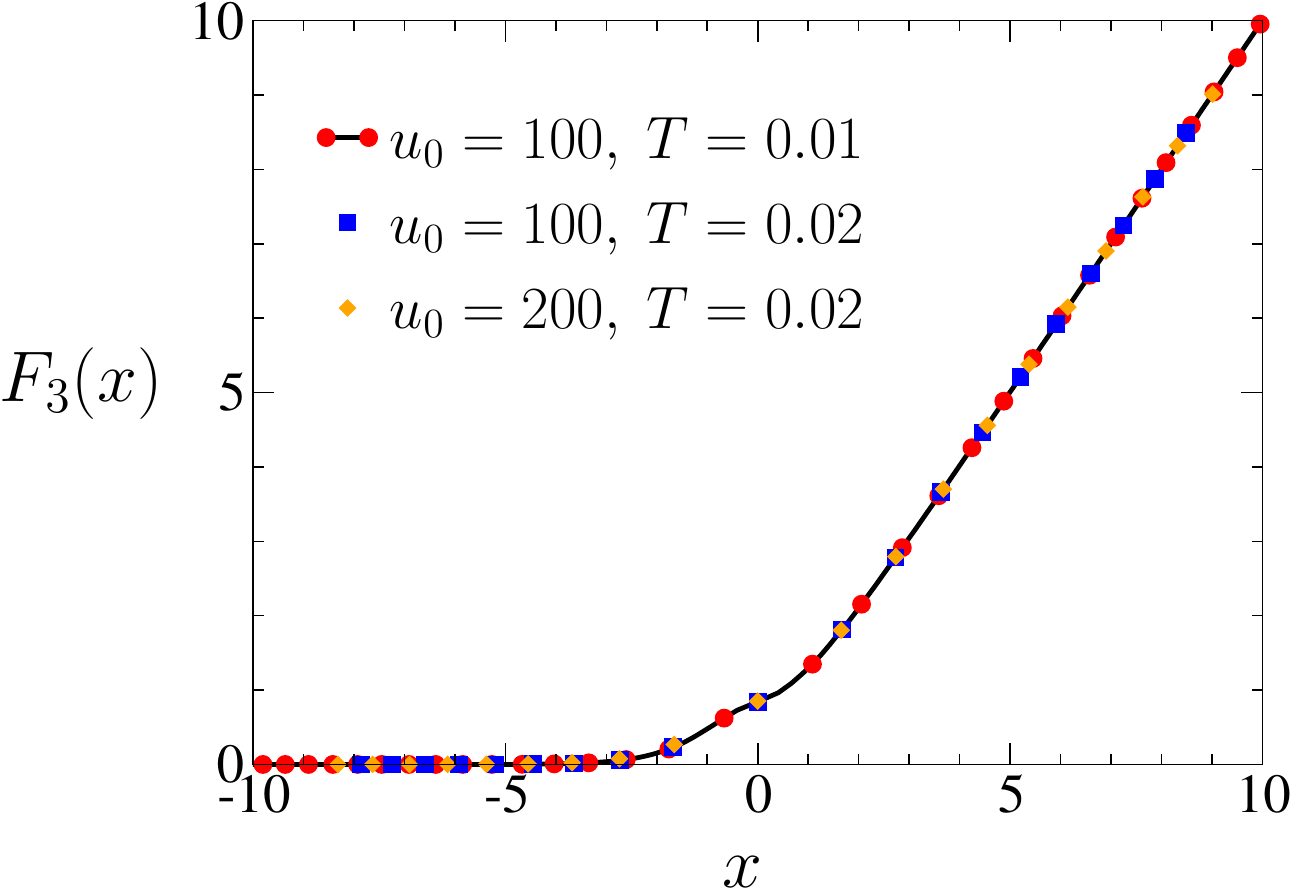}}
\caption{(Color online) Universal scaling function $F_3(x)$ [Eq.~(\ref{gap6})] computed for various values of the microscopic parameters ($\Lambda=100$ and $c_0=1$). }
\label{fig_F_univ}
\centerline{\includegraphics[width=6.5cm]{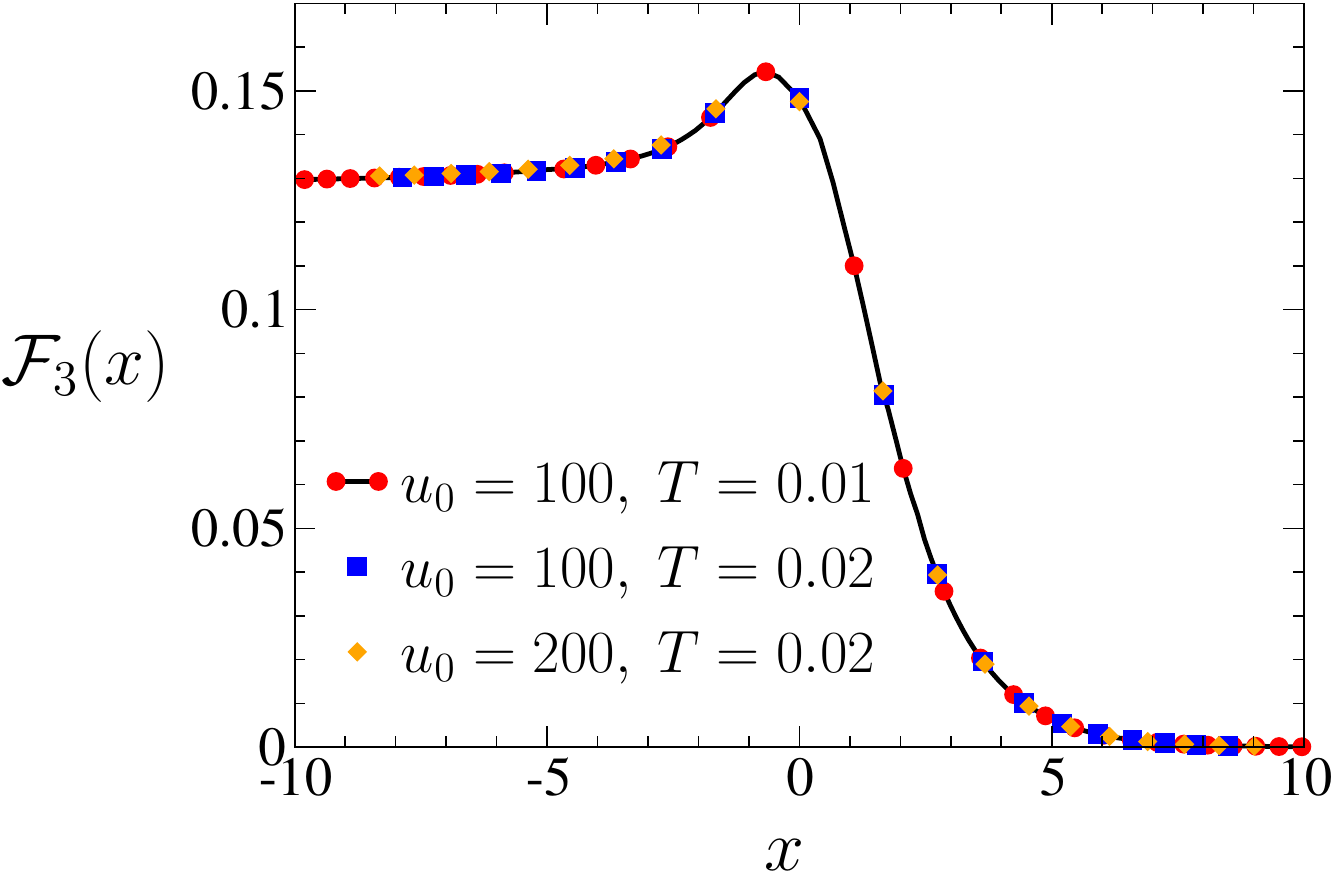}}
\caption{(Color online) Same as Fig.~\ref{fig_F_univ} but for the universal scaling function $\calF_3(x)$ [Eq.~(\ref{pressure0})].}
\label{fig_calF_univ}
\centerline{\hspace{-0.1cm}\includegraphics[width=6.4cm]{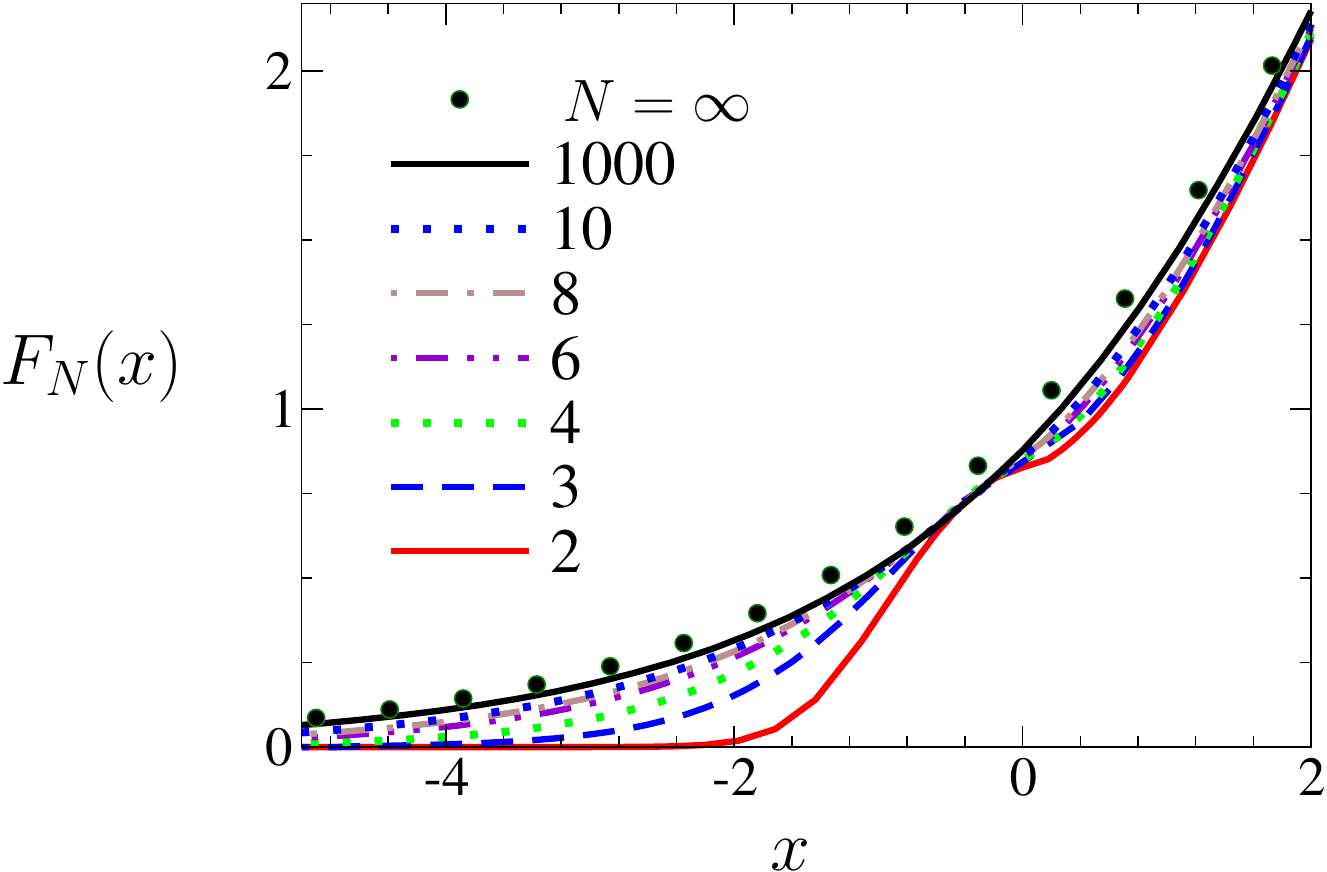}}
\centerline{\includegraphics[width=6.5cm]{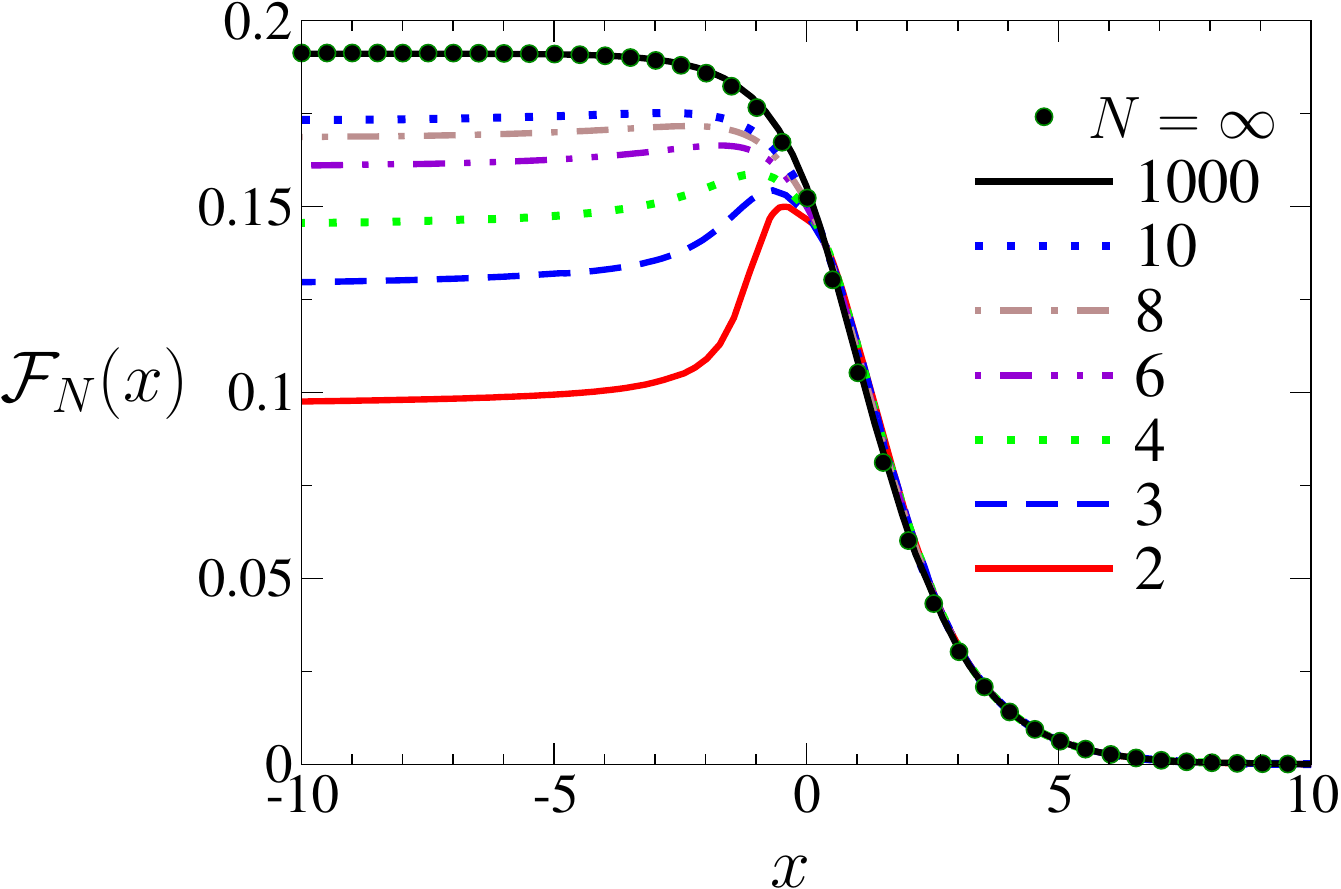}}
\caption{(Color online) Universal scaling functions $F_N$ and $\calF_N$ for various values of $N$ obtained from the NPRG. The black points show the analytic results~\eqref{Finf} and \eqref{calFinf} in the limit  $N\to\infty$.}
\label{fig_F}
\end{figure}

We first solve the equations at $T=0$ to determine $r_{0c}$ and obtain the critical exponents $\nu$ and $\eta$ as well as the characteristic energy scale $\Delta\equiv\Delta(r_0)$. For $N=3$ we find $\nu\simeq0.699$ and $\eta=0.0507$, to be compared with the best estimates for the three-dimensional O(3) model obtained from resummed perturbative calculations~\cite{Pogorelov08} ($\nu\simeq 0.7060$, $\eta=0.0333$), Monte Carlo simulations~\cite{Campostrini02} ($\nu\simeq 0.7112$, $\eta=0.0375$), or the NPRG in the BMW approximation~\cite{Benitez12} ($\nu\simeq 0.715$, $\eta=0.040$). For $N=2$, our results $\nu\simeq 0.613$ and $\eta=0.0582$ should be compared with the critical exponents of the three-dimensional O(2) model: resummed perturbative calculations~\cite{Pogorelov08} ($\nu\simeq 0.6700$, $\eta=0.0334$), Monte Carlo simulations~\cite{Campostrini06} ($\nu\simeq 0.6717$, $\eta=0.0381$), NPRG-BMW~\cite{Benitez12} ($\nu\simeq 0.674$, $\eta=0.041$). Note that the rather poor 
estimate of $\eta$ 
is a well-known limitation of the LPA'; a much better result can be obtained by considering the full derivative expansion to order $\calO(\partial^2)$~\cite{Canet03a}. At finite temperatures, the two-dimensional relativistic O(2) model exhibits a BKT phase transition. Although, {\it stricto sensu}, the NPRG does not capture this transition, most universal properties of the latter are nevertheless correctly reproduced~\cite{Graeter95,Gersdorff01}. In particular, recent work on the two-dimensional Bose gas has shown that the thermodynamics can be accurately computed using the NPRG~\cite{Rancon12b}. The BKT transition is further discussed in Sec.~\ref{subsec_bkt}. 

Once the QCP is located and the energy scale $\Delta$ determined as a function of $r_0-r_{0c}$, we compute the gap $m(T)$ and the pressure $P(T)$, and deduce the universal scaling functions $F_N(x)$ and $\calF_N(x)$ [Eqs.~(\ref{gap6},\ref{pressure0})]. To ensure that we are in the universal (critical) regime, we solve the NPRG equations for various values of the ultraviolet cutoff $\Lambda$, interaction strength $u_0$ or temperature $T$, and verify that the final results for $F_N$ and $\calF_N$ remain unchanged (Figs.~\ref{fig_F_univ} and \ref{fig_calF_univ}). Only at sufficiently low temperatures and close enough to the QCP ($T,|\Delta|\ll ck_G$) do the universal scaling forms~(\ref{pressure0},\ref{gap6}) hold. 

\begin{table}
\renewcommand{\arraystretch}{1.5}
\begin{center}
\begin{tabular}{cccccccc}
\hline \hline
$N$ & 1000 & 10 & 8 & 6 & 4 & 3 & 2 \\
\hline
$\rho_s/(N|\Delta|)$ & 0.0838 & 0.0853 & 0.0864 & 0.0891 & 0.0965 & 0.1059 & 0.1321  \\
\hline \hline
\end{tabular}
\end{center}
\caption{Universal ratio $\rho_s/(N|\Delta|)$ in the zero-temperature ordered phase. The exact result in the limit $N\to\infty$ is $1/4\pi\simeq 0.080$.}
\label{table_ratio}
\end{table}

Figure~\ref{fig_F} shows the universal scaling functions $F_N$ and $\calF_N$ for various values of $N$ (Table~\ref{table_ratio} shows the universal ratio $\rho_s/(N|\Delta|)$). In the limit $N\to\infty$, the truncated LPA' slightly differs from the exact result for the excitation gap $m(T)$ but turns out to be extremely accurate for the computation of the pressure $P(T)$ (the LPA' would be exact without the truncation of $U_k(\rho)$~\cite{note10}). For smaller values of $N$, $F_N$ and $\calF_N$ differ significantly from the $N\to\infty$ limit. While the large-$N$ result remains a good approximation in the quantum disordered regime, it becomes inaccurate in the quantum critical and renormalized classical regimes. In particular, it misses the nonmonotonic behavior of $\calF_N(x)$ in the quantum critical regime ($|x|\lesssim 1$) for $N\lesssim 10$. The possibility of such a nonmonotonic behavior is discussed in Ref.~\cite{Neto93}.

\begin{figure}
\centerline{\includegraphics[width=6.5cm]{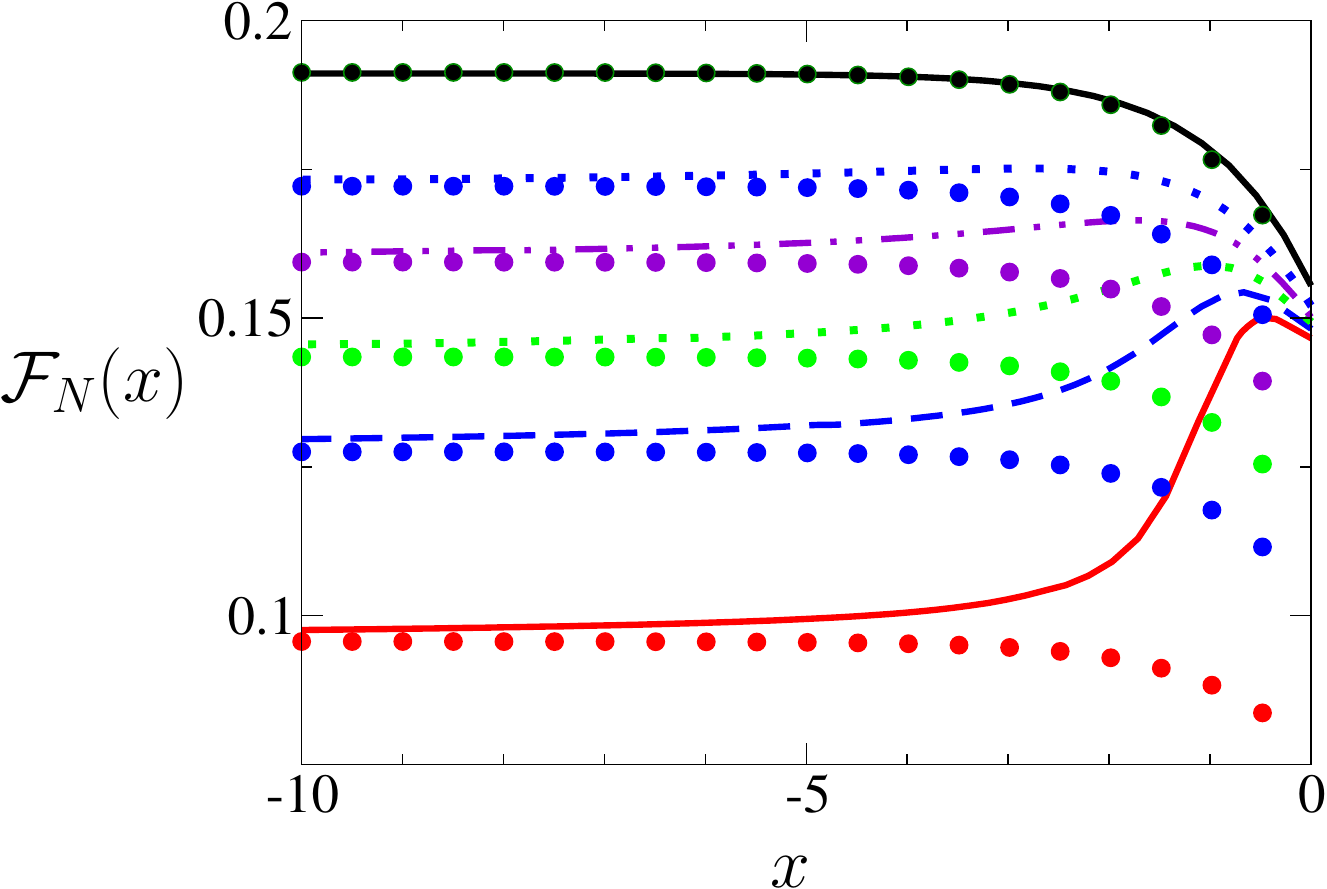}}
\caption{(Color online) Same as Fig.~\ref{fig_F} but for negative values of $x$ only. The dots show the large-$N$ result rescaled by the factor $(N-1)/N$.} 
\label{fig_calF_bis}  
\end{figure}

In the renormalized classical regime, it is possible to reinterpret the large-$N$ result so that it becomes consistent with the NPRG approach even for small values of $N$. Since the correlation length $\xi$ is exponentially large, we expect the thermodynamics to be dominated by the $N-1$ modes corresponding to transverse  fluctuations to the local order. In the NPRG approach, these modes show up as Goldstone modes as long as $\rho_{0,k}>0$ (i.e. $k\gtrsim \xi^{-1}$) and dominate the RG flow as in the large-$N$ approach (see the discussion in Sec.~\ref{subsubsec_RC_regime} below). Since $N-1$ is identified with $N$ in the large-$N$ approach, the latter overestimates the pressure, and therefore the scaling function $\calF_N$, by a factor $N/(N-1)$. In Fig.~\ref{fig_calF_bis}, we show that the large-$N$ result, when rescaled by a factor $(N-1)/N$, is indeed consistent with the NPRG approach. This shows that in the renormalized classical regime
\begin{equation}
P(T) \simeq (N-1) \frac{\zeta(3)}{2\pi} \frac{T^3}{c^2} ,
\label{Prc}
\end{equation}
which is nothing but the pressure of $N-1$ free bosonic modes with dispersion $\w=c|\q|$~\cite{note11}. The very small excitation gap of the transverse fluctuations ($m\ll T$) does not influence the thermodynamics. For $N=2$ and $N=3$, Eq.~(\ref{Prc}) agrees with a RG analysis of the non-linear sigma model~\cite{Hofmann13,Hofmann10}.

\begin{table}
\renewcommand{\arraystretch}{1.5}
\begin{center}
\begin{tabular}{cccccccc}
\hline \hline
$N$ & 1000 & 10 & 8 & 6 & 4 & 3 & 2 \\
\hline
$\tilde\calC_N/N$ to $\calO(1/N)$ & 0.800 & 0.767 & 0.758 & 0.744 & 0.716 & 0.689 & 0.633 \\
\hline
$\tilde\calC_N/N$ (NPRG) & 0.812 & 0.796 & 0.793 & 0.788 & 0.781 & 0.775 &  0.767 \\ 
\hline \hline
\end{tabular}
\end{center}
\caption{$\tilde\calC_N/N$ as obtained from the NPRG and the large-$N$ approach [Eq.~(\ref{CNlargeN})].}
\label{table}
\end{table}

Of particular interest is the temperature variation of the pressure at the QCP ($\Delta=0$). Following Ref.~\cite{Sachdev93}, we express the pressure as 
\begin{equation}
P(T) = P(0) + \frac{\zeta(3)}{2\pi} \tilde\calC_N \frac{T^3}{c^2}  
\label{calC_N}
\end{equation}
for $\Delta=0$, where $\tilde\calC_N=N 2\pi\calF_N(0)/\zeta(3)$. In the large-$N$ limit~\cite{Chubukov94},
\begin{equation}
\frac{\tilde\calC_N}{N} \simeq \frac{4}{5} - \frac{0.3344}{N} + \calO\left(\frac{1}{N^2}\right) 
\label{CNlargeN} 
\end{equation}
(the leading-order term $4/5$ is given by Eq.~(\ref{calF1})). The agreement between the $\calO(1/N)$ result and the NPRG one rapidly deteriorates for $N\lesssim 10$ (Table~\ref{table}). 

\begin{figure}
\centerline{\includegraphics[width=6cm]{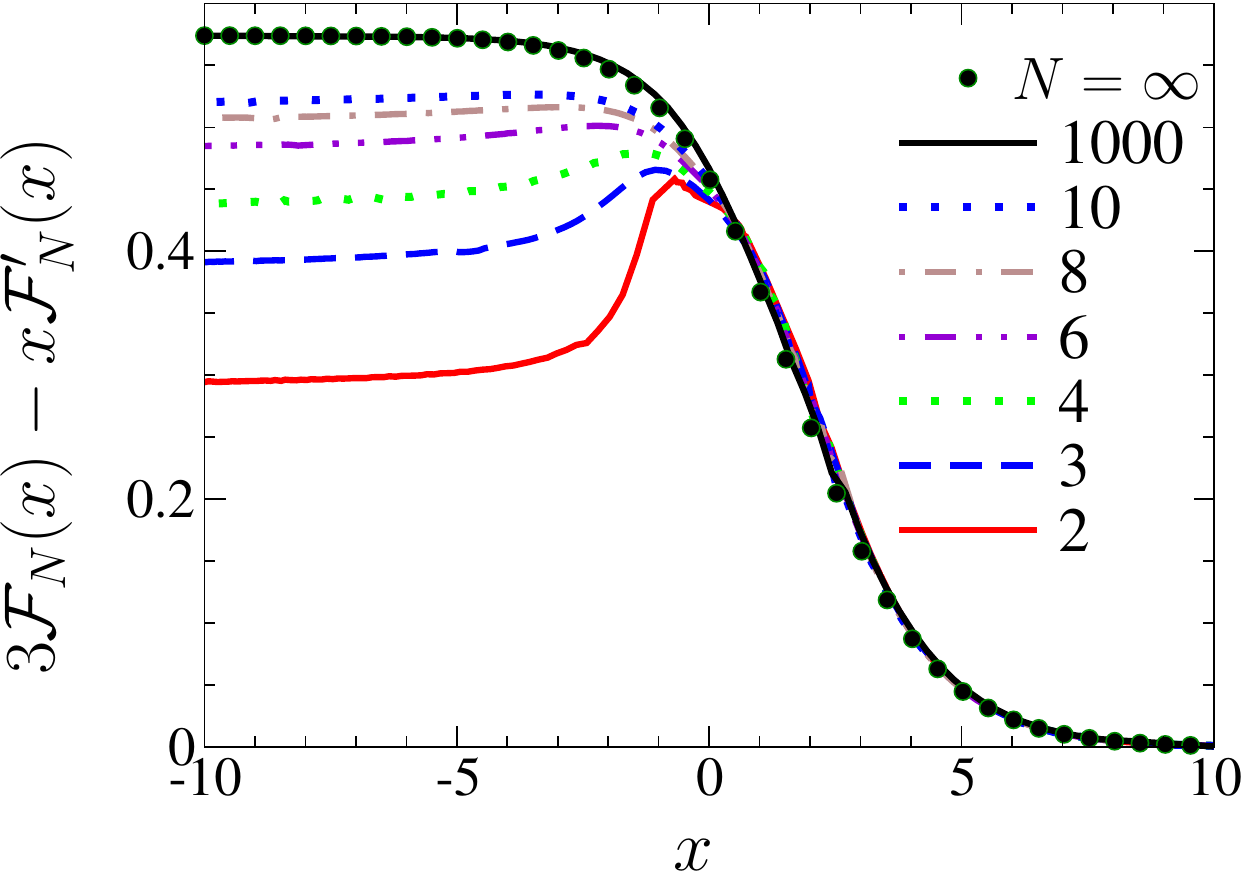}}
\caption{(Color online) Universal scaling function $3\calF_N(x)-x\calF'(x)$ for the entropy per unit volume [Eq.~(\ref{entropy})].} 
\label{fig_entropy}  
\end{figure}

The entropy per unit volume is equal to the temperature derivative $\partial P/\partial T$ of the pressure, 
\begin{equation}
\frac{S(T)}{V} = N \frac{T^2}{c^2} \left[ 3 \calF_N\left(\frac{\Delta}{T}\right) - \frac{\Delta}{T} \calF_N'\left(\frac{\Delta}{T}\right) \right] . 
\label{entropy}
\end{equation}
Up to the factor $NT^2/c^2$, it is entirely determined by the universal scaling function $3\calF_N(x)-x\calF_N'(x)$. The latter is nonmonotonic in the quantum critical regime (Fig.~\ref{fig_entropy}). 

\subsection{RG flows} 
\label{subsec_rgflow}

\begin{figure}
\centerline{\hspace{0.1cm}\includegraphics[width=6cm]{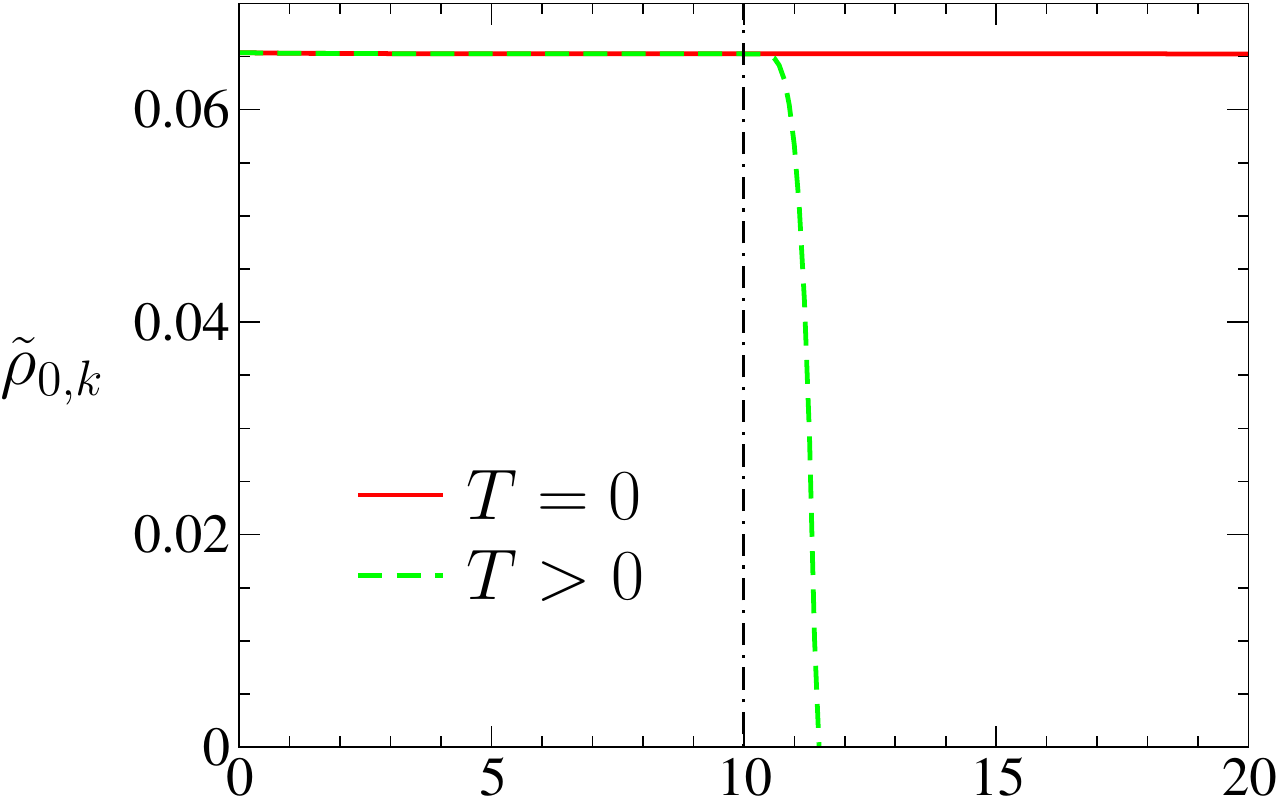}}
\centerline{\hspace{0.4cm}\includegraphics[width=5.7cm]{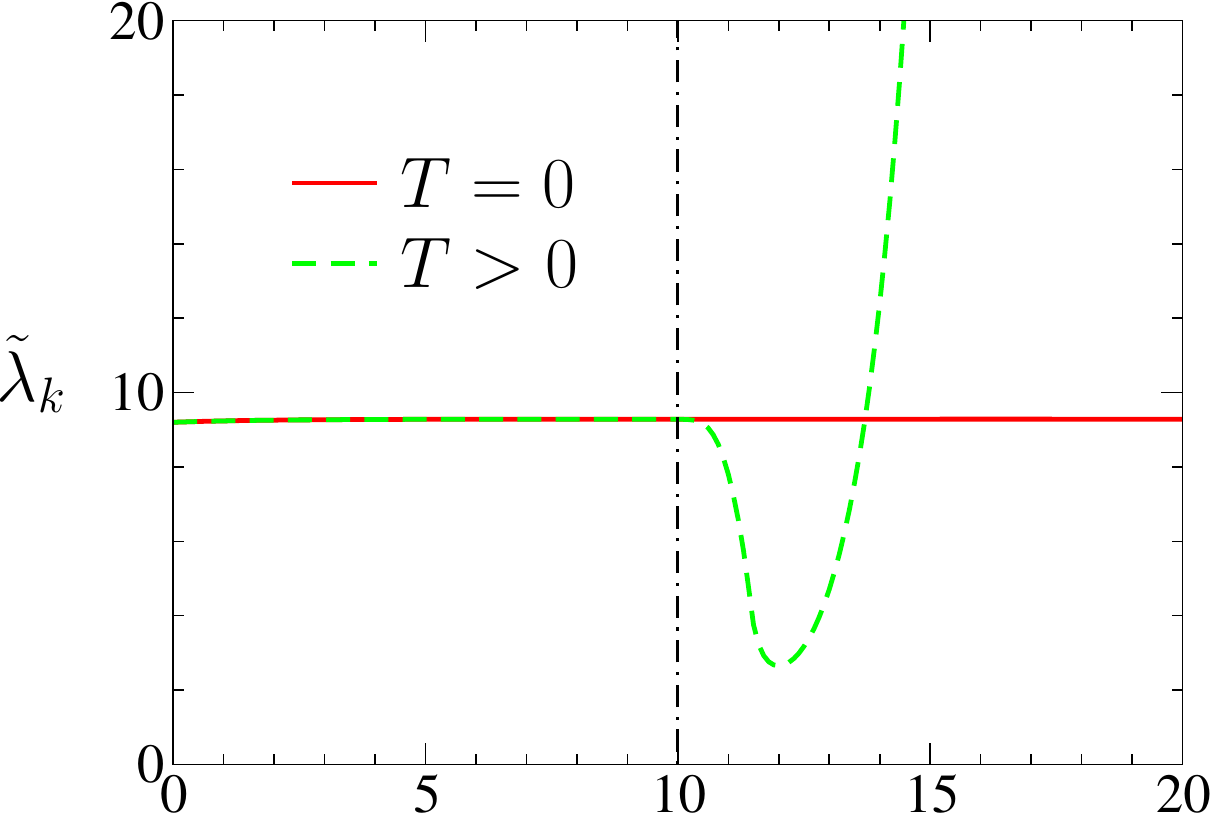}}
\centerline{\includegraphics[width=6.2cm]{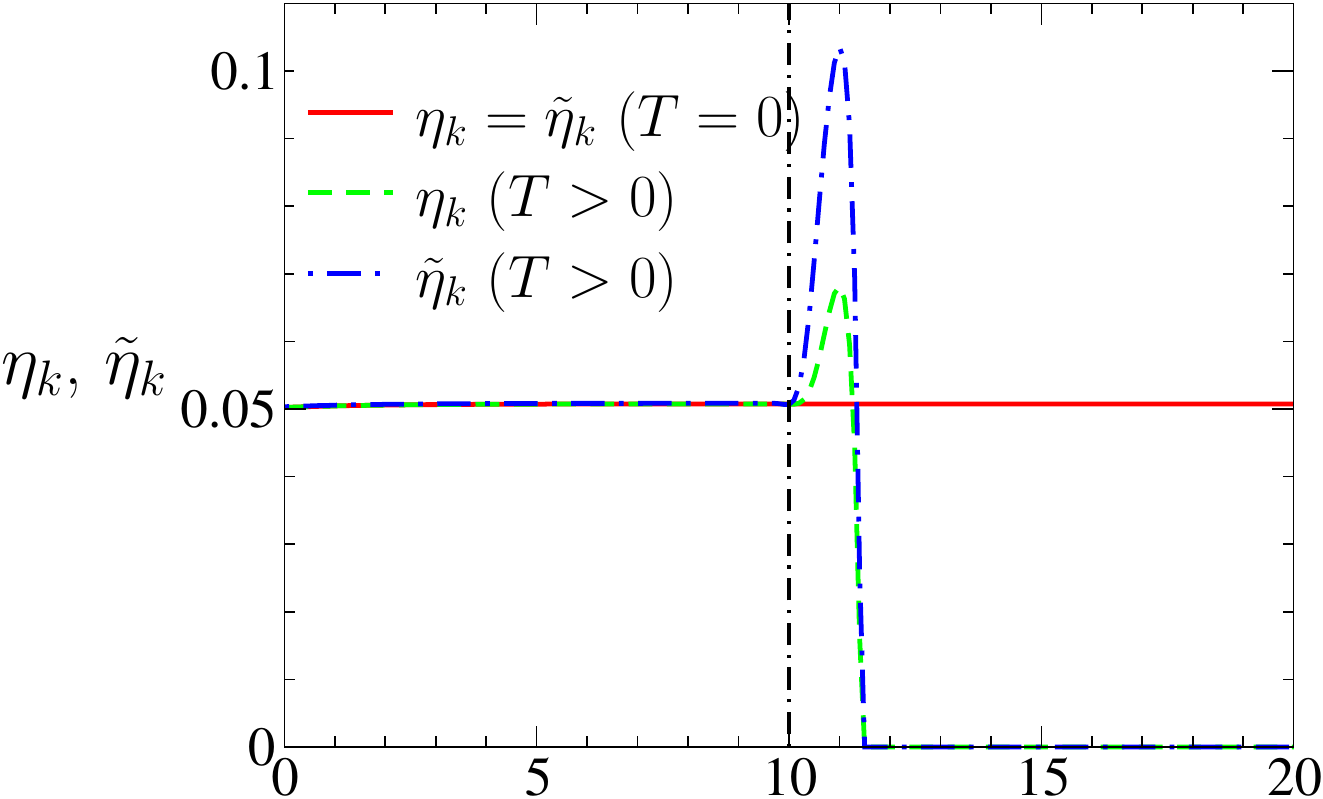}}
\centerline{\includegraphics[width=6.25cm]{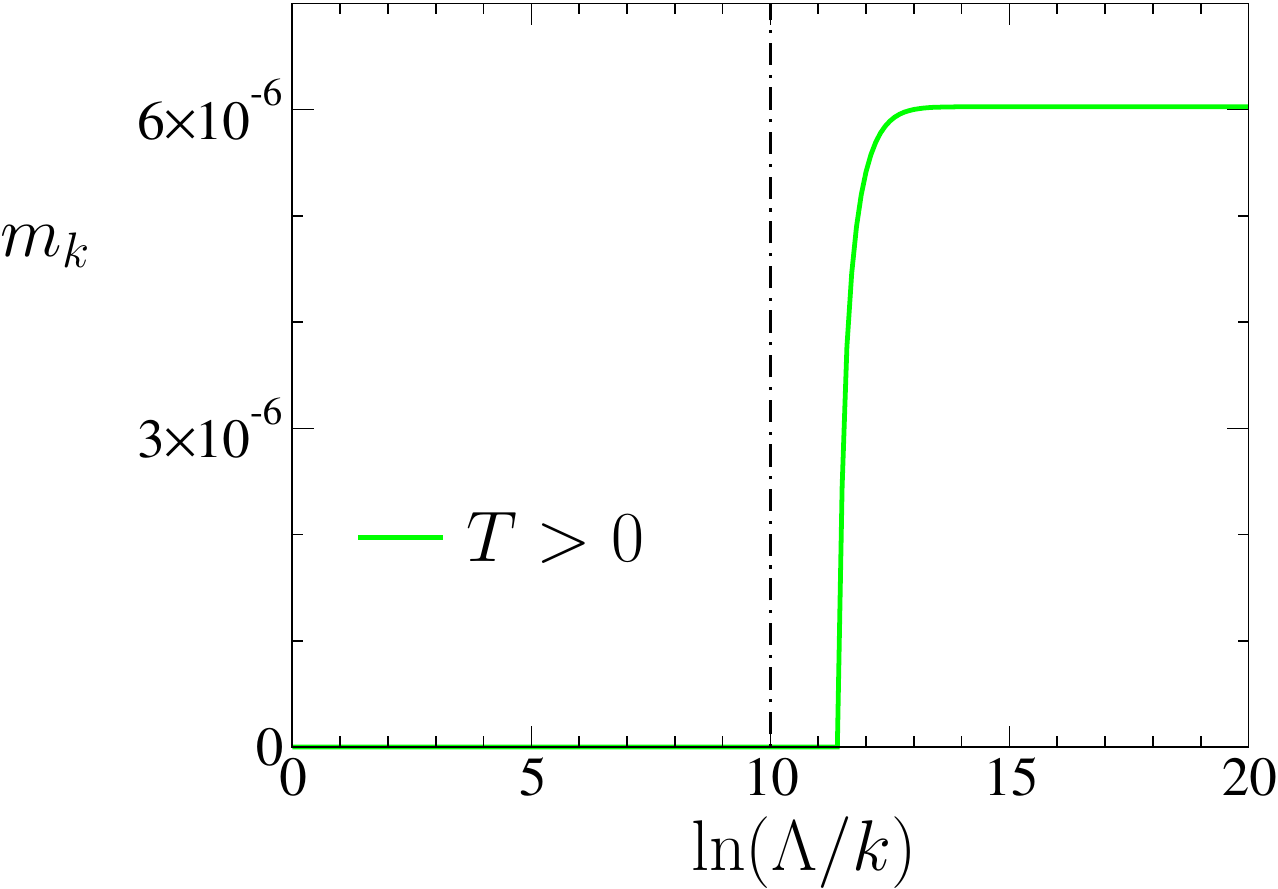}}
\caption{(Color online) RG flows in the quantum critical regime ($d=2$ and $N=3$). The vertical dash-dotted line shows the thermal momentum scale $k_T=\Lambda e^{-10}$. [$\Lambda=c_0=1$, $u_0=27.6$, $r_{0c}\simeq 0.065355$.]}
\label{fig_flow_qc}
\end{figure}

In this section, we qualitatively discuss the RG flows in the various regimes of the phase diagram in the vicinity of the QCP for $N\geq 3$ (Fig.~\ref{fig_phase_dia}). We use the dimensionless variables
\begin{equation}
\begin{split}
\trho_{0,k} &= k^{1-d} (Z_{A,k}V_{A,k})^{1/2} \rho_{0,k} , \\ 
\tdelta_k &= (Z_{A,k}k^2)^{-1} \delta_k , \\ 
\tlamb_k &= k^{d-3} Z_{A,k}^{-3/2} V_{A,k}^{-1/2} \lamb_k , \\ 
\end{split}
\label{vardim}
\end{equation}
and the corresponding RG equations
\begin{equation}
\begin{split}
\dt \trho_{0,k} ={}& (1-d-\eta_k/2-\teta_k/2) \trho_{0,k} \\ & - \frac{3}{2} \tIl  - \frac{N-1}{2} \tIt \quad \mbox{if} \quad \trho_{0,k} > 0, \\ 
\dt \tdelta_k ={}& (\eta_k-2) \tdelta_k + \frac{\tlamb_k}{2} (N+2)\tIl \quad \mbox{if} \quad \trho_{0,k}=0 , \\ 
\dt \tlamb_k ={}& (d-3+3\eta_k/2+\teta_k/2) \tlamb_k \\ 
& - \tlamb_k^2 [9 \tJll(0) + (N-1) \tJtt(0) ] , 
\end{split}
\end{equation}
where
\begin{equation}
\begin{split}
\eta_k ={}& 2 \tlamb_k^2 \trho_{0,k} \frac{\partial}{\partial y} [\tJlt(\tilde p) + \tJtl(\tilde p) ] \bigl|_{\tilde p=0} , \\  
\teta_k ={}& 2 \tlamb_k^2 \trho_{0,k} \frac{\partial}{\partial \twn^2} [\tJlt(\tilde p) + \tJtl(\tilde p) ] \bigl|_{\tilde p=0} ,
\end{split}
\label{rgeqdim}
\end{equation}
with $\tilde p=(\p/k,i\twn)$, $y=\p^2/k^2$, $\twn=\wn/c_kk=2\pi\tilde T_kn$ ($\tilde T_k=T/c_kk$). The dimensionless threshold functions $\tilde I_{k,\alpha}$ and $\tilde J_{k,\alpha\beta}(\tilde p)$ are defined in Appendix~\ref{subsec__threshold_dim}. For the sake of generality, we consider an arbitrary space dimension $d$. 

In the zero-temperature limit, using the results of Appendix~\ref{subsec_thresholdzero} for the threshold functions, we recover the flow equations of the $(d+1)$-dimensional (classical) O($N$) model in the LPA'. At the QCP ($r_0=r_{0c}$), critical fluctuations develop below the Ginzburg momentum scale $k_G$. In the following, we discuss only the universal part of the flow $k\ll k_G$. Deviations from criticality are characterized by two momentum scales. The first one, $k_\Delta=|\Delta|/c_0$, is associated to the detuning from the QCP. In the $T=0$ disordered phase, $k_\Delta^{-1}$ is nothing but the correlation length. In the $T=0$ ordered phase, $k_\Delta\sim k_J$ is related to the Josephson momentum scale $k_J=\rho_s/c_0$. The latter separates the critical regime $k_J\ll k\ll k_G$ from the Goldstone regime $k\ll k_J$ dominated by the Goldstone modes. The second characteristic momentum scale is the thermal scale $k_T=2\pi T/c_0$ associated to the crossover between the quantum ($k\gg k_T$) 
and classical ($k\ll k_T$) 
regimes. The three regimes of the phase diagram (Fig.~\ref{fig_phase_dia}) are defined by $k_T\gg k_\Delta$ (quantum critical), $k_T\ll k_\Delta$ and $r_0>r_{0c}$ (quantum disordered), $k_T\ll k_\Delta$ and $r_0<r_{0c}$ (renormalized classical).

\subsubsection{Quantum critical regime}
\label{subsubsec_QC} 

The RG flow in the quantum critical regime is shown in Fig.~\ref{fig_flow_qc} for $N=3$. The parameters of the microscopic action~(\ref{action1}) are chosen such that the initial value $\Lambda$ of the momentum cutoff is of the order of the Ginzburg scale $k_G$. At the QCP ($r_0=r_{0c}$) and for $T=0$, we observe plateaus characteristic of critical behavior: $\trho_{0,k}\sim \trho^*_{\rm crit}$, $\tlamb_k\sim \tlamb^*_{\rm crit}$, and $\eta_k=\teta_k=\eta$ (with $\eta$ the anomalous dimension at the three-dimensional Wilson-Fisher fixed point). At finite temperatures, the flow is modified when $k$ becomes smaller than the thermal scale $k_T$: $\trho_{0,k}$ and $\eta_k,\teta_k$ rapidly vanish while $\tlamb_k$ diverges; the (dimensionful) order parameter $\rho_{0,k}$ vanishes and $m_k=\sqrt{\delta_k/V_{A,k}}$ takes a finite value (indicating that the system is in a disordered phase). Near $k_T$, $\eta_k$ and $\teta_k$ differ, implying a breakdown of Lorentz invariance. It is however difficult to estimate the 
renormalized value of the velocity. At finite temperature, $V_{A,k}$ gives the coefficient of the $\wn^2$ term in the expansion of the vertex $\Gamma_{k,\rm t}^{(2)}(q;\rho_{0,k})$ in powers of $\wn^2$. $c_k=\sqrt{Z_{A,k}/V_{A,k}}$ can be identified with the velocity of the (transverse) fluctuations only when $k\gg k_T$. For $k\ll k_T$, the flow is classical (the propagator is dominated by its $\w_{n=0}=0$ component) and $V_{A,k}$ does not enter the RG equations anymore. In this regime the actual value $\bar c_k$ of the velocity should be obtained from the retarded vertex $\Gamma_{k,\rm t}^{(2)}(\q,\w;\rho_{0,k}) = U'_k(\rho_{0,k}) + Z_{A,k}(\q^2-\w^2/\bar c_k^2) + \cdots$ (with $\w$ a real frequency)~\cite{note16}.

\subsubsection{Quantum disordered regime}

\begin{figure}
\centerline{\includegraphics[width=6cm]{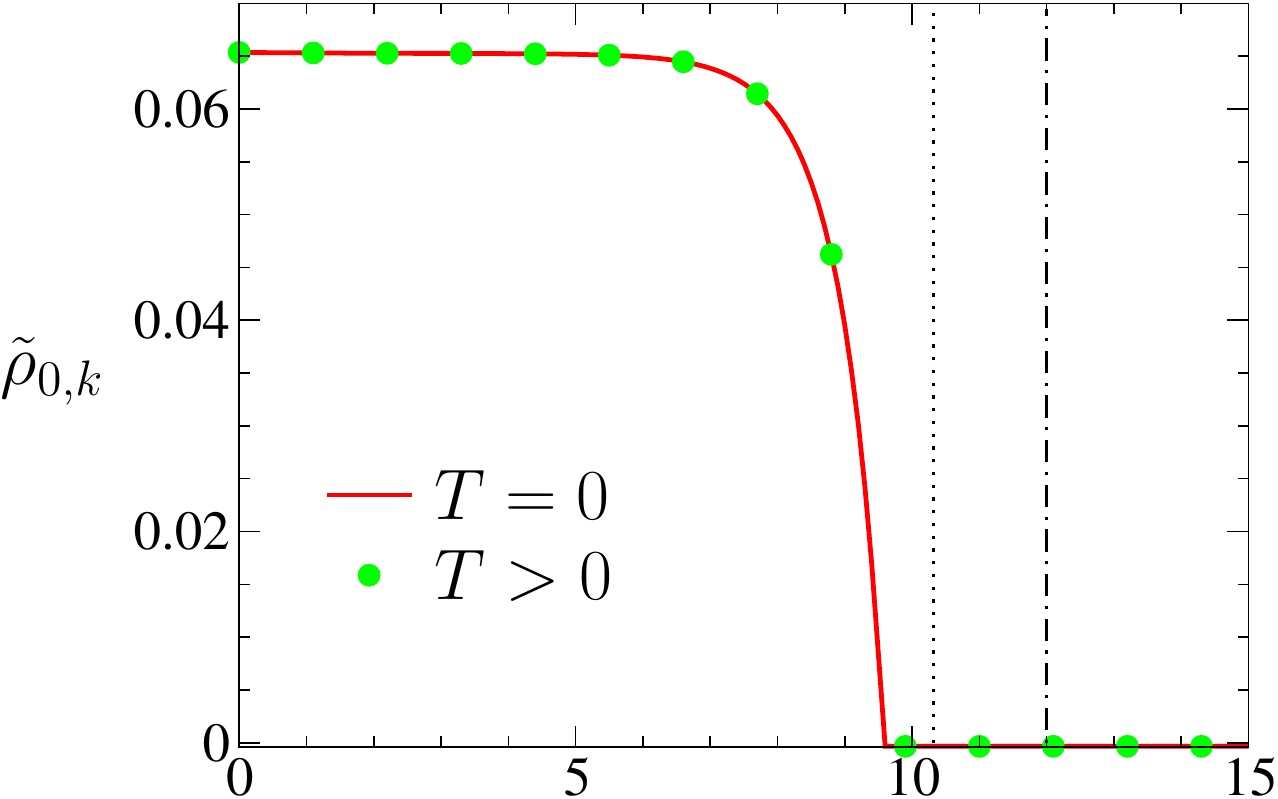}}
\centerline{\hspace{0.35cm}\includegraphics[width=5.65cm]{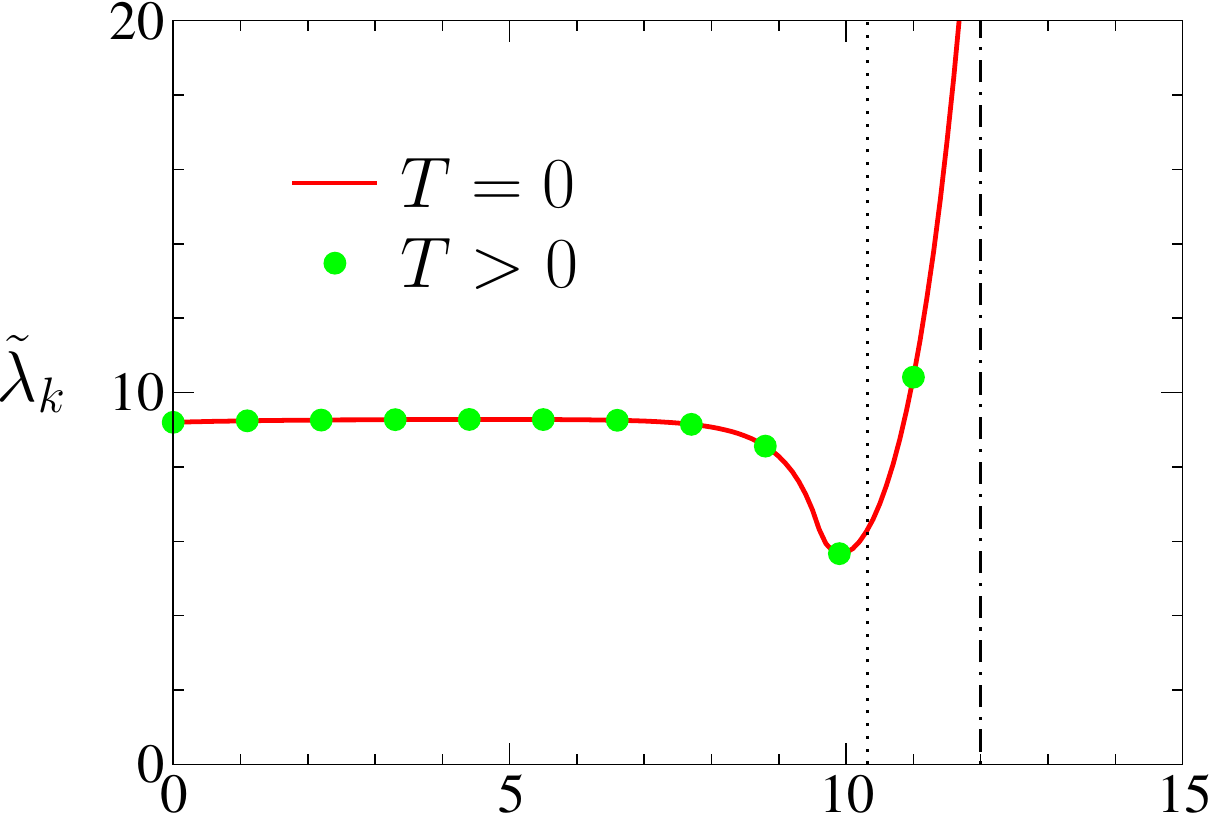}}
\centerline{\includegraphics[width=6.2cm]{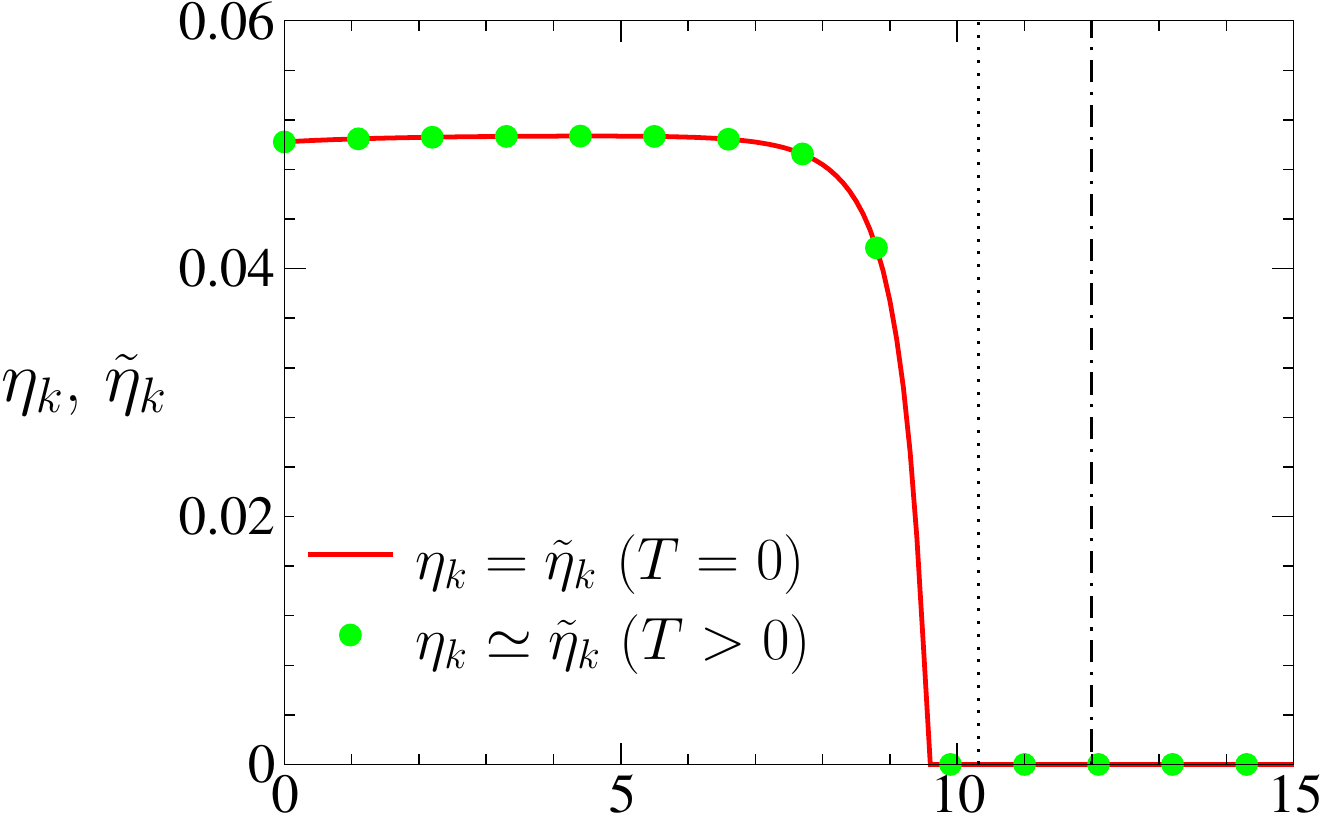}}
\centerline{\includegraphics[width=6.2cm]{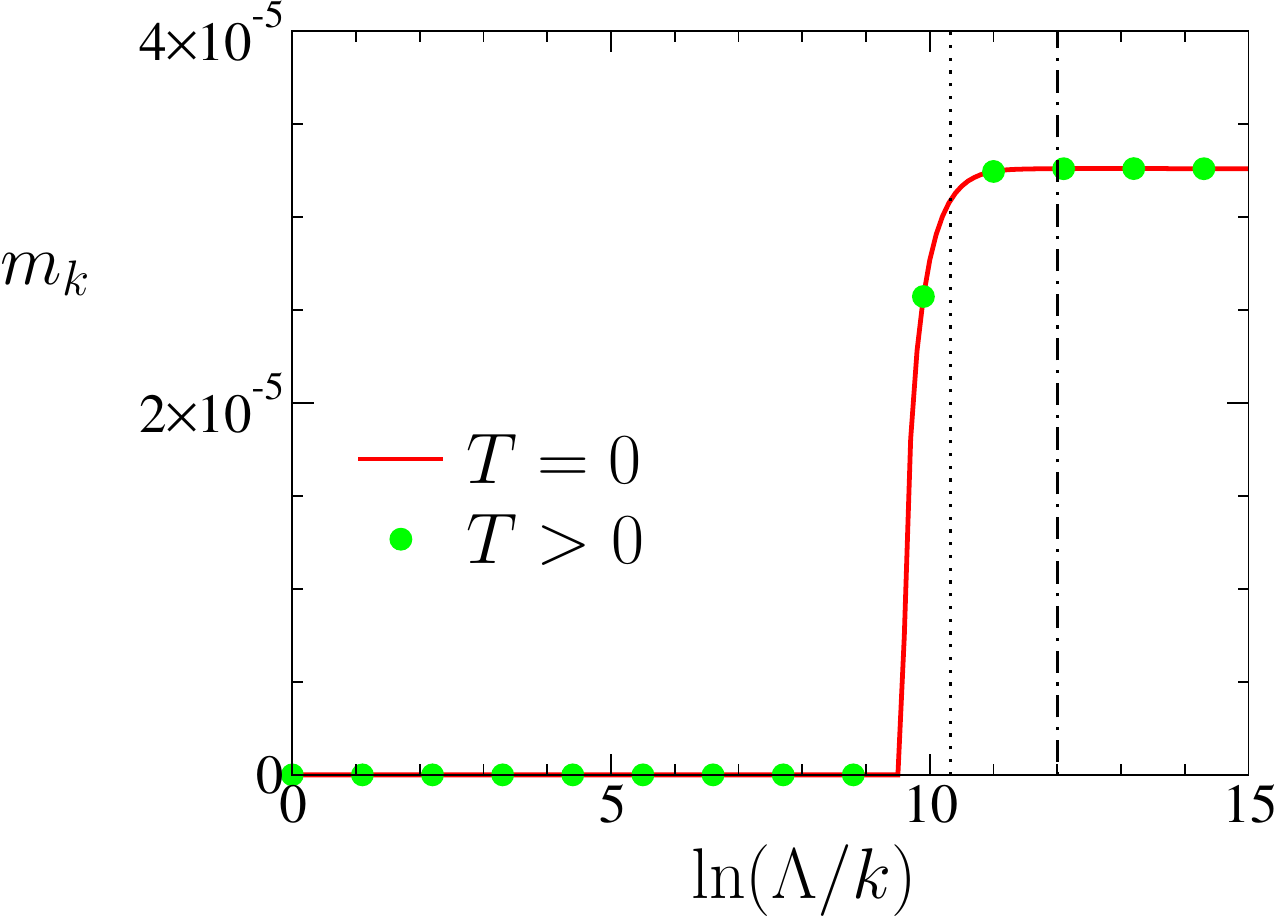}}
\caption{(Color online) Same as Fig.~\ref{fig_flow_qc} but in quantum disordered regime ($r_0=r_{0c}(1-10^{-6})$). The dotted and dash-dotted vertical lines show the momentum scales $k_\Delta$ and $k_T=\Lambda e^{-12}$, respectively.} 
\label{fig_flow_qd}
\end{figure}

Figure~\ref{fig_flow_qd} shows the flow in the quantum disordered regime. At $T=0$, the critical flow terminates at $k\sim k_\Delta$. For $k\lesssim k_\Delta$, $\trho_{0,k}$ and $\eta_k=\teta_k$ vanish while $\tlamb_k$ diverges; the (dimensionful) order parameter $\rho_{0,k}$ vanishes and $m_k=\sqrt{\delta_k/V_{A,k}}$ takes a finite value. As expected, a finite temperature has hardly any effect on the flow when $k_T\ll k_\Delta$. Only for $k_T\sim k_\Delta$ (i.e. near the crossover to the quantum critical regime) do we observe a modification of the $T=0$ flow.

\subsubsection{Renormalized classical regime} 
\label{subsubsec_RC_regime} 

\begin{figure}
\centerline{\includegraphics[width=6cm]{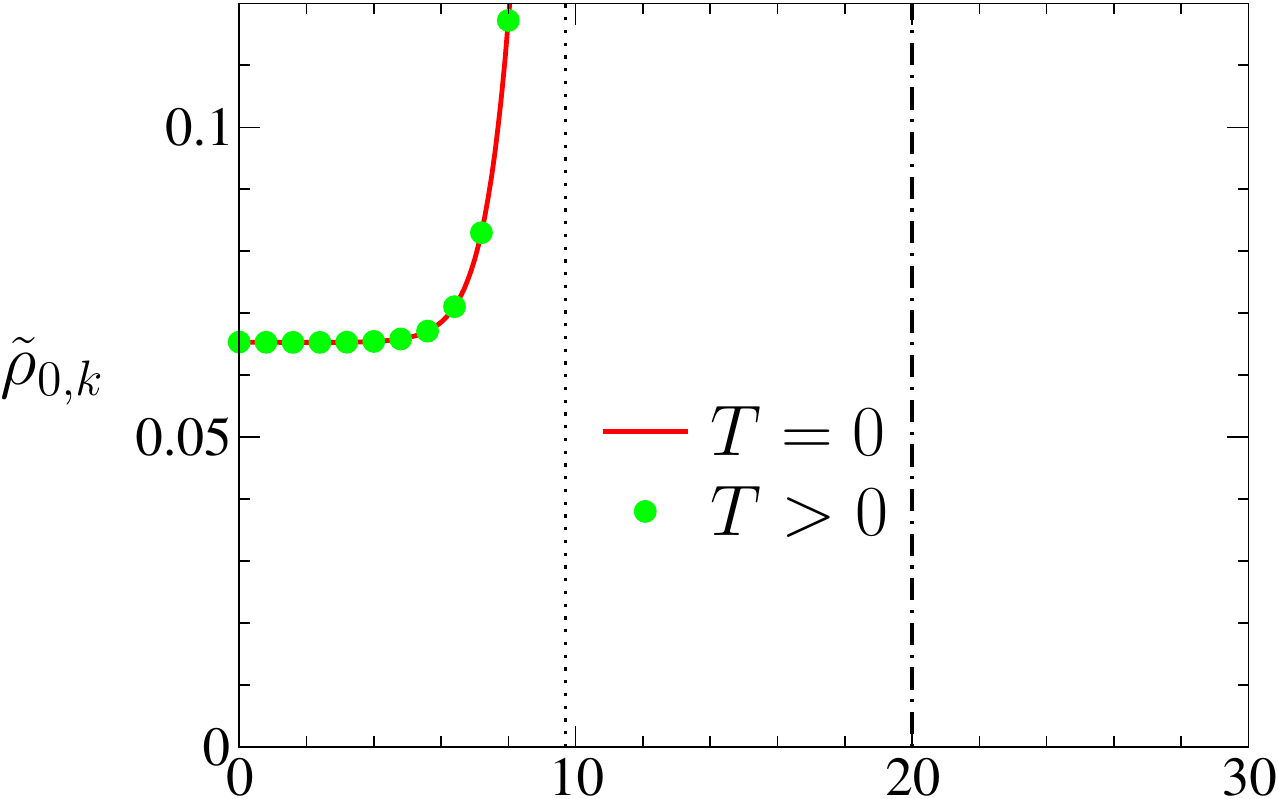}}
\centerline{\hspace{0.35cm}\includegraphics[width=5.7cm]{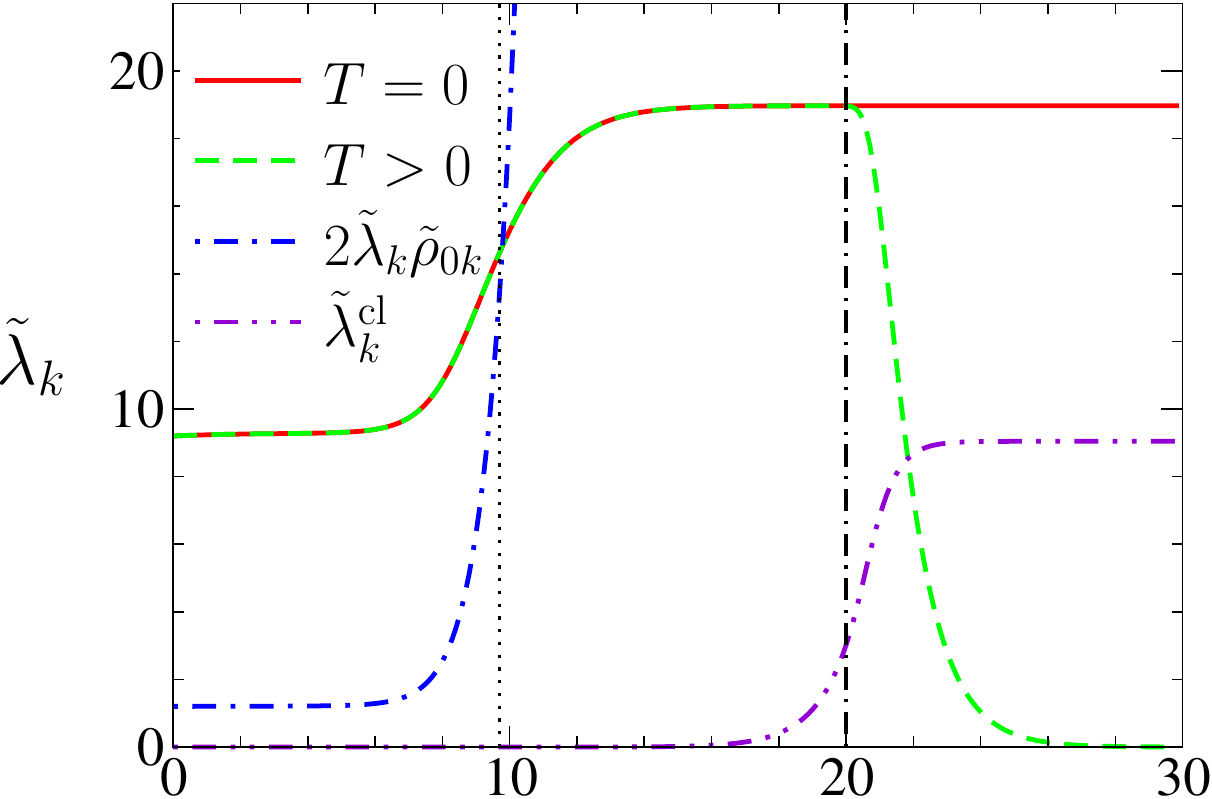}}
\centerline{\hspace{-0.1cm}\includegraphics[width=6.2cm]{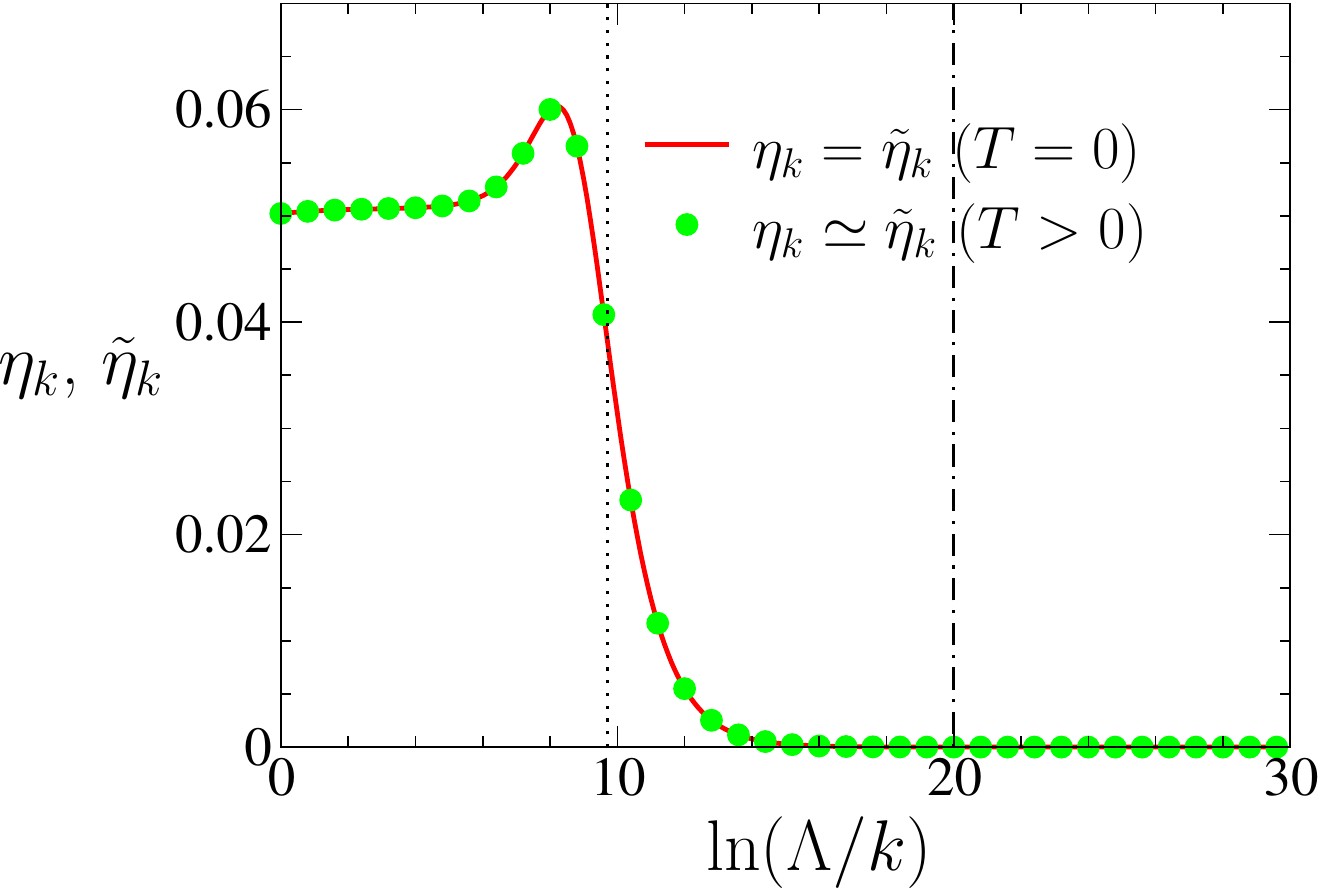}}
\caption{(Color online) Same as Fig.~\ref{fig_flow_qc} but in renormalized classical regime ($r_0=r_{0c}(1+10^{-5})$). The dotted and dash-dotted vertical lines show the momentum scales $k_J$ and $k_T=\Lambda e^{-20}$, respectively.} 
\label{fig_flow_rc}
\end{figure}

The flow in the renormalized classical regime is shown in Fig.~\ref{fig_flow_rc}. The plateaus observed for $k\gg k_J\sim\Lambda e^{-10}$ in $\trho_{0,k}$, $\tlamb_k$ and $\eta_k,\teta_k$ show that the behavior of the system at sufficiently high energies (or short distances) is critical. This critical regime terminates at the Josephson scale $k_J$. For $k\ll k_J$, longitudinal fluctuations are suppressed (the (dimensionless) mass $2\tlamb_k\trho_{0,k}$ of the longitudinal mode is much larger than unity) and the flow is dominated by the Goldstone modes. The anomalous dimensions $\eta_k$ and $\teta_k$ then nearly vanish and the dimensionless interaction $\tlamb_k$ exhibits a second plateau whose physical meaning is discussed below. At finite temperatures, this plateau terminates at the thermal scale $k_T$ with $\tlamb_k$ vanishing for $k\ll k_T$. 

In the Goldstone regime, the flow equations simplify into~\cite{note7}  
\begin{equation}
\begin{split}
\dt \trho_{0,k} &= \left(1-d - \frac{\eta_k+\teta_k}{2} \right) \trho_{0,k} - \frac{N-1}{2} \tIt , \\ 
\dt \tlamb_k &= (d-3) \tlamb_k - \tlamb_k^2 (N-1) \tJtt(0) , 
\end{split}
\label{rgeqrc}
\end{equation}
where the threshold functions are given in Appendix~\ref{subsec_threshold_goldstone}. $\trho_{0,k}$ becomes very large in the renormalized classical regime. For $k$ smaller than the inverse correlation length $\xi^{-1}$, $\trho_{0,k}$ will ultimately vanish, but $\xi$ being exponentially large this is not seen in Fig.~\ref{fig_flow_rc}. In Sec.~\ref{subsec_nlsm} we discuss in more detail the behavior of $\trho_{0,k}$ and make the connection with the quantum NL$\sigma$M. 

For $k_T\ll k\ll k_J$ (quantum Goldstone regime), we can take the $T=0$ limit of the threshold functions (Appendix~\ref{subsubsec_threshold_quantum_Goldstone}). We then find the fixed-point value
\begin{equation}
\tlamb^* = \frac{d-3}{(N-1)\tJtt(0)} = \frac{4\pi^{3/2}}{(N-1)(2-\sqrt{2})} , 
\end{equation}
where the last value is obtained with the exponential cutoff $r(Y)=1/(e^Y-1)$ [Eq.~(\ref{R1})] and for $d=2$. This fixed-point value shows up as a plateau $\tlamb_k\simeq \tlamb^*$ for $k_T\ll k \ll k_J$, which should not be confused with the plateau $\tlamb_k\simeq\tlamb^*_{\rm crit}$ corresponding to the critical regime $k_J\ll k\ll k_G$ (Fig.~\ref{fig_flow_rc}). As discussed in detail in Ref.~\cite{Dupuis11}, the constant value $\tlamb_k\simeq \tlamb^*$ and the diverging $\trho_{0,k}\sim k^{1-d}$ (corresponding to a constant value of the order parameter $\rho_{0,k}$) imply a vanishing of the longitudinal propagator in the infrared limit: $G_{\rm l}(p)\sim 1/(\wn^2+c^2\p^2)^{(3-d)/2}$ for $k_T\ll |\p|,|\wn|/c\ll k_J$ and $d<3$ (the vanishing is logarithmic for $d=3$).

In the classical Goldstone regime $k\ll k_T$, the flow is dominated by classical ($\wn=0$) transverse fluctuations. As a result, the threshold function $\tJtt$ becomes proportional to $\tilde T_k$ (Appendix~\ref{subsubsec_threshold_classical_Goldstone}) and $\tlamb_k$ vanishes linearly with $k$. Since only the classical component $\phibf(\r)\equiv \phibf(\r,i\wn=0)$ of the field matters, the effective action~(\ref{ansatz1}) becomes
\begin{equation}
\Gamma_k^{\rm cl}[\phibf] = \beta \int d^dr  \biggl\lbrace \frac{Z_{A,k}}{2} (\nablabf\phibf)^2 + \frac{\lamb_k}{2} (\rho-\rho_{0,k})^2 \biggr\rbrace 
\end{equation}
for $\rho_{0,k}>0$. Rescaling the field $\phibf\to\sqrt{T}\phibf$, we obtain the usual form 
\begin{equation}
\Gamma_k^{\rm cl}[\phibf] = \int d^dr  \biggl\lbrace \frac{Z_{A,k}}{2} (\nablabf\phibf)^2 + \frac{\lamb^{\rm cl}_k}{2} (\rho-\rho_{0,k})^2 \biggr\rbrace 
\end{equation}
of the effective action for a classical model in the LPA', with the coupling constant $\lamb_k^{\rm cl} = T \lamb_k$. The appropriate dimensionless variable 
\begin{equation}
\tlamb_k^{\rm cl} = \lamb_k^{\rm cl} Z_{A,k}^{-2} k^{d-4} 
\end{equation}
satisfies the RG equation 
\begin{equation}
\dt \tlamb_k^{\rm cl} = (d-4) \tlamb_k^{\rm cl} - (N-1) \tilde J^{\rm cl}_{k,\rm tt}(0) (\tlamb_k^{\rm cl})^2 
\end{equation}
with the threshold function 
\begin{equation}
\tilde J^{\rm cl}_{k,\rm tt} = \tilde T_k^{-1} \tJtt .
\end{equation}
This equation admits the fixed-point value 
\begin{equation}
\tlamb^{\rm cl}{}^* = \frac{d-4}{(N-1)\tilde J^{\rm cl}_{k,\rm tt}(0)} = \frac{4\pi}{(N-1)\ln 2} , 
\end{equation}
where the last value is obtained with the exponential cutoff and for $d=2$.

\subsection{Goldstone regime and NL$\sigma$M} 
\label{subsec_nlsm} 

In the Goldstone regime, the behavior of the system is governed by the Goldstone modes and we expect a description based on an effective NL$\sigma$M to be possible. To identify the coupling  constant of the effective NL$\sigma$M~\cite{Delamotte04}, we consider the following microscopic action 
\begin{align}
S[\varphibf] = \inttau \intr \biggl\lbrace & \frac{Z_A}{2} (\nablabf\varphibf)^2 + \frac{V_A}{2} (\dtau \varphibf)^2 \nonumber \\ & + \frac{\lamb}{2} (\rho-\rho_0)^2 \biggr\rbrace 
\end{align}
($\rho=\varphibf^2/2$), which is analog to the ansatz~(\ref{ansatz1},\ref{Ueff}) for the effective action $\Gamma_k[\phibf]$. This action can be written in the dimensionless form
\begin{equation}
S[\tvarphibf] = \int_0^{\tilde\beta} d\tilde\tau \int d^d\tilde r \biggl\lbrace \half (\nablabf_{\tilde r}\tvarphibf)^2 + \half (\partial_{\tilde\tau} \tvarphibf)^2 + \frac{\tlamb}{2} (\trho-\trho_{0})^2 \biggr\rbrace , 
\end{equation}
where $\tilde \r=k\r$, $\tilde\tau=\sqrt{Z_A/V_A}k\tau$, and $\tvarphibf$ is defined in the same way as $\tphibf$ in Eq.~(\ref{vardim}). Rescaling the field $\tvarphibf\to \sqrt{2\trho_{0}}\tvarphibf$, we then obtain
\begin{equation}
S[\tvarphibf] = \trho_0\int_0^{\tilde\beta} d\tilde\tau \int d^d\tilde r \biggl\lbrace  (\nablabf_{\tilde r}\tvarphibf)^2 + (\partial_{\tilde\tau} \tvarphibf)^2 + \frac{\tlamb\trho_0}{2} (\tvarphibf^2-1)^2 \biggr\rbrace .
\label{Stilde}
\end{equation}
In the limit $\tlamb\trho_0\gg 1$, the last term in~(\ref{Stilde}) imposes the constraint $\tvarphibf^2=1$, and we obtain a quantum NL$\sigma$M  with dimensionless coupling constant $\tilde g=1/2\trho_0$. 

In the NPRG approach, the coupling constant $\tilde g_k=1/2\trho_{0,k}$ satisfies the RG equation 
\begin{equation}
\dt \tilde g_k = - \left( 1-d - \frac{\eta_k+\teta_k}{2} \right) \tilde g_k  +(N-1) \tIt \tilde g_k^2, 
\label{geq}
\end{equation}
which can be deduced from~(\ref{rgeqrc}). Equation~(\ref{geq}) should be considered together with the RG equation of the dimensionless temperature $\tilde T_k=T/c_kk$, 
\begin{equation}
\dt \tilde T_k = - \left(1- \frac{\eta_k-\teta_k}{2} \right) \tilde T_k . 
\end{equation}
Following Chakravarty {\it et al.}~\cite{Chakravarty89}, we consider the coupling constants $\tilde g_k$ and $\tilde t_k=\tilde g_k\tilde T_k$ (rather than $\tilde g_k$ and $\tilde T_k$), with
\begin{equation}
\dt \tilde t_k = -(2-d-\eta_k) \tilde t_k + (N-1) \tIt \tilde g_k \tilde t_k . 
\end{equation}

In the quantum Goldstone regime $k\gg k_T$, the system is effectively in the zero-temperature limit  since $\tilde T_k\ll 1$. Let us first consider the theta cutoff function~\cite{note5} 
\begin{equation}
R_k(q)=Z_{A,k} \left(k^2-\q^2-\frac{\wn^2}{c_k^2} \right) \Theta \left(k^2-\q^2-\frac{\wn^2}{c_k^2} \right) .
\label{thetacut}
\end{equation}
In that case, one has $2\eta_k\trho_{0,k}=-\tIt$ (see Appendix~\ref{subsubsec_theta}), so that $\tilde g_k$ satisfies the flow equation
\begin{equation}
\dt \tilde g_k = - (1-d) \tilde g_k  - 2\frac{K_{d+1}}{d+1} (N-2) \tilde g_k^2
\label{geq1}
\end{equation}
We have used $\eta_k=\teta_k$ (which holds for $\tilde T_k\to 0$)  and $\tIt= -2K_{d+1}/(d+1)$ (with $K_d(2\pi)^d=4v_d(2\pi)^d$ the surface of the $d$-dimensional unit sphere). Equation~(\ref{geq1}) is nothing but the flow equation of the coupling constant in the $d+1$-dimensional classical NL$\sigma$M~\cite{Chakravarty89,Nelson77,Polyakov75}. It agrees with the result of Chakravarty {\it et al.} to order $\tilde g_k^2$~\cite{Chakravarty89}. In particular, we find that the $\calO(\tilde g_k^2)$ term vanishes for the O(2) model ($N=2$) as expected. Note however that the coefficient of $(N-2)\tilde g_k^2$ depends on the RG scheme and may therefore differ from the result of Ref.~\cite{Chakravarty89} obtained with a sharp cutoff. For an arbitrary function $R_k(q)$, the equality $2\eta_k\trho_{0,k}=-\tIt$ is in general violated, which does not allow us to recover the factor $N-2$ in the RG equation $\dt \tilde g_k$. For example, with the exponential cutoff $r(Y)=1/(e^Y-1)$, $-2\eta_k\trho_{0,k}/\tIt\simeq 1.11$ 
for $d=2$. This failure is clearly an 
artifact of the LPA'. In the regime where Eq.~(\ref{geq1}) is valid, the $\calO(\tilde g_k^2)$ term is small and the LPA' remains nevertheless a very good approximation to the flow equation of $\tilde g_k$. In one dimension, $\dt \tilde g_k$ is independent of the choice of the cutoff function $R_k$ to $\calO(\tilde g_k^2)$, in agreement with the fact that the beta function of the classical two-dimensional NL$\sigma$M is universal (i.e. independent of the RG scheme) to two-loop order~\cite{Friedan85}. 

There is a quantum classical crossover when $k\sim k_T$, and for $k\ll k_T$ the system is governed by an effective classical NL$\sigma$M with coupling constant $\tilde t_k$~\cite{Chakravarty89}. With the theta cutoff~(\ref{thetacut}), using $2\eta_k\trho_{0,k}=-\tIt$ and $\tIt=-2(K_d/d)\tilde t_k/\tilde g_k$ (Appendix~\ref{subsubsec_theta}), we recover the beta function 
\begin{equation}
\dt \tilde t_k = -(2-d) \tilde t_k - 2\frac{K_d}{d} (N-2) \tilde t_k^2 
\end{equation}
of the $d$-dimensional classical NL$\sigma$M~\cite{Chakravarty89}. As discussed above, for an arbitrary cutoff function $R_k(q)$, the coefficient $N-2$ is not exactly reproduced except for $d=2$ due to the universality of the beta function $\dt\tilde t_k$ to two-loop order in that case.

\subsection{O(2) model: BKT transition temperature} 
\label{subsec_bkt}

In the O(2) model, there is a finite-temperature BKT transition for $r_0\leq r_{0c}$. For the classical O(2) model, the NPRG reproduces most of the universal properties of the BKT transition~\cite{Graeter95,Gersdorff01}. In particular one finds a value $\tilde\rho_0^*$ of the dimensionless order parameter (the spin-wave ``stiffness'') such that the beta function $\beta( \tilde\rho_{0,k})=\dt \tilde\rho_{0,k}$ nearly vanishes for $\tilde\rho_{0,k}\geq \tilde\rho_0^*$. This implies the existence of a line of quasi-fixed points and enables to identify a low-temperature phase ($T<\Tkt$) where the running of the stiffness $\tilde\rho_{0,k}$, after a transient regime, becomes very slow, implying a very large (although not strictly infinite as expected in the low-temperature phase of the BKT transition) correlation length $\xi$. In this low-temperature phase, the anomalous dimension $\eta_k$ depends on the (slowly varying) stiffness $\tilde\rho_{0,k}$. It takes its largest value $\sim 1/4$ when the RG flow crosses 
over to the 
disordered (long-distance) regime (for $\tilde\rho_{0,k}\sim\tilde\rho_0^*$ and $k\sim\xi^{-1}$), and is then rapidly suppressed as $\tilde\rho_{0,k}$ further decreases. On the other hand, the beta function is well approximated by $\beta(\tilde\rho_{0,k})=\const\times(\tilde\rho_0^*-\tilde\rho_{0,k})^{3/2}$ for $\tilde\rho_{0,k}\leq\tilde\rho_0^*$, and the essential scaling $\xi\sim e^{\const/(T-\Tkt)^{1/2}}$ of the correlation length above the BKT transition temperature $\Tkt$ is reproduced~\cite{Gersdorff01}. Thus, although the NPRG approach does not yield a low-temperature phase with an infinite correlation length, it nevertheless allows us to estimate the BKT transition temperature from the value of $\tilde\rho^*_0$. A reasonable estimate of the BKT transition temperature in the two-dimensional XY model has been obtained using the NPRG~\cite{Machado10}. The same method has been used to determine $\Tkt$ in a two-dimensional Bose gas~\cite{Rancon12b} in very good agreement with Monte Carlo 
simulations~\cite{Prokofev01,Prokofev02}. We refer to Ref.~\cite{Rancon12b} for more details about the determination of the BKT transition temperature in the NPRG approach. 

\begin{figure}
\centerline{\includegraphics[width=7cm]{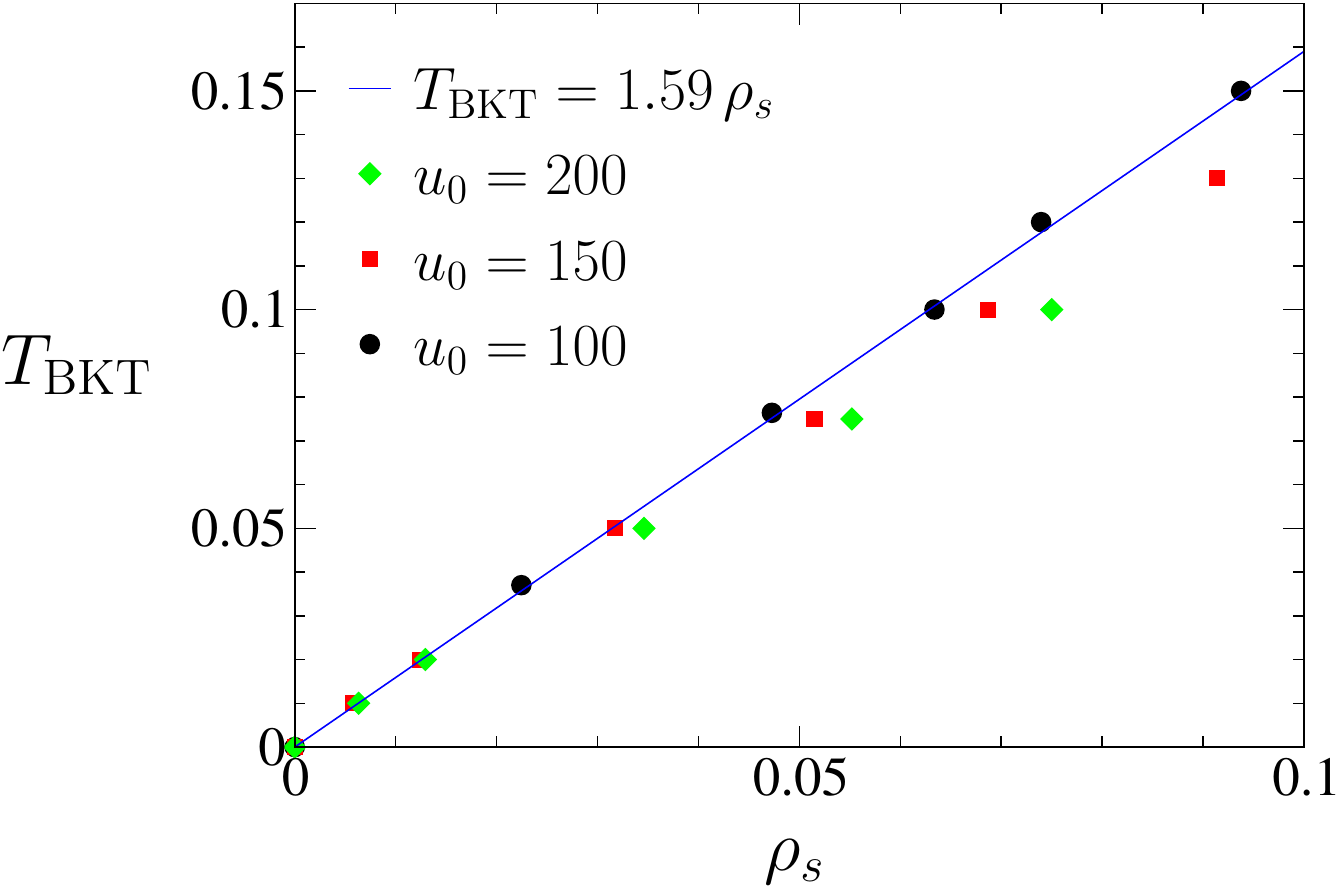}}
\caption{(Color online) BKT transition temperature $\Tkt$ vs the zero-temperature stiffness $\rho_s$ for various values of $u_0$ [$\Lambda=100$ and $c_0=1$]. Sufficiently close to the quantum critical point, the ratio $\Tkt/\rho_s\simeq 1.59$ is universal (as shown here by its independence with respect to $u_0$).}
\label{fig_Tkt} 
\end{figure}

The BKT transition temperature corresponds to an essential singularity in the scaling function $\calF_2(x)$. Since $\calF_2$ is universal, the ratios $\Tkt/|\Delta|$ and $\Tkt/\rho_s$ are also universal in the vicinity of the QCP (recall that $\rho_s\equiv\rho_s(T=0)$ is the stiffness in the zero-temperature ordered phase).
The NPRG approach predicts $\Tkt/\rho_s\simeq 1.59$ with the exponential cutoff (Fig.~\ref{fig_Tkt})
and $\Tkt/\rho_s\simeq 1.5$ with a theta cutoff $R_k(\q)=Z_{A,k}(k^2-\q^2)\Theta(k^2-\q^2)$ acting only on momenta. On the other hand, the ratio $\Tkt/\rho_s(\Tkt^-) = \pi/2$ is universal anywhere on the transition line, where $\rho_s(\Tkt^-)$ denotes the stiffness jump at the transition~\cite{Nelson77a,note13}. While our determination of $\Tkt$ is not precise enough to yield an accurate estimate of $\Tkt/\rho_s$, the latter is close to $\pi/2$, which implies that $\rho_s(\Tkt^-)$ is only slightly reduced with respect to the zero-temperature stiffness $\rho_s$.

More generally, in the low-temperature phase near the QCP we can write the stiffness in the scaling form $\rho_s(T)=\rho_s\calJ(T/\rho_s)$ with $\calJ(x)$ a universal scaling function satisfying $\calJ(0)=1$. The weak suppression of $\rho_s$ by thermal fluctuations for $T<\Tkt$ implies that $\calJ(x)$ remains close to unity for $x< \pi\rho_s(\Tkt^-)/2\rho_s$.

\section{Conclusion} 

Using a NPRG approach, we have obtained the universal function $\calF_N$ which determines the scaling form of the pressure near a relativistic QCP with O($N$) symmetry [Eq.~(\ref{pressure0})]. For $N\lesssim 10$, the results are in strong disagreement with the large-$N$ approach both in the renormalized classical and quantum critical regimes. If the large-$N$ approach is properly interpreted, its results in the renormalized classical regime can be reconciled with those of the NPRG approach. It fails however to describe the nonmonotonic behavior of the scaling function $\calF_N(x)$ in the quantum critical regime ($|x|\lesssim 1$) as predicted by the NPRG approach. A similar nonmonotonic behavior is observed in the scaling function of the entropy. 

We have also shown how the NPRG allows us to obtain a complete picture of the quantum O($N$) model in the vicinity of the zero-temperature QCP when $N\geq 3$. The characteristic momentum scales $k_T$ and $k_\Delta$, associated with temperature and detuning from the QCP, show up very clearly and yield distinctive RG flows in the quantum critical, quantum disordered and renormalized classical regimes. In the renormalized classical regime, where the physics is dominated by the $N-1$ Goldstone modes of zero-temperature  broken-symmetry phase, the NPRG equations reproduce those of the quantum O($N$) NL$\sigma$M~\cite{Chakravarty89}.

In the quantum O(2) model, the ratio between the BKT transition temperature $\Tkt$ and the zero-temperature stiffness $\rho_s(0)$ is universal near the QCP. The NPRG results show that $\Tkt/\rho_s(0)$ is close to the universal ratio $\Tkt/\rho_s(\Tkt^-)=\pi/2$, implying that the stiffness $\rho_s(\Tkt^-)$ at the transition is only slightly reduced with respect to $\rho_s(0)$. 

The superfluid--Mott-insulator (at constant density) of a Bose gas in an optical lattice provides us with a well controlled experimental realization of a relativistic QCP with a two-component (complex) field. Recent experiments have shown that it should be possible in the near future to observe quantum criticality associated with this QCP~\cite{Zhang12}. A measure of the temperature dependence of the pressure in the quantum critical regime would give an experimental estimate of the universal number $\calF_2(0)$ and enable a comparison with our theoretical result $\calF_2(0)\simeq 0.147$ (i.e. $\tilde\calC_2/2\simeq 0.767$). Whether the full scaling function $\calF_2(x)$ can be determined in the present experimental conditions requires a detailed study of the relativistic O(2) QCP in the Bose-Hubbard model which will be reported elsewhere.  

\begin{acknowledgments}
We would like to thank N. Wschebor and A. Ipp for useful discussions or correspondence. 
\end{acknowledgments} 


\appendix

\section{Calculation of $\Tr\ln g^{-1}$} 
\label{app_det} 
In this appendix, we compute 
\begin{equation}
\frac{1}{\beta V} \Tr\ln g^{-1} = \int_q \ln (\q^2+\wn^2/c^2+m^2/c^2) . 
\end{equation} 
Since $\Tr\ln g^{-1}$ is divergent, we subtract an infinite constant and consider
\begin{align}
D(m^2) ={}& \int_q \ln (\q^2+\wn^2/c^2+m^2/c^2) \nonumber \\ &
-\int_\q \int_\w \ln(\q^2+\w^2/c^2) .
\end{align}
It is convenient to write $D(m^2)=D_0(m^2)+D_1(m^2)$, where
\begin{align}
D_0(m^2) ={}& \int_\q \int_\w \bigl[ \ln(\q^2+\w^2/c^2+m^2/c^2) \nonumber \\ & 
- \ln(\q^2+\w^2/c^2) \bigr] ,
\end{align}
and 
\begin{align}
D_1(m^2) ={}& \int_q \ln(\q^2+\wn^2/c^2+m^2/c^2) \nonumber \\ & - \int_\q \int_\w \ln(\q^2+\w^2/c^2+m^2/c^2) .
\end{align}
Note that $D_1(m^2)$ vanishes at zero temperature.

\subsection{$D_0(m^2)$} 

Using $D_0(0)=0$ and 
\begin{equation}
D_0'(m^2) = \int_\q \int_\w \frac{1}{\w^2+c^2\q^2+m^2} 
= - \frac{6 r_{0c}}{u_0c^2} - \frac{m}{4\pi c^2} , 
\end{equation}
we obtain 
\begin{equation}
D_0(m^2) = - \frac{6 r_{0c}m^2}{u_0c^2} - \frac{m^3}{6\pi c^2} , 
\label{app4}
\end{equation}

\subsection{$D_1(m^2)$} 

Given that 
\begin{equation}
\frac{1}{\beta}\sum_{\wn} \frac{1}{\wn^2+a^2} - \int \frac{d\w}{2\pi} \frac{1}{\w^2+a^2} = \frac{1}{a} \frac{1}{e^{\beta a}-1} 
\end{equation}
($a>0$), we deduce 
\begin{equation}
D_1'(m^2) = \int_\q \frac{1}{\sqrt{c^2\q^2+m^2}} \frac{1}{\exp \left(\beta \sqrt{c^2\q^2+m^2}\right) -1}.
\end{equation}
Performing the momentum integral, we then obtain
\begin{align}
D_1'(m^2) &= \frac{1}{2\pi\beta c^2} \bigl[ -\beta (c\Lambda-m) + \ln (e^{\beta c\Lambda}-1) \nonumber \\ & \hspace{1.5cm} - \ln (e^{\beta m}-1) \bigr] \nonumber \\ 
&\simeq \frac{1}{2\pi\beta c^2} \left[ \beta m - \ln (e^{\beta m}-1) \right] 
\end{align}
for $T\ll c\Lambda$. From 
\begin{align}
D_1'(m^2) &= - \frac{1}{2\pi\beta c^2} \ln\left(1-e^{-\beta m}\right) \nonumber \\ 
&= \frac{1}{2\pi\beta c^2} \sum_{k=1}^\infty \frac{e^{-k\beta m}}{k} , 
\end{align}
we deduce 
\begin{align}
D_1(m^2) &= C - \frac{1}{\pi\beta^3 c^2} \sum_{k=1}^\infty \left( \frac{\beta m}{k^2} + \frac{1}{k^3} \right) e^{-k\beta m} \nonumber \\
&= C - \frac{1}{\pi\beta^3 c^2} \left[ \beta m \Li_2\bigl(e^{-\beta m}\bigr) + \Li_3\bigl(e^{-\beta m}\bigr) \right] ,
\label{app3}
\end{align}
where $\Li_s(z)$ is a polylogarithm [Eq.~(\ref{polylog})]. The integration constant $C$ is fixed by requiring $\lim_{m^2\to\infty} D_1(m^2) = 0$. Since $\Li_s(z)\simeq z$ for $|z|\to 0$, this gives $C=0$.

\subsection{$D(m^2)$} 

From Eqs.~(\ref{app4},\ref{app3}), we finally obtain 
\begin{multline}
D(m^2) = - \frac{6 r_{0c}m^2}{u_0c^2} - \frac{m^3}{6\pi c^2} \\ - \frac{1}{\pi\beta^3 c^2} \left[ \beta m \Li_2\bigl(e^{-\beta m}\bigr) + \Li_3\bigl(e^{-\beta m}\bigr) \right] . 
\label{app5}
\end{multline} 
Using the same method, we can compute $D(m^2)$ in the fermionic case (which amounts to replacing the bosonic Matsubara frequencies $\wn$ by fermionic ones). We have verified that we then reproduce the result of Ref.~\cite{Chamati11} obtained by a different method. The fermionic result differs from the bosonic one only by the sign of the argument of the polylogarithm functions.

\section{Dimensionless threshold functions} 

\subsection{Definition} 
\label{subsec__threshold_dim}

The dimensionless threshold functions [Eq.~(\ref{rgeqdim})] are defined by 
\begin{widetext}
\begin{equation}
\begin{split}
\tilde I_{k,\alpha} ={}& 2 v_d \int_{\tilde p} y^{d/2-1} [\eta_k Y r + 2Y^2r' + (\teta_k-\eta_k)(r+Yr')\twn^2 ] \tilde A_\alpha^{-2} , \\ 
\tilde J_{k,\alpha\beta}(0) ={}& 2 v_d \int_{\tilde p} y^{d/2-1} [\eta_k Y r + 2Y^2r' + (\teta_k-\eta_k)(r+Yr')\twn^2 ] \tilde A_\alpha^{-2} \tilde A_\beta^{-1} , \\
\frac{\partial}{\partial y} \tilde J_{k,\alpha\beta}(\tilde p) \biggl|_{\tilde p=0} ={}& - 4 \frac{v_d}{d} 
\int_{\tilde p} y^{d/2} \bigl\lbrace 2 \bigl[\eta_k Y r + 2Y^2r' + (\teta_k-\eta_k)(r+Yr')\twn^2 \bigr] \tilde A'_\alpha \tilde A_\alpha^{-3} \\ 
& - \bigl[\eta_k r + (\eta_k+4)Yr' + 2 Y^2 r'' + (\teta_k-\eta_k) (2r'+Yr'') \twn^2 \bigr] \tilde A_\alpha^{-2} \bigr\rbrace \tilde A'_\beta \tilde A_\beta^{-2} , \\ 
\frac{\partial}{\partial \twn^2} \tilde J_{k,\alpha\beta}(\tilde p) \biggl|_{\tilde p=0} ={}& 2 v_d \int_{\tilde p} y^{d/2-1} \bigl[\eta_k Y r + 2Y^2r' + (\teta_k-\eta_k)(r+Yr')\twn^2 \bigr] (\tilde A_1^2-\tilde A_2 \tilde A_\beta)\tilde A_\alpha^{-2} \tilde A_\beta^{-3} , 
\end{split} 
\end{equation}
\end{widetext} 
where 
\begin{equation}
\begin{gathered}
\tilde A_{\rm t} = Y(1+r) + \tdelta_k , \quad 
\tilde A_{\rm l} = \tilde A_{\rm t} + 2 \tlamb_k\trho_{0,k} , \\ 
\tilde A'_{\rm l}  = \tilde A'_{\rm t} = 1 + r +Yr' , 
\end{gathered}
\end{equation}
and 
\begin{equation}
\begin{split}
\tilde A_1 &= 2\twn (1+r+Yr') , \\ 
\tilde A_2 &= 1+r+Yr' + 2\twn^2 (2r'+Yr'') .
\end{split}
\end{equation}
We use the notations $v_d^{-1} = 2^{d+1} \pi^{d/2} \Gamma(d/2)$, $Y=y+\twn^2$, $r\equiv r(Y)$, $r'\equiv r'(Y)$, $r''\equiv r''(Y)$ and
\begin{equation}
\int_{\tilde p} \equiv  \tilde T_k \sum_{\twn}  \int_0^\infty dy .
\end{equation}

\subsection{Zero-temperature limit} 
\label{subsec_thresholdzero}

For $T=0$, Lorentz invariance implies that $Z_{A,k}=V_{A,k}$ and $\eta_k=\teta_k$. Using 
\begin{align}
v_d & \int_0^\infty dy\, y^{d/2-1} \intinf \frac{d\tw}{2\pi} f(Y) \nonumber \\ 
&= v_{d+1} \int_0^\infty dY \, Y^{(d+1)/2-1} f(Y) 
\label{app6}
\end{align}
for any function $f(Y)=f(y+\twn^2)$, we obtain 
\begin{equation}
\begin{split}
\tilde I_{k,\alpha} ={}& 2 v_{d+1} \int_0^\infty dY\,  Y^{(d+1)/2} (\eta_k r + 2Yr') \tilde A_\alpha^{-2} , \\ 
\tilde J_{k,\alpha\beta}(0) ={}& 2 v_{d+1} \int_0^\infty dY\,  Y^{(d+1)/2} (\eta_k r + 2Yr') \tilde A_\alpha^{-2} \tilde A_\beta^{-1} .
\end{split}
\label{app7}
\end{equation}
Equation~(\ref{app6}) implies  
\begin{equation}
\int_0^\infty dy\, y^{d/2} \intinf \frac{d\tw}{2\pi} f(Y) = \frac{v_{d+3}}{v_{d+2}} \int_0^\infty dY \, Y^{(d+1)/2} f(Y) .
\end{equation}
Using 
\begin{equation}
\frac{v_{d+3}}{v_{d+2}} = \frac{d}{d+1} \frac{v_{d+1}}{v_d} ,
\end{equation}
we obtain 
\begin{multline}
\frac{\partial}{\partial y } \bigl[ \tJlt(\tilde p) + \tJtl(\tilde p) \bigr] \Bigl|_{\tilde p=0} \\ 
= - 8 \frac{v_{d+1}}{d+1} \int_0^\infty dY \, Y^{(d+1)/2} (1+r+Yr') \tilde A_{\rm l}^{-2}\tilde A_{\rm t}^{-2} \\  
\times  \bigl[ Y (\eta_k r + 2Yr')(1+r+Yr') (\tilde A_{\rm l}^{-1} + \tilde A_{\rm t}^{-1}) \\ 
-\eta_k r - (\eta_k+4)Yr' - 2Y^2 r'' \bigr] .
\label{app8}
\end{multline} 
Equations~(\ref{app7}) and (\ref{app8}) yield the known RG equations of the $(d+1)$-dimensional (classical) O($N$) model in the LPA'. 

\subsection{Goldstone regime} 
\label{subsec_threshold_goldstone}

In the Goldstone regime $2\tlamb_k\trho_{0,k}\gg 1$, longitudinal fluctuations are subleading with respect to the transverse one. This yields $\tIl = \tJll = 0$, 
\begin{equation}
\begin{split}
\tIt &= 4 v_d \int_{\tilde p}  y^{d/2-1} Y^2 r' \tilde A_{\rm t}^{-2} , \\
\tJtt &= 4 v_d \int_{\tilde p}  y^{d/2-1} Y^2 r' \tilde A_{\rm t}^{-3} , 
\end{split}
\label{app9a}
\end{equation}
and
\begin{equation}
\begin{split} 
\frac{\partial}{\partial y} \tJlt(\tilde p)\Bigl|_{\tilde p=0} &= \frac{2}{\tlamb_k^2\trho_{0,k}^2} \frac{v_d}{d} \int_{\tilde p} y^{d/2}(2r'+Yr'') \frac{1+r+Yr'}{Y(1+r)^2} , \\  
\frac{\partial}{\partial y} \tJtl(\tilde p)\Bigl|_{\tilde p=0} &= -\frac{2}{\tlamb_k^2\trho_{0,k}^2} \frac{v_d}{d} \int_{\tilde p} y^{d/2} \frac{1+r+Yr'}{(1+r)^3} \\ & \hspace{2cm}  \times [2r'^2-(1+r)r''] 
\end{split}
\label{app9b}
\end{equation}
to leading order (we have set $\eta_k=\teta_k=0$ on the rhs of Eqs.~(\ref{app9a},\ref{app9b})). 

\subsubsection{Quantum Goldstone regime $k_T\ll k$} 
\label{subsubsec_threshold_quantum_Goldstone}

For $k_T\ll k$ (i.e. $T \ll c_kk$), we can take the zero-temperature limit in~(\ref{app9a},\ref{app9b}), which gives 
\begin{equation}
\begin{split}
\tIt &= 4 v_{d+1} \int_0^\infty dY \,Y^{(d-1)/2}\frac{r'}{(1+r)^2} , \\ 
\tJtt &= 4 v_{d+1} \int_0^\infty dY \,Y^{(d-3)/2} \frac{r'}{(1+r)^3} .
\end{split} 
\label{app10a}
\end{equation}
and
\begin{equation}
\begin{split} 
\frac{\partial}{\partial y} \tJlt(\tilde p)\Bigl|_{\tilde p=0} ={}& \frac{2}{\tlamb_k^2\trho_{0,k}^2} \frac{v_{d+1}}{d+1} \int_0^\infty dY \,Y^{(d-1)/2} \\ & \hspace{0cm} \times \frac{1+r+Yr'}{(1+r)^2} (2r'+Yr'') , \\  
\frac{\partial}{\partial y} \tJtl(\tilde p)\Bigl|_{\tilde p=0} ={}& -\frac{2}{\tlamb_k^2\trho_{0,k}^2} \frac{v_{d+1}}{d+1} \int_0^\infty dY \,Y^{(d+1)/2} \\ & \hspace{0cm} \times  \frac{1+r+Yr'}{(1+r)^3}[2r'^2-(1+r)r''] .
\end{split}
\label{app10b}
\end{equation}
Equations~(\ref{app10a},\ref{app10b}) can also be deduced from~(\ref{app7}) and (\ref{app8}) in the limit $2\tlamb_k\trho_{0,k}\gg 1$ and with $\tdelta_k=0$. 

\subsubsection{Classical Goldstone regime $k\ll k_T$} 
\label{subsubsec_threshold_classical_Goldstone}

For $k\ll k_T$, the Matsubara sums in~(\ref{app9a},\ref{app9b}) are dominated by the zero-frequency term $\twn=0$, which gives 
\begin{equation}
\begin{split}
\tIt &= 4 v_d \tilde T_k \int_0^\infty dy \, y^{d/2-1} \frac{r'}{(1+r)^2} , \\ 
\tJtt &= 4 v_d \tilde T_k \int_0^\infty dy \, y^{d/2-2} \frac{r'}{(1+r)^3} , 
\end{split}
\end{equation}
and 
\begin{equation}
\begin{split}
\frac{\partial}{\partial y} \tJlt(\tilde p)\Bigl|_{\tilde p=0} &= \frac{2}{\tlamb_k^2\trho_{0,k}^2} \frac{v_d}{d} \tilde T_k \int_0^\infty dy\, y^{d/2-1} \frac{1+r+yr'}{(1+r)^2} \\ & \hspace{2cm} \times (2r'+yr'')\\ 
\frac{\partial}{\partial y} \tJtl(\tilde p)\Bigl|_{\tilde p=0} &= -\frac{2}{\tlamb_k^2\trho_{0,k}^2} \frac{v_d}{d} \tilde T_k \int_0^\infty dy\, y^{d/2} \frac{1+r+yr'}{(1+r)^3} \\ & \hspace{2cm} \times [2r'{}^2-(1+r)r''].
\end{split}
\end{equation}

\subsubsection{Theta cutoff} 
\label{subsubsec_theta}

In this section, we show that with the theta cutoff~(\ref{thetacut}) the relation $2\eta_k\trho_{0,k}=-\tIt$ is satisfied in the Goldstone regime (to leading order in $1/\tlamb_k\trho_{0,k}$). Equation~(\ref{thetacut}) implies 
\begin{align}
r(Y) &= \frac{1-Y}{Y} \Theta(1-Y) , \nonumber \\ 
r'(Y) &= -\frac{1}{Y^2} \Theta(1-Y) - \frac{1-Y}{Y} \delta(1-Y) , \\ 
r''(Y) &= \frac{2}{Y^3} \Theta(1-Y) + \frac{2}{Y^2} \delta(1-Y) + \frac{1-Y}{Y} \delta'(1-Y) . \nonumber
\end{align}

At zero temperature, from Eq.~(\ref{app10a}) we deduce 
\begin{equation}
\tIt = - 8 \frac{v_{d+1}}{d+1} . 
\label{app11}
\end{equation}
Since $r(1+r+Yr')=r'(1+r+Yr')=0$, Eqs.~(\ref{app10b}) simplify into
\begin{multline} 
\frac{\partial}{\partial y} \tJlt(\tilde p)\Bigl|_{\tilde p=0} = \frac{\partial}{\partial y} \tJtl(\tilde p)\Bigl|_{\tilde p=0} \\ = 
\frac{2}{\tlamb_k^2\trho_{0,k}^2} \frac{v_{d+1}}{d+1} \int_0^\infty dY\, Y^{(d+1)/2}r'' \frac{1+r+Yr'}{(1+r)^2} .
\end{multline}
The product $r''(1+r+Yr')$ is ill-defined with the theta cutoff because of the derivative $\delta'(1-Y)$ of the Dirac function in $r''$. To circumvent this difficulty, we integrate by part, 
\begin{align}
& \frac{\partial}{\partial y} \tJlt(\tilde p)\Bigl|_{\tilde p=0} = \frac{\partial}{\partial y} \tJtl(\tilde p)\Bigl|_{\tilde p=0} \nonumber \\ 
={}& - \frac{2}{\tlamb_k^2\trho_{0,k}^2} \frac{v_{d+1}}{d+1} \int_0^1 dY\, \biggl\lbrace r' \frac{\partial}{\partial Y} \frac{Y^{(d+1)/2}}{1+r} \nonumber \\ & \hspace{1.3cm} + \frac{r'{}^2}{2} 
\frac{\partial}{\partial Y} \frac{Y^{(d+3)/2}}{(1+r)^2} \biggr\rbrace  
=  \frac{1}{\tlamb_k^2\trho_{0,k}^2}\frac{v_{d+1}}{d+1} ,
\label{app12}
\end{align}
where we have used $Y(1+r)=1$ when $0\leq Y\leq 1$. From~(\ref{app11}) and (\ref{app12}), we deduce $2\eta_k\trho_{0,k}=-\tIt$. 

In the classical Goldstone regime, we retain only the zero-frequency term $\twn=0$ in the Matsubara sums. The calculation of $\tIt$ and $\eta_k$ is similar to the $T=0$ limit with the $(d+1)$-dimensional integrals over $Y$ replaced by $d$-dimensional integrals over $y$. Again we find $2\eta_k\trho_{0,k}=-\tIt$, with
\begin{equation}
\tIt = 4 v_d \tilde T_k \int_0^\infty dy \, y^{d/2-1} \frac{r'(y)}{[1+r(y)]^2} = - 8\tilde T_k \frac{v_d}{d} . 
\end{equation}


\begin{thebibliography}{72}%
\makeatletter
\providecommand \@ifxundefined [1]{%
 \@ifx{#1\undefined}
}%
\providecommand \@ifnum [1]{%
 \ifnum #1\expandafter \@firstoftwo
 \else \expandafter \@secondoftwo
 \fi
}%
\providecommand \@ifx [1]{%
 \ifx #1\expandafter \@firstoftwo
 \else \expandafter \@secondoftwo
 \fi
}%
\providecommand \natexlab [1]{#1}%
\providecommand \enquote  [1]{``#1''}%
\providecommand \bibnamefont  [1]{#1}%
\providecommand \bibfnamefont [1]{#1}%
\providecommand \citenamefont [1]{#1}%
\providecommand \href@noop [0]{\@secondoftwo}%
\providecommand \href [0]{\begingroup \@sanitize@url \@href}%
\providecommand \@href[1]{\@@startlink{#1}\@@href}%
\providecommand \@@href[1]{\endgroup#1\@@endlink}%
\providecommand \@sanitize@url [0]{\catcode `\\12\catcode `\$12\catcode
  `\&12\catcode `\#12\catcode `\^12\catcode `\_12\catcode `\%12\relax}%
\providecommand \@@startlink[1]{}%
\providecommand \@@endlink[0]{}%
\providecommand \url  [0]{\begingroup\@sanitize@url \@url }%
\providecommand \@url [1]{\endgroup\@href {#1}{\urlprefix }}%
\providecommand \urlprefix  [0]{URL }%
\providecommand \Eprint [0]{\href }%
\providecommand \doibase [0]{http://dx.doi.org/}%
\providecommand \selectlanguage [0]{\@gobble}%
\providecommand \bibinfo  [0]{\@secondoftwo}%
\providecommand \bibfield  [0]{\@secondoftwo}%
\providecommand \translation [1]{[#1]}%
\providecommand \BibitemOpen [0]{}%
\providecommand \bibitemStop [0]{}%
\providecommand \bibitemNoStop [0]{.\EOS\space}%
\providecommand \EOS [0]{\spacefactor3000\relax}%
\providecommand \BibitemShut  [1]{\csname bibitem#1\endcsname}%
\let\auto@bib@innerbib\@empty
\bibitem [{\citenamefont {Sachdev}(2011)}]{Sachdev_book}%
  \BibitemOpen
  \bibfield  {author} {\bibinfo {author} {\bibfnamefont {S.}~\bibnamefont
  {Sachdev}},\ }\href@noop {} {\emph {\bibinfo {title} {Quantum Phase
  Transitions}}},\ \bibinfo {edition} {2nd}\ ed.\ (\bibinfo  {publisher}
  {Cambridge University Press},\ \bibinfo {address} {Cambridge, England},\
  \bibinfo {year} {2011})\BibitemShut {NoStop}%
\bibitem [{\citenamefont {Podolsky}\ and\ \citenamefont
  {Sachdev}(2012)}]{Podolsky12}%
  \BibitemOpen
  \bibfield  {author} {\bibinfo {author} {\bibfnamefont {D.}~\bibnamefont
  {Podolsky}}\ and\ \bibinfo {author} {\bibfnamefont {S.}~\bibnamefont
  {Sachdev}},\ }\href {\doibase 10.1103/PhysRevB.86.054508} {\bibfield
  {journal} {\bibinfo  {journal} {Phys. Rev. B}\ }\textbf {\bibinfo {volume}
  {86}},\ \bibinfo {pages} {054508} (\bibinfo {year} {2012})}\BibitemShut
  {NoStop}%
\bibitem [{\citenamefont {Jaksch}\ \emph {et~al.}(1998)\citenamefont {Jaksch},
  \citenamefont {Bruder}, \citenamefont {Cirac}, \citenamefont {Gardiner},\
  and\ \citenamefont {Zoller}}]{Jaksch98}%
  \BibitemOpen
  \bibfield  {author} {\bibinfo {author} {\bibfnamefont {D.}~\bibnamefont
  {Jaksch}}, \bibinfo {author} {\bibfnamefont {C.}~\bibnamefont {Bruder}},
  \bibinfo {author} {\bibfnamefont {J.~I.}\ \bibnamefont {Cirac}}, \bibinfo
  {author} {\bibfnamefont {C.~W.}\ \bibnamefont {Gardiner}}, \ and\ \bibinfo
  {author} {\bibfnamefont {P.}~\bibnamefont {Zoller}},\ }\href {\doibase
  10.1103/PhysRevLett.81.3108} {\bibfield  {journal} {\bibinfo  {journal}
  {Phys. Rev. Lett.}\ }\textbf {\bibinfo {volume} {81}},\ \bibinfo {pages}
  {3108} (\bibinfo {year} {1998})}\BibitemShut {NoStop}%
\bibitem [{\citenamefont {Greiner}\ \emph {et~al.}(2002)\citenamefont
  {Greiner}, \citenamefont {Mandel}, \citenamefont {Esslinger}, \citenamefont
  {H\"ansch},\ and\ \citenamefont {Bloch}}]{Greiner02}%
  \BibitemOpen
  \bibfield  {author} {\bibinfo {author} {\bibfnamefont {M.}~\bibnamefont
  {Greiner}}, \bibinfo {author} {\bibfnamefont {O.}~\bibnamefont {Mandel}},
  \bibinfo {author} {\bibfnamefont {T.}~\bibnamefont {Esslinger}}, \bibinfo
  {author} {\bibfnamefont {T.~W.}\ \bibnamefont {H\"ansch}}, \ and\ \bibinfo
  {author} {\bibfnamefont {I.}~\bibnamefont {Bloch}},\ }\href {\doibase
  doi:10.1038/415039a} {\bibfield  {journal} {\bibinfo  {journal} {Nature}\
  }\textbf {\bibinfo {volume} {415}},\ \bibinfo {pages} {39} (\bibinfo {year}
  {2002})}\BibitemShut {NoStop}%
\bibitem [{\citenamefont {St\"oferle}\ \emph {et~al.}(2004)\citenamefont
  {St\"oferle}, \citenamefont {Moritz}, \citenamefont {Schori}, \citenamefont
  {K\"ohl},\ and\ \citenamefont {Esslinger}}]{Stoferle04}%
  \BibitemOpen
  \bibfield  {author} {\bibinfo {author} {\bibfnamefont {T.}~\bibnamefont
  {St\"oferle}}, \bibinfo {author} {\bibfnamefont {H.}~\bibnamefont {Moritz}},
  \bibinfo {author} {\bibfnamefont {C.}~\bibnamefont {Schori}}, \bibinfo
  {author} {\bibfnamefont {M.}~\bibnamefont {K\"ohl}}, \ and\ \bibinfo {author}
  {\bibfnamefont {T.}~\bibnamefont {Esslinger}},\ }\href {\doibase
  10.1103/PhysRevLett.92.130403} {\bibfield  {journal} {\bibinfo  {journal}
  {Phys. Rev. Lett.}\ }\textbf {\bibinfo {volume} {92}},\ \bibinfo {pages}
  {130403} (\bibinfo {year} {2004})}\BibitemShut {NoStop}%
\bibitem [{\citenamefont {Spielman}\ \emph {et~al.}(2007)\citenamefont
  {Spielman}, \citenamefont {Phillips},\ and\ \citenamefont
  {Porto}}]{Spielman07}%
  \BibitemOpen
  \bibfield  {author} {\bibinfo {author} {\bibfnamefont {I.~B.}\ \bibnamefont
  {Spielman}}, \bibinfo {author} {\bibfnamefont {W.~D.}\ \bibnamefont
  {Phillips}}, \ and\ \bibinfo {author} {\bibfnamefont {J.~V.}\ \bibnamefont
  {Porto}},\ }\href {\doibase 10.1103/PhysRevLett.98.080404} {\bibfield
  {journal} {\bibinfo  {journal} {Phys. Rev. Lett.}\ }\textbf {\bibinfo
  {volume} {98}},\ \bibinfo {pages} {080404} (\bibinfo {year}
  {2007})}\BibitemShut {NoStop}%
\bibitem [{\citenamefont {Fisher}\ \emph {et~al.}(1989)\citenamefont {Fisher},
  \citenamefont {Weichman}, \citenamefont {Grinstein},\ and\ \citenamefont
  {Fisher}}]{Fisher89}%
  \BibitemOpen
  \bibfield  {author} {\bibinfo {author} {\bibfnamefont {M.~P.~A.}\
  \bibnamefont {Fisher}}, \bibinfo {author} {\bibfnamefont {P.~B.}\
  \bibnamefont {Weichman}}, \bibinfo {author} {\bibfnamefont {G.}~\bibnamefont
  {Grinstein}}, \ and\ \bibinfo {author} {\bibfnamefont {D.~S.}\ \bibnamefont
  {Fisher}},\ }\href {\doibase 10.1103/PhysRevB.40.546} {\bibfield  {journal}
  {\bibinfo  {journal} {Phys. Rev. B}\ }\textbf {\bibinfo {volume} {40}},\
  \bibinfo {pages} {546} (\bibinfo {year} {1989})}\BibitemShut {NoStop}%
\bibitem [{\citenamefont {Ran\c{c}on}\ and\ \citenamefont
  {Dupuis}(2011{\natexlab{a}})}]{Rancon11b}%
  \BibitemOpen
  \bibfield  {author} {\bibinfo {author} {\bibfnamefont {A.}~\bibnamefont
  {Ran\c{c}on}}\ and\ \bibinfo {author} {\bibfnamefont {N.}~\bibnamefont
  {Dupuis}},\ }\href {\doibase 10.1103/PhysRevB.84.174513} {\bibfield
  {journal} {\bibinfo  {journal} {Phys. Rev. B}\ }\textbf {\bibinfo {volume}
  {84}},\ \bibinfo {pages} {174513} (\bibinfo {year}
  {2011}{\natexlab{a}})}\BibitemShut {NoStop}%
\bibitem [{\citenamefont {Podolsky}\ \emph {et~al.}(2011)\citenamefont
  {Podolsky}, \citenamefont {Auerbach},\ and\ \citenamefont
  {Arovas}}]{Podolsky11}%
  \BibitemOpen
  \bibfield  {author} {\bibinfo {author} {\bibfnamefont {D.}~\bibnamefont
  {Podolsky}}, \bibinfo {author} {\bibfnamefont {A.}~\bibnamefont {Auerbach}},
  \ and\ \bibinfo {author} {\bibfnamefont {D.~P.}\ \bibnamefont {Arovas}},\
  }\href {\doibase 10.1103/PhysRevB.84.174522} {\bibfield  {journal} {\bibinfo
  {journal} {Phys. Rev. B}\ }\textbf {\bibinfo {volume} {84}},\ \bibinfo
  {pages} {174522} (\bibinfo {year} {2011})}\BibitemShut {NoStop}%
\bibitem [{\citenamefont {Pollet}\ and\ \citenamefont
  {Prokof'ev}(2012)}]{Pollet12}%
  \BibitemOpen
  \bibfield  {author} {\bibinfo {author} {\bibfnamefont {L.}~\bibnamefont
  {Pollet}}\ and\ \bibinfo {author} {\bibfnamefont {N.}~\bibnamefont
  {Prokof'ev}},\ }\href {\doibase 10.1103/PhysRevLett.109.010401} {\bibfield
  {journal} {\bibinfo  {journal} {Phys. Rev. Lett.}\ }\textbf {\bibinfo
  {volume} {109}},\ \bibinfo {pages} {010401} (\bibinfo {year}
  {2012})}\BibitemShut {NoStop}%
\bibitem [{\citenamefont {Gazit}\ \emph {et~al.}(2012)\citenamefont {Gazit},
  \citenamefont {Podolsky},\ and\ \citenamefont {Auerbach}}]{Gazit12}%
  \BibitemOpen
  \bibfield  {author} {\bibinfo {author} {\bibfnamefont {S.}~\bibnamefont
  {Gazit}}, \bibinfo {author} {\bibfnamefont {D.}~\bibnamefont {Podolsky}}, \
  and\ \bibinfo {author} {\bibfnamefont {A.}~\bibnamefont {Auerbach}},\
  }\href@noop {} \Eprint
  {http://arxiv.org/abs/arXiv:1212.3759} {arXiv:1212.3759} \BibitemShut
  {NoStop}%
\bibitem [{\citenamefont {Chen}\ \emph {et~al.}(2013)\citenamefont {Chen},
  \citenamefont {Liu}, \citenamefont {Deng}, \citenamefont {Pollet},\ and\
  \citenamefont {Prokof'ev}}]{Chen13}%
  \BibitemOpen
  \bibfield  {author} {\bibinfo {author} {\bibfnamefont {K.}~\bibnamefont
  {Chen}}, \bibinfo {author} {\bibfnamefont {L.}~\bibnamefont {Liu}}, \bibinfo
  {author} {\bibfnamefont {Y.}~\bibnamefont {Deng}}, \bibinfo {author}
  {\bibfnamefont {L.}~\bibnamefont {Pollet}}, \ and\ \bibinfo {author}
  {\bibfnamefont {N.}~\bibnamefont {Prokof'ev}},\ }\href@noop {} 
  \Eprint {http://arxiv.org/abs/arXiv:1301.3139} {arXiv:1301.3139} \BibitemShut
  {NoStop}%
\bibitem [{\citenamefont {Endres}\ \emph {et~al.}(2012)\citenamefont {Endres},
  \citenamefont {Fukuhara}, \citenamefont {Pekker}, \citenamefont {Cheneau},
  \citenamefont {Schauß}, \citenamefont {Gross}, \citenamefont {Demler},
  \citenamefont {Kuhr},\ and\ \citenamefont {Bloch}}]{Endres12}%
  \BibitemOpen
  \bibfield  {author} {\bibinfo {author} {\bibfnamefont {M.}~\bibnamefont
  {Endres}}, \bibinfo {author} {\bibfnamefont {T.}~\bibnamefont {Fukuhara}},
  \bibinfo {author} {\bibfnamefont {D.}~\bibnamefont {Pekker}}, \bibinfo
  {author} {\bibfnamefont {M.}~\bibnamefont {Cheneau}}, \bibinfo {author}
  {\bibfnamefont {P.}~\bibnamefont {Schauß}}, \bibinfo {author} {\bibfnamefont
  {C.}~\bibnamefont {Gross}}, \bibinfo {author} {\bibfnamefont
  {E.}~\bibnamefont {Demler}}, \bibinfo {author} {\bibfnamefont
  {S.}~\bibnamefont {Kuhr}}, \ and\ \bibinfo {author} {\bibfnamefont
  {I.}~\bibnamefont {Bloch}},\ }\href {\doibase doi:10.1038/nature11255}
  {\bibfield  {journal} {\bibinfo  {journal} {Nature}\ }\textbf {\bibinfo
  {volume} {487}},\ \bibinfo {pages} {454} (\bibinfo {year}
  {2012})}\BibitemShut {NoStop}%
\bibitem [{\citenamefont {Chubukov}\ \emph {et~al.}(1994)\citenamefont
  {Chubukov}, \citenamefont {Sachdev},\ and\ \citenamefont {Ye}}]{Chubukov94}%
  \BibitemOpen
  \bibfield  {author} {\bibinfo {author} {\bibfnamefont {A.~V.}\ \bibnamefont
  {Chubukov}}, \bibinfo {author} {\bibfnamefont {S.}~\bibnamefont {Sachdev}}, \
  and\ \bibinfo {author} {\bibfnamefont {J.}~\bibnamefont {Ye}},\ }\href
  {\doibase 10.1103/PhysRevB.49.11919} {\bibfield  {journal} {\bibinfo
  {journal} {Phys. Rev. B}\ }\textbf {\bibinfo {volume} {49}},\ \bibinfo
  {pages} {11919} (\bibinfo {year} {1994})}\BibitemShut {NoStop}%
\bibitem [{\citenamefont {Berges}\ \emph {et~al.}(2002)\citenamefont {Berges},
  \citenamefont {Tetradis},\ and\ \citenamefont {Wetterich}}]{Berges02}%
  \BibitemOpen
  \bibfield  {author} {\bibinfo {author} {\bibfnamefont {J.}~\bibnamefont
  {Berges}}, \bibinfo {author} {\bibfnamefont {N.}~\bibnamefont {Tetradis}}, \
  and\ \bibinfo {author} {\bibfnamefont {C.}~\bibnamefont {Wetterich}},\ }\href
  {\doibase doi:10.1016/S0370-1573(01)00098-9} {\bibfield  {journal} {\bibinfo
  {journal} {Phys. Rep.}\ }\textbf {\bibinfo {volume} {363}},\ \bibinfo {pages}
  {223} (\bibinfo {year} {2002})}\BibitemShut {NoStop}%
\bibitem [{\citenamefont {Delamotte}(2007)}]{Delamotte07}%
  \BibitemOpen
  \bibfield  {author} {\bibinfo {author} {\bibfnamefont {B.}~\bibnamefont
  {Delamotte}},\ }\href@noop {} \Eprint
  {http://arxiv.org/abs/cond-mat/0702365} {cond-mat/0702365} \BibitemShut
  {NoStop}%
\bibitem [{\citenamefont {Kopietz}\ \emph {et~al.}(2010)\citenamefont
  {Kopietz}, \citenamefont {Bartosch},\ and\ \citenamefont
  {Sch\"utz}}]{Kopietz_book}%
  \BibitemOpen
  \bibfield  {author} {\bibinfo {author} {\bibfnamefont {P.}~\bibnamefont
  {Kopietz}}, \bibinfo {author} {\bibfnamefont {L.}~\bibnamefont {Bartosch}}, \
  and\ \bibinfo {author} {\bibfnamefont {F.}~\bibnamefont {Sch\"utz}},\
  }\href@noop {} {\emph {\bibinfo {title} {Introduction to the Functional
  Renormalization Group}}}\ (\bibinfo  {publisher} {Springer},\ \bibinfo
  {address} {Berlin},\ \bibinfo {year} {2010})\BibitemShut {NoStop}%
\bibitem [{\citenamefont {Chakravarty}\ \emph {et~al.}(1989)\citenamefont
  {Chakravarty}, \citenamefont {Halperin},\ and\ \citenamefont
  {Nelson}}]{Chakravarty89}%
  \BibitemOpen
  \bibfield  {author} {\bibinfo {author} {\bibfnamefont {S.}~\bibnamefont
  {Chakravarty}}, \bibinfo {author} {\bibfnamefont {B.~I.}\ \bibnamefont
  {Halperin}}, \ and\ \bibinfo {author} {\bibfnamefont {D.~R.}\ \bibnamefont
  {Nelson}},\ }\href {\doibase 10.1103/PhysRevB.39.2344} {\bibfield  {journal}
  {\bibinfo  {journal} {Phys. Rev. B}\ }\textbf {\bibinfo {volume} {39}},\
  \bibinfo {pages} {2344} (\bibinfo {year} {1989})}\BibitemShut {NoStop}%
\bibitem [{\citenamefont {Berezinskii}(1970)}]{Berezinskii70}%
  \BibitemOpen
  \bibfield  {author} {\bibinfo {author} {\bibfnamefont {V.~L.}\ \bibnamefont
  {Berezinskii}},\ }\href@noop {} {\bibfield  {journal} {\bibinfo  {journal}
  {Sov. Phys. JETP}\ }\textbf {\bibinfo {volume} {32}},\ \bibinfo {pages} {493}
  (\bibinfo {year} {1970})}\BibitemShut {NoStop}%
\bibitem [{\citenamefont {Berezinskii}(1971)}]{Berezinskii71}%
  \BibitemOpen
  \bibfield  {author} {\bibinfo {author} {\bibfnamefont {V.~L.}\ \bibnamefont
  {Berezinskii}},\ }\href@noop {} {\bibfield  {journal} {\bibinfo  {journal}
  {Sov. Phys. JETP}\ }\textbf {\bibinfo {volume} {34}},\ \bibinfo {pages} {610}
  (\bibinfo {year} {1971})}\BibitemShut {NoStop}%
\bibitem [{\citenamefont {Kosterlitz}\ and\ \citenamefont
  {Thouless}(1973)}]{Kosterlitz73}%
  \BibitemOpen
  \bibfield  {author} {\bibinfo {author} {\bibfnamefont {J.~M.}\ \bibnamefont
  {Kosterlitz}}\ and\ \bibinfo {author} {\bibfnamefont {D.~J.}\ \bibnamefont
  {Thouless}},\ }\href {\doibase doi:10.1088/0022-3719/6/7/010} {\bibfield
  {journal} {\bibinfo  {journal} {J. of Phys. C}\ }\textbf {\bibinfo {volume}
  {6}},\ \bibinfo {pages} {1181} (\bibinfo {year} {1973})}\BibitemShut
  {NoStop}%
\bibitem [{\citenamefont {Kosterlitz}\ and\ \citenamefont
  {Thouless}(1974)}]{Kosterlitz74}%
  \BibitemOpen
  \bibfield  {author} {\bibinfo {author} {\bibfnamefont {J.~M.}\ \bibnamefont
  {Kosterlitz}}\ and\ \bibinfo {author} {\bibfnamefont {D.~J.}\ \bibnamefont
  {Thouless}},\ }\href {\doibase doi:10.1088/0022-3719/7/6/005} {\bibfield
  {journal} {\bibinfo  {journal} {J. Phys. C}\ }\textbf {\bibinfo {volume}
  {7}},\ \bibinfo {pages} {1046} (\bibinfo {year} {1974})}\BibitemShut
  {NoStop}%
\bibitem [{not({\natexlab{a}})}]{note4}%
  \BibitemOpen
  \href@noop {} {} \bibinfo {note} {Hyperscaling implies
  that the pressure can be written as $P(T)=P_{\rm reg} + N(T^3/c^2)
  \bar\calF_N(\Delta/T)$ with $\bar\calF_N$ a universal scaling function. Since
  the regular part $P_{\rm reg}$ is nearly temperature independent in the
  critical regime, one obtains Eq.~(\ref{pressure0}) with
  $\calF_N(x)=\bar\calF_N(x)-x^3 \lim_{y\to\infty\times \sgn(x)}
  y^{-3}\bar\calF_N(y)$.}\BibitemShut {Stop}%
\bibitem [{not({\natexlab{b}})}]{note2}%
  \BibitemOpen
  \href@noop {} {} \bibinfo {note} {If the ultraviolet
  cutoff respects the Lorentz invariance of the action~(\ref{action1}), then
  the velocity $c$ is equal to the bare velocity $c_0$.}\BibitemShut {Stop}%
\bibitem [{\citenamefont {Zinn-Justin}(2007)}]{Zinn_book_2}%
  \BibitemOpen
  \bibfield  {author} {\bibinfo {author} {\bibfnamefont {J.}~\bibnamefont
  {Zinn-Justin}},\ }\href@noop {} {\emph {\bibinfo {title} {Phase Transitions
  and Renormalisation Group}}}\ (\bibinfo  {publisher} {Oxford University
  Press},\ \bibinfo {address} {Oxford},\ \bibinfo {year} {2007})\BibitemShut
  {NoStop}%
\bibitem [{\citenamefont {Dupuis}(2011)}]{Dupuis11}%
  \BibitemOpen
  \bibfield  {author} {\bibinfo {author} {\bibfnamefont {N.}~\bibnamefont
  {Dupuis}},\ }\href {\doibase 10.1103/PhysRevE.83.031120} {\bibfield
  {journal} {\bibinfo  {journal} {Phys. Rev. E}\ }\textbf {\bibinfo {volume}
  {83}},\ \bibinfo {pages} {031120} (\bibinfo {year} {2011})}\BibitemShut
  {NoStop}%
\bibitem [{not({\natexlab{c}})}]{note3}%
  \BibitemOpen
  \href@noop {} {} \bibinfo {note} {Note that the Ginzburg
  momentum scale $k_G$ can be obtained directly from dimensional
  analysis.}\BibitemShut {Stop}%
\bibitem [{not({\natexlab{d}})}]{note6}%
  \BibitemOpen
  \href@noop {} {} \bibinfo {note} {The stiffness $\rho_s$
  is defined by the increase $\Delta E=\half \rho_s \intr (\nablabf\n)^2$ of
  the energy when the direction $\n$ of the order parameter slowly varies in
  space. Equivalently, $\rho_s$ can be defined from the propagator $G_{\rm
  t}(\q,i\wn=0)=|\mean{\varphibf}|^2/\rho_s\q^2$ of the $N-1$ transverse modes
  [$G_{\rm t}(q)\equiv g(q)$ in the large-$N$ approach,
  Eq.~(\ref{ginverse})].}\BibitemShut {Stop}%
\bibitem [{\citenamefont {Sachdev}(1993)}]{Sachdev93}%
  \BibitemOpen
  \bibfield  {author} {\bibinfo {author} {\bibfnamefont {S.}~\bibnamefont
  {Sachdev}},\ }\href {\doibase 10.1016/0370-2693(93)90935-B} {\bibfield
  {journal} {\bibinfo  {journal} {Phys. Lett. B}\ }\textbf {\bibinfo {volume}
  {309}},\ \bibinfo {pages} {285 } (\bibinfo {year} {1993})}\BibitemShut
  {NoStop}%
\bibitem [{\citenamefont {Wetterich}(1993)}]{Wetterich93}%
  \BibitemOpen
  \bibfield  {author} {\bibinfo {author} {\bibfnamefont {C.}~\bibnamefont
  {Wetterich}},\ }\href {\doibase doi:10.1016/0370-2693(93)90726-X} {\bibfield
  {journal} {\bibinfo  {journal} {Phys. Lett. B}\ }\textbf {\bibinfo {volume}
  {301}},\ \bibinfo {pages} {90} (\bibinfo {year} {1993})}\BibitemShut
  {NoStop}%
\bibitem [{\citenamefont {D'Attanasio}\ and\ \citenamefont
  {Morris}(1997)}]{DAttanasio97}%
  \BibitemOpen
  \bibfield  {author} {\bibinfo {author} {\bibfnamefont {M.}~\bibnamefont
  {D'Attanasio}}\ and\ \bibinfo {author} {\bibfnamefont {T.~R.}\ \bibnamefont
  {Morris}},\ }\href {\doibase http://dx.doi.org/10.1016/S0370-2693(97)00866-6}
  {\bibfield  {journal} {\bibinfo  {journal} {Phys. Lett. B}\ }\textbf
  {\bibinfo {volume} {409}},\ \bibinfo {pages} {363} (\bibinfo {year}
  {1997})}\BibitemShut {NoStop}%
\bibitem [{not({\natexlab{e}})}]{note15}%
  \BibitemOpen
  \href@noop {} {} \bibinfo {note} {The difference between
  $P(T)$ and $P(0)$ is very small (typically of order $0.1 T^3/c^2$). To obtain
  the pressure at low temperatures, it is therefore necessary to compute $P(T)$
  with a very high precision, which is quite difficult numerically when dealing
  with the full potential.}\BibitemShut {Stop}%
\bibitem [{\citenamefont {Blaizot}\ \emph {et~al.}(2007)\citenamefont
  {Blaizot}, \citenamefont {Ipp}, \citenamefont {Méndez-Galain},\ and\
  \citenamefont {Wschebor}}]{Blaizot07a}%
  \BibitemOpen
  \bibfield  {author} {\bibinfo {author} {\bibfnamefont {J.-P.}\ \bibnamefont
  {Blaizot}}, \bibinfo {author} {\bibfnamefont {A.}~\bibnamefont {Ipp}},
  \bibinfo {author} {\bibfnamefont {R.}~\bibnamefont {Méndez-Galain}}, \ and\
  \bibinfo {author} {\bibfnamefont {N.}~\bibnamefont {Wschebor}},\ }\href
  {\doibase 10.1016/j.nuclphysa.2006.11.139} {\bibfield  {journal} {\bibinfo
  {journal} {Nucl. Phys. A}\ }\textbf {\bibinfo {volume} {784}},\ \bibinfo
  {pages} {376 } (\bibinfo {year} {2007})}\BibitemShut {NoStop}%
\bibitem [{\citenamefont {Blaizot}\ \emph {et~al.}(2011)\citenamefont
  {Blaizot}, \citenamefont {Ipp},\ and\ \citenamefont {Wschebor}}]{Blaizot11}%
  \BibitemOpen
  \bibfield  {author} {\bibinfo {author} {\bibfnamefont {J.-P.}\ \bibnamefont
  {Blaizot}}, \bibinfo {author} {\bibfnamefont {A.}~\bibnamefont {Ipp}}, \ and\
  \bibinfo {author} {\bibfnamefont {N.}~\bibnamefont {Wschebor}},\ }\href
  {\doibase 10.1016/j.nuclphysa.2010.10.007} {\bibfield  {journal} {\bibinfo
  {journal} {Nucl. Phys. A}\ }\textbf {\bibinfo {volume} {849}},\ \bibinfo
  {pages} {165 } (\bibinfo {year} {2011})}\BibitemShut {NoStop}%
\bibitem [{\citenamefont {Blaizot}\ \emph {et~al.}(2006)\citenamefont
  {Blaizot}, \citenamefont {M\'endez-Galain},\ and\ \citenamefont
  {Wschebor}}]{Blaizot06}%
  \BibitemOpen
  \bibfield  {author} {\bibinfo {author} {\bibfnamefont {J.-P.}\ \bibnamefont
  {Blaizot}}, \bibinfo {author} {\bibfnamefont {R.}~\bibnamefont
  {M\'endez-Galain}}, \ and\ \bibinfo {author} {\bibfnamefont {N.}~\bibnamefont
  {Wschebor}},\ }\href {\doibase 10.1016/j.physletb.2005.10.086} {\bibfield
  {journal} {\bibinfo  {journal} {Phys. Lett. B}\ }\textbf {\bibinfo {volume}
  {632}},\ \bibinfo {pages} {571} (\bibinfo {year} {2006})}\BibitemShut
  {NoStop}%
\bibitem [{\citenamefont {Benitez}\ \emph {et~al.}(2009)\citenamefont
  {Benitez}, \citenamefont {Blaizot}, \citenamefont {Chat\'e}, \citenamefont
  {Delamotte}, \citenamefont {M\'endez-Galain},\ and\ \citenamefont
  {Wschebor}}]{Benitez09}%
  \BibitemOpen
  \bibfield  {author} {\bibinfo {author} {\bibfnamefont {F.}~\bibnamefont
  {Benitez}}, \bibinfo {author} {\bibfnamefont {J.~P.}\ \bibnamefont
  {Blaizot}}, \bibinfo {author} {\bibfnamefont {H.}~\bibnamefont {Chat\'e}},
  \bibinfo {author} {\bibfnamefont {B.}~\bibnamefont {Delamotte}}, \bibinfo
  {author} {\bibfnamefont {R.}~\bibnamefont {M\'endez-Galain}}, \ and\ \bibinfo
  {author} {\bibfnamefont {N.}~\bibnamefont {Wschebor}},\ }\href {\doibase
  10.1103/PhysRevE.80.030103} {\bibfield  {journal} {\bibinfo  {journal} {Phys.
  Rev. E}\ }\textbf {\bibinfo {volume} {80}},\ \bibinfo {pages} {030103(R)}
  (\bibinfo {year} {2009})}\BibitemShut {NoStop}%
\bibitem [{\citenamefont {Benitez}\ \emph {et~al.}(2012)\citenamefont
  {Benitez}, \citenamefont {Blaizot}, \citenamefont {Chat\'e}, \citenamefont
  {Delamotte}, \citenamefont {M\'endez-Galain},\ and\ \citenamefont
  {Wschebor}}]{Benitez12}%
  \BibitemOpen
  \bibfield  {author} {\bibinfo {author} {\bibfnamefont {F.}~\bibnamefont
  {Benitez}}, \bibinfo {author} {\bibfnamefont {J.-P.}\ \bibnamefont
  {Blaizot}}, \bibinfo {author} {\bibfnamefont {H.}~\bibnamefont {Chat\'e}},
  \bibinfo {author} {\bibfnamefont {B.}~\bibnamefont {Delamotte}}, \bibinfo
  {author} {\bibfnamefont {R.}~\bibnamefont {M\'endez-Galain}}, \ and\ \bibinfo
  {author} {\bibfnamefont {N.}~\bibnamefont {Wschebor}},\ }\href {\doibase
  10.1103/PhysRevE.85.026707} {\bibfield  {journal} {\bibinfo  {journal} {Phys.
  Rev. E}\ }\textbf {\bibinfo {volume} {85}},\ \bibinfo {pages} {026707}
  (\bibinfo {year} {2012})}\BibitemShut {NoStop}%
\bibitem [{\citenamefont {Canet}\ \emph {et~al.}(2003)\citenamefont {Canet},
  \citenamefont {Delamotte}, \citenamefont {Mouhanna},\ and\ \citenamefont
  {Vidal}}]{Canet03a}%
  \BibitemOpen
  \bibfield  {author} {\bibinfo {author} {\bibfnamefont {L.}~\bibnamefont
  {Canet}}, \bibinfo {author} {\bibfnamefont {B.}~\bibnamefont {Delamotte}},
  \bibinfo {author} {\bibfnamefont {D.}~\bibnamefont {Mouhanna}}, \ and\
  \bibinfo {author} {\bibfnamefont {J.}~\bibnamefont {Vidal}},\ }\href
  {\doibase 10.1103/PhysRevD.67.065004} {\bibfield  {journal} {\bibinfo
  {journal} {Phys. Rev. D}\ }\textbf {\bibinfo {volume} {67}},\ \bibinfo
  {pages} {065004} (\bibinfo {year} {2003})}\BibitemShut {NoStop}%
\bibitem [{not({\natexlab{f}})}]{note10}%
  \BibitemOpen
  \href@noop {} {} \bibinfo {note} {In the $T=0$ ordered
  phase, the truncation of the effective potential about $\rho_{0,k}$ gives the
  exact result for the pressure in the limit $N\to\infty$~\cite{DAttanasio97}.
  In the renormalized classical regime, where the RG flow remains in the
  ordered phase down to exponential small values of $k$, the truncated LPA' is
  also nearly exact for the computation of the pressure.}\BibitemShut {Stop}%
\bibitem [{\citenamefont {Ran\c{c}on}\ and\ \citenamefont
  {Dupuis}(2011{\natexlab{b}})}]{Rancon11a}%
  \BibitemOpen
  \bibfield  {author} {\bibinfo {author} {\bibfnamefont {A.}~\bibnamefont
  {Ran\c{c}on}}\ and\ \bibinfo {author} {\bibfnamefont {N.}~\bibnamefont
  {Dupuis}},\ }\href {\doibase 10.1103/PhysRevB.83.172501} {\bibfield
  {journal} {\bibinfo  {journal} {Phys. Rev. B}\ }\textbf {\bibinfo {volume}
  {83}},\ \bibinfo {pages} {172501} (\bibinfo {year}
  {2011}{\natexlab{b}})}\BibitemShut {NoStop}%
\bibitem [{\citenamefont {Ran\c{c}on}\ and\ \citenamefont
  {Dupuis}(2012{\natexlab{a}})}]{Rancon12d}%
  \BibitemOpen
  \bibfield  {author} {\bibinfo {author} {\bibfnamefont {A.}~\bibnamefont
  {Ran\c{c}on}}\ and\ \bibinfo {author} {\bibfnamefont {N.}~\bibnamefont
  {Dupuis}},\ }\href {\doibase 10.1103/PhysRevA.86.043624} {\bibfield
  {journal} {\bibinfo  {journal} {Phys. Rev. A}\ }\textbf {\bibinfo {volume}
  {86}},\ \bibinfo {pages} {043624} (\bibinfo {year}
  {2012}{\natexlab{a}})}\BibitemShut {NoStop}%
\bibitem [{\citenamefont {Capogrosso-Sansone}\ \emph
  {et~al.}(2007)\citenamefont {Capogrosso-Sansone}, \citenamefont {Prokof'ev},\
  and\ \citenamefont {Svistunov}}]{Capogrosso07}%
  \BibitemOpen
  \bibfield  {author} {\bibinfo {author} {\bibfnamefont {B.}~\bibnamefont
  {Capogrosso-Sansone}}, \bibinfo {author} {\bibfnamefont {N.~V.}\ \bibnamefont
  {Prokof'ev}}, \ and\ \bibinfo {author} {\bibfnamefont {B.~V.}\ \bibnamefont
  {Svistunov}},\ }\href {\doibase 10.1103/PhysRevB.75.134302} {\bibfield
  {journal} {\bibinfo  {journal} {Phys. Rev. B}\ }\textbf {\bibinfo {volume}
  {75}},\ \bibinfo {pages} {134302} (\bibinfo {year} {2007})}\BibitemShut
  {NoStop}%
\bibitem [{\citenamefont {Capogrosso-Sansone}\ \emph
  {et~al.}(2010)\citenamefont {Capogrosso-Sansone}, \citenamefont {Giorgini},
  \citenamefont {Pilati}, \citenamefont {Pollet}, \citenamefont {Prokof'ev},
  \citenamefont {Svistunov},\ and\ \citenamefont {Troyer}}]{Capogrosso10}%
  \BibitemOpen
  \bibfield  {author} {\bibinfo {author} {\bibfnamefont {B.}~\bibnamefont
  {Capogrosso-Sansone}}, \bibinfo {author} {\bibfnamefont {S.}~\bibnamefont
  {Giorgini}}, \bibinfo {author} {\bibfnamefont {S.}~\bibnamefont {Pilati}},
  \bibinfo {author} {\bibfnamefont {L.}~\bibnamefont {Pollet}}, \bibinfo
  {author} {\bibfnamefont {N.}~\bibnamefont {Prokof'ev}}, \bibinfo {author}
  {\bibfnamefont {B.}~\bibnamefont {Svistunov}}, \ and\ \bibinfo {author}
  {\bibfnamefont {M.}~\bibnamefont {Troyer}},\ }\href {\doibase
  10.1088/1367-2630/12/4/043010} {\bibfield  {journal} {\bibinfo  {journal}
  {New J. Phys.}\ }\textbf {\bibinfo {volume} {12}},\ \bibinfo {pages} {043010}
  (\bibinfo {year} {2010})}\BibitemShut {NoStop}%
\bibitem [{\citenamefont {Tetradis}\ and\ \citenamefont
  {Wetterich}(1993)}]{Tetradis93}%
  \BibitemOpen
  \bibfield  {author} {\bibinfo {author} {\bibfnamefont {N.}~\bibnamefont
  {Tetradis}}\ and\ \bibinfo {author} {\bibfnamefont {C.}~\bibnamefont
  {Wetterich}},\ }\href {\doibase 10.1016/0550-3213(93)90608-R} {\bibfield
  {journal} {\bibinfo  {journal} {Nucl. Phys. B}\ }\textbf {\bibinfo {volume}
  {398}},\ \bibinfo {pages} {659 } (\bibinfo {year} {1993})}\BibitemShut
  {NoStop}%
\bibitem [{\citenamefont {Reuter}\ \emph {et~al.}(1993)\citenamefont {Reuter},
  \citenamefont {Tetradis},\ and\ \citenamefont {Wetterich}}]{Reuter93}%
  \BibitemOpen
  \bibfield  {author} {\bibinfo {author} {\bibfnamefont {M.}~\bibnamefont
  {Reuter}}, \bibinfo {author} {\bibfnamefont {N.}~\bibnamefont {Tetradis}}, \
  and\ \bibinfo {author} {\bibfnamefont {C.}~\bibnamefont {Wetterich}},\ }\href
  {\doibase 10.1016/0550-3213(93)90314-F} {\bibfield  {journal} {\bibinfo
  {journal} {Nucl. Phys. B}\ }\textbf {\bibinfo {volume} {401}},\ \bibinfo
  {pages} {567 } (\bibinfo {year} {1993})}\BibitemShut {NoStop}%
\bibitem [{not({\natexlab{g}})}]{note8}%
  \BibitemOpen
  \href@noop {} {} \bibinfo {note} {See Sec.~IV.A
  in~\cite{Rancon11b}.}\BibitemShut {Stop}%
\bibitem [{not({\natexlab{h}})}]{note9}%
  \BibitemOpen
  \href@noop {} {} \bibinfo {note} {Note that $I_{k,\rm
  t}=I_{k,\rm l}$ when $\rho_{0,k}=0$.}\BibitemShut {Stop}%
\bibitem [{not({\natexlab{i}})}]{note14}%
  \BibitemOpen
  \href@noop {} {} \bibinfo {note} {In practice, we use the
  dimensionless equations introduced in Sec.~\ref{subsec_rgflow}.}\BibitemShut
  {Stop}%
\bibitem [{\citenamefont {Pogorelov}\ and\ \citenamefont
  {Suslov}(2008)}]{Pogorelov08}%
  \BibitemOpen
  \bibfield  {author} {\bibinfo {author} {\bibfnamefont {A.~A.}\ \bibnamefont
  {Pogorelov}}\ and\ \bibinfo {author} {\bibfnamefont {I.~M.}\ \bibnamefont
  {Suslov}},\ }\href@noop {} {\bibfield  {journal} {\bibinfo  {journal} {Sov.
  Phys. JETP}\ }\textbf {\bibinfo {volume} {106}},\ \bibinfo {pages} {1118}
  (\bibinfo {year} {2008})}\BibitemShut {NoStop}%
\bibitem [{\citenamefont {Campostrini}\ \emph {et~al.}(2002)\citenamefont
  {Campostrini}, \citenamefont {Hasenbusch}, \citenamefont {Pelissetto},
  \citenamefont {Rossi},\ and\ \citenamefont {Vicari}}]{Campostrini02}%
  \BibitemOpen
  \bibfield  {author} {\bibinfo {author} {\bibfnamefont {M.}~\bibnamefont
  {Campostrini}}, \bibinfo {author} {\bibfnamefont {M.}~\bibnamefont
  {Hasenbusch}}, \bibinfo {author} {\bibfnamefont {A.}~\bibnamefont
  {Pelissetto}}, \bibinfo {author} {\bibfnamefont {P.}~\bibnamefont {Rossi}}, \
  and\ \bibinfo {author} {\bibfnamefont {E.}~\bibnamefont {Vicari}},\ }\href
  {\doibase 10.1103/PhysRevB.65.144520} {\bibfield  {journal} {\bibinfo
  {journal} {Phys. Rev. B}\ }\textbf {\bibinfo {volume} {65}},\ \bibinfo
  {pages} {144520} (\bibinfo {year} {2002})}\BibitemShut {NoStop}%
\bibitem [{\citenamefont {Campostrini}\ \emph {et~al.}(2006)\citenamefont
  {Campostrini}, \citenamefont {Hasenbusch}, \citenamefont {Pelissetto},\ and\
  \citenamefont {Vicari}}]{Campostrini06}%
  \BibitemOpen
  \bibfield  {author} {\bibinfo {author} {\bibfnamefont {M.}~\bibnamefont
  {Campostrini}}, \bibinfo {author} {\bibfnamefont {M.}~\bibnamefont
  {Hasenbusch}}, \bibinfo {author} {\bibfnamefont {A.}~\bibnamefont
  {Pelissetto}}, \ and\ \bibinfo {author} {\bibfnamefont {E.}~\bibnamefont
  {Vicari}},\ }\href {\doibase 10.1103/PhysRevB.74.144506} {\bibfield
  {journal} {\bibinfo  {journal} {Phys. Rev. B}\ }\textbf {\bibinfo {volume}
  {74}},\ \bibinfo {pages} {144506} (\bibinfo {year} {2006})}\BibitemShut
  {NoStop}%
\bibitem [{\citenamefont {Gr\"ater}\ and\ \citenamefont
  {Wetterich}(1995)}]{Graeter95}%
  \BibitemOpen
  \bibfield  {author} {\bibinfo {author} {\bibfnamefont {M.}~\bibnamefont
  {Gr\"ater}}\ and\ \bibinfo {author} {\bibfnamefont {C.}~\bibnamefont
  {Wetterich}},\ }\href {\doibase 10.1103/PhysRevLett.75.378} {\bibfield
  {journal} {\bibinfo  {journal} {Phys. Rev. Lett.}\ }\textbf {\bibinfo
  {volume} {75}},\ \bibinfo {pages} {378} (\bibinfo {year} {1995})}\BibitemShut
  {NoStop}%
\bibitem [{\citenamefont {Gersdorff}\ and\ \citenamefont
  {Wetterich}(2001)}]{Gersdorff01}%
  \BibitemOpen
  \bibfield  {author} {\bibinfo {author} {\bibfnamefont {G.~V.}\ \bibnamefont
  {Gersdorff}}\ and\ \bibinfo {author} {\bibfnamefont {C.}~\bibnamefont
  {Wetterich}},\ }\href {\doibase 10.1103/PhysRevB.64.054513} {\bibfield
  {journal} {\bibinfo  {journal} {Phys. Rev. B}\ }\textbf {\bibinfo {volume}
  {64}},\ \bibinfo {pages} {054513} (\bibinfo {year} {2001})}\BibitemShut
  {NoStop}%
\bibitem [{\citenamefont {Ran\c{c}on}\ and\ \citenamefont
  {Dupuis}(2012{\natexlab{b}})}]{Rancon12b}%
  \BibitemOpen
  \bibfield  {author} {\bibinfo {author} {\bibfnamefont {A.}~\bibnamefont
  {Ran\c{c}on}}\ and\ \bibinfo {author} {\bibfnamefont {N.}~\bibnamefont
  {Dupuis}},\ }\href {\doibase 10.1103/PhysRevA.85.063607} {\bibfield
  {journal} {\bibinfo  {journal} {Phys. Rev. A}\ }\textbf {\bibinfo {volume}
  {85}},\ \bibinfo {pages} {063607} (\bibinfo {year}
  {2012}{\natexlab{b}})}\BibitemShut {NoStop}%
\bibitem [{\citenamefont {Neto}\ and\ \citenamefont {Fradkin}(1993)}]{Neto93}%
  \BibitemOpen
  \bibfield  {author} {\bibinfo {author} {\bibfnamefont {A.~C.}\ \bibnamefont
  {Neto}}\ and\ \bibinfo {author} {\bibfnamefont {E.}~\bibnamefont {Fradkin}},\
  }\href {\doibase 10.1016/0550-3213(93)90414-K} {\bibfield  {journal}
  {\bibinfo  {journal} {Nucl. Phys. B}\ }\textbf {\bibinfo {volume} {400}},\
  \bibinfo {pages} {525 } (\bibinfo {year} {1993})}\BibitemShut {NoStop}%
\bibitem [{not({\natexlab{j}})}]{note11}%
  \BibitemOpen
  \href@noop {} {} \bibinfo {note} {The argument leading to
  Eq.~(\ref{Prc}) assumes that the velocity in the renormalized classical
  regime is the same as at the QCP (see Sec.~\ref{subsubsec_QC}).}\BibitemShut
  {Stop}%
\bibitem [{\citenamefont {Hofmann}(2013)}]{Hofmann13}%
  \BibitemOpen
  \bibfield  {author} {\bibinfo {author} {\bibfnamefont {C.~P.}\ \bibnamefont
  {Hofmann}},\ }\href@noop {} \Eprint {http://arxiv.org/abs/arXiv:1306.1944}
  {arXiv:1306.1944} \BibitemShut {NoStop}%
\bibitem [{\citenamefont {Hofmann}(2010)}]{Hofmann10}%
  \BibitemOpen
  \bibfield  {author} {\bibinfo {author} {\bibfnamefont {C.~P.}\ \bibnamefont
  {Hofmann}},\ }\href {\doibase 10.1103/PhysRevB.81.014416} {\bibfield
  {journal} {\bibinfo  {journal} {Phys. Rev. B}\ }\textbf {\bibinfo {volume}
  {81}},\ \bibinfo {pages} {014416} (\bibinfo {year} {2010})}\BibitemShut
  {NoStop}%
\bibitem [{not({\natexlab{k}})}]{note16}%
  \BibitemOpen
  \href@noop {} {} \bibinfo {note} {While a calculation of
  $\bar c_k$ is difficult, as it requires one to analytically continue
  $\Gamma_{k,\rm t}^{(2)}(q;\rho_{0,k})$ to real frequencies, we note that
  $c_k=\sqrt{Z_{A,k}/V_{A,k}}$ varies only weakly when $k$ becomes of the order
  of $k_T$ (only for $k\ll k_T$ does $c_k$ significantly differ from $c_0$). If
  we take $c_{k\sim k_T}$ as an estimate of the renormalized value $\bar
  c_{k=0}$ of the velocity, we conclude that the latter differs only slightly
  from $c_0$ in the quantum critical regime. While there is no doubt that $\bar
  c_{k=0} \simeq c_0$ in the quantum disordered regime, the agreement of
  Eq.~(\ref{Prc}) with the numerical solution of the flow equations shows that
  this conclusion also holds in the renormalized classical regime (see also
  Refs.~\cite{Hofmann10,Hofmann13}).}\BibitemShut {Stop}%
\bibitem [{not({\natexlab{l}})}]{note7}%
  \BibitemOpen
  \href@noop {} {} \bibinfo {note} {Although $\eta_k$ and
  $\teta_k$ are very small in the Goldstone regime, the term
  $(\eta_k+\teta_k)\trho_{0,k}$ is of order 1 and must be kept. It plays a
  crucial role in the derivation of the quantum NL$\sigma$M
  (Sec.~\ref{subsec_nlsm}).}\BibitemShut {Stop}%
\bibitem [{\citenamefont {Delamotte}\ \emph {et~al.}(2004)\citenamefont
  {Delamotte}, \citenamefont {Mouhanna},\ and\ \citenamefont
  {Tissier}}]{Delamotte04}%
  \BibitemOpen
  \bibfield  {author} {\bibinfo {author} {\bibfnamefont {B.}~\bibnamefont
  {Delamotte}}, \bibinfo {author} {\bibfnamefont {D.}~\bibnamefont {Mouhanna}},
  \ and\ \bibinfo {author} {\bibfnamefont {M.}~\bibnamefont {Tissier}},\ }\href
  {\doibase 10.1103/PhysRevB.69.134413} {\bibfield  {journal} {\bibinfo
  {journal} {Phys. Rev. B}\ }\textbf {\bibinfo {volume} {69}},\ \bibinfo
  {pages} {134413} (\bibinfo {year} {2004})}\BibitemShut {NoStop}%
\bibitem [{not({\natexlab{m}})}]{note5}%
  \BibitemOpen
  \href@noop {} {} \bibinfo {note} {At finite temperature,
  the quantification of the Matsubara frequencies $\wn=2\pi Tn$ makes the theta
  cutoff~(\ref{thetacut}) ill-suited, except in the regime $2\pi T>c_kk$ where
  only classical fluctuations ($\wn=0$) contribute to the flow. Here we use the
  theta cutoff only for illustrative purpose in the $T=0$ and $T\gg c_kk$
  limits.}\BibitemShut {Stop}%
\bibitem [{\citenamefont {Nelson}\ and\ \citenamefont
  {Pelcovits}(1977)}]{Nelson77}%
  \BibitemOpen
  \bibfield  {author} {\bibinfo {author} {\bibfnamefont {D.~R.}\ \bibnamefont
  {Nelson}}\ and\ \bibinfo {author} {\bibfnamefont {R.~A.}\ \bibnamefont
  {Pelcovits}},\ }\href {\doibase 10.1103/PhysRevB.16.2191} {\bibfield
  {journal} {\bibinfo  {journal} {Phys. Rev. B}\ }\textbf {\bibinfo {volume}
  {16}},\ \bibinfo {pages} {2191} (\bibinfo {year} {1977})}\BibitemShut
  {NoStop}%
\bibitem [{\citenamefont {Polyakov}(1975)}]{Polyakov75}%
  \BibitemOpen
  \bibfield  {author} {\bibinfo {author} {\bibfnamefont {A.~M.}\ \bibnamefont
  {Polyakov}},\ }\href {\doibase
  http://dx.doi.org/10.1016/0370-2693(75)90161-6} {\bibfield  {journal}
  {\bibinfo  {journal} {Phys. Lett. B}\ }\textbf {\bibinfo {volume} {59}},\
  \bibinfo {pages} {79} (\bibinfo {year} {1975})}\BibitemShut {NoStop}%
\bibitem [{\citenamefont {Friedan}(1985)}]{Friedan85}%
  \BibitemOpen
  \bibfield  {author} {\bibinfo {author} {\bibfnamefont {D.~H.}\ \bibnamefont
  {Friedan}},\ }\href {\doibase http://dx.doi.org/10.1016/0003-4916(85)90384-7}
  {\bibfield  {journal} {\bibinfo  {journal} {Ann. Phys. (N.Y.)}\ }\textbf
  {\bibinfo {volume} {163}},\ \bibinfo {pages} {318} (\bibinfo {year}
  {1985})}\BibitemShut {NoStop}%
\bibitem [{\citenamefont {Machado}\ and\ \citenamefont
  {Dupuis}(2010)}]{Machado10}%
  \BibitemOpen
  \bibfield  {author} {\bibinfo {author} {\bibfnamefont {T.}~\bibnamefont
  {Machado}}\ and\ \bibinfo {author} {\bibfnamefont {N.}~\bibnamefont
  {Dupuis}},\ }\href {\doibase 10.1103/PhysRevE.82.041128} {\bibfield
  {journal} {\bibinfo  {journal} {Phys. Rev. E}\ }\textbf {\bibinfo {volume}
  {82}},\ \bibinfo {pages} {041128} (\bibinfo {year} {2010})}\BibitemShut
  {NoStop}%
\bibitem [{\citenamefont {Prokof'ev}\ \emph {et~al.}(2001)\citenamefont
  {Prokof'ev}, \citenamefont {Ruebenacker},\ and\ \citenamefont
  {Svistunov}}]{Prokofev01}%
  \BibitemOpen
  \bibfield  {author} {\bibinfo {author} {\bibfnamefont {N.}~\bibnamefont
  {Prokof'ev}}, \bibinfo {author} {\bibfnamefont {O.}~\bibnamefont
  {Ruebenacker}}, \ and\ \bibinfo {author} {\bibfnamefont {B.}~\bibnamefont
  {Svistunov}},\ }\href {\doibase 10.1103/PhysRevLett.87.270402} {\bibfield
  {journal} {\bibinfo  {journal} {Phys. Rev. Lett.}\ }\textbf {\bibinfo
  {volume} {87}},\ \bibinfo {pages} {270402} (\bibinfo {year}
  {2001})}\BibitemShut {NoStop}%
\bibitem [{\citenamefont {Prokof'ev}\ and\ \citenamefont
  {Svistunov}(2002)}]{Prokofev02}%
  \BibitemOpen
  \bibfield  {author} {\bibinfo {author} {\bibfnamefont {N.}~\bibnamefont
  {Prokof'ev}}\ and\ \bibinfo {author} {\bibfnamefont {B.}~\bibnamefont
  {Svistunov}},\ }\href {\doibase 10.1103/PhysRevA.66.043608} {\bibfield
  {journal} {\bibinfo  {journal} {Phys. Rev. A}\ }\textbf {\bibinfo {volume}
  {66}},\ \bibinfo {pages} {043608} (\bibinfo {year} {2002})}\BibitemShut
  {NoStop}%
\bibitem [{\citenamefont {Nelson}\ and\ \citenamefont
  {Kosterlitz}(1977)}]{Nelson77a}%
  \BibitemOpen
  \bibfield  {author} {\bibinfo {author} {\bibfnamefont {D.~R.}\ \bibnamefont
  {Nelson}}\ and\ \bibinfo {author} {\bibfnamefont {J.~M.}\ \bibnamefont
  {Kosterlitz}},\ }\href {\doibase 10.1103/PhysRevLett.39.1201} {\bibfield
  {journal} {\bibinfo  {journal} {Phys. Rev. Lett.}\ }\textbf {\bibinfo
  {volume} {39}},\ \bibinfo {pages} {1201} (\bibinfo {year}
  {1977})}\BibitemShut {NoStop}%
\bibitem [{not({\natexlab{n}})}]{note13}%
  \BibitemOpen
  \href@noop {} {} \bibinfo {note} {The NPRG approach
  (within the standard approximations used to solve the flow equations) does
  not allow us to obtain a reliable estimate of $\rho_s(\Tkt^-)$ and therefore
  the ratio $\Tkt/\rho_s(\Tkt^-)$.}\BibitemShut {Stop}%
\bibitem [{\citenamefont {Zhang}\ \emph {et~al.}(2012)\citenamefont {Zhang},
  \citenamefont {Hung}, \citenamefont {Tung},\ and\ \citenamefont
  {Chin}}]{Zhang12}%
  \BibitemOpen
  \bibfield  {author} {\bibinfo {author} {\bibfnamefont {X.}~\bibnamefont
  {Zhang}}, \bibinfo {author} {\bibfnamefont {C.-L.}\ \bibnamefont {Hung}},
  \bibinfo {author} {\bibfnamefont {S.-K.}\ \bibnamefont {Tung}}, \ and\
  \bibinfo {author} {\bibfnamefont {C.}~\bibnamefont {Chin}},\ }\href {\doibase
  10.1126/science.1217990} {\bibfield  {journal} {\bibinfo  {journal}
  {Science}\ }\textbf {\bibinfo {volume} {335}},\ \bibinfo {pages} {1070}
  (\bibinfo {year} {2012})}\BibitemShut {NoStop}%
\bibitem [{\citenamefont {Chamati}\ and\ \citenamefont
  {Tonchev}(2011)}]{Chamati11}%
  \BibitemOpen
  \bibfield  {author} {\bibinfo {author} {\bibfnamefont {H.}~\bibnamefont
  {Chamati}}\ and\ \bibinfo {author} {\bibfnamefont {N.~S.}\ \bibnamefont
  {Tonchev}},\ }\href {\doibase doi:10.1209/0295-5075/95/40005} {\bibfield
  {journal} {\bibinfo  {journal} {Europhys. Lett.}\ }\textbf {\bibinfo {volume}
  {95}},\ \bibinfo {pages} {40005} (\bibinfo {year} {2011})}\BibitemShut
  {NoStop}%
\end{thebibliography}

%

\end{document}